\documentclass[]{aastex631}
\usepackage{graphicx}

\shorttitle{The HII-Regions' Molecular Law of Star Formation}
\shortauthors{Calzetti et al.}

\graphicspath{{./}{figures/}}

\begin{document}

\title{The HII Regions' Molecular Law of Star Formation}

\author[0000-0002-5189-8004]{Daniela Calzetti}
\affiliation{Department of Astronomy, University of Massachusetts Amherst, 710 North Pleasant Street, Amherst, MA 01003, USA}

\author[0009-0009-5509-4706]{Drew Lapeer}
\affiliation{Department of Astronomy, University of Massachusetts Amherst, 710 North Pleasant Street, Amherst, MA 01003, USA}

\author[0000-0001-5448-1821]{Robert C. Kennicutt}
\affiliation{Department of Physics and Astronomy, Texas A\&M University, 578 University Drive, College Station, TX 77843-4242, USA}
 \affiliation{Steward Observatory, University of Arizona, 933 N Cherry Avenue, Tucson, AZ 85721, USA}
 
 \author[0000-0002-1723-6330]{Bruce Elmegreen}
\affiliation{Katonah, NY 10536, USA}

\author[0000-0002-1000-6081]{Sean T. Linden}
\affiliation{Steward Observatory, University of Arizona, 933 N Cherry Avenue, Tucson, AZ 85721, USA}

\author[0000-0003-3893-854X]{Mark R. Krumholz}
\affiliation{Research School of Astronomy and Astrophysics, Australian National University, 233 Mount Stromlo Road, Stromlo ACT 2611, Australia}

\author[0000-0002-8192-8091]{Angela Adamo}
\affiliation{Department of Astronomy, The Oskar Klein Centre, Stockholm University, AlbaNova, SE-10691 Stockholm, Sweden}

\author[0000-0002-5782-9093]{Daniel~A.~Dale}
\affiliation{Department of Physics and Astronomy, University of Wyoming, Laramie, WY 82071, USA}

\author[0000-0002-4378-8534]{Karin Sandstrom}
\affiliation{Department of Astronomy \& Astrophysics, University of California, San Diego, 9500 Gilman Drive, La Jolla, CA 92093}

\author[0009-0003-6182-8928]{Giacomo Bortolini}
\affiliation{Department of Astronomy, The Oskar Klein Centre, Stockholm University, AlbaNova, SE-10691 Stockholm, Sweden}

\author[0000-0001-6291-6813]{Michele Cignoni}
\affiliation{Department of Physics - University of Pisa, Largo B. Pontecorvo 3, 56127 Pisa, Italy }
\affiliation{INFN, Largo B. Pontecorvo 3, 56127 Pisa, Italy }
\affiliation{INAF - Osservatorio di Astrofisica e Scienza dello Spazio di Bologna, Via Gobetti 93/3, I-40129 Bologna, Italy }

\author[0000-0001-6464-3257]{Matteo Correnti}
\affiliation{INAF Osservatorio Astronomico di Roma, Via Frascati 33, 00078, Monteporzio Catone, Rome, Italy}
\affiliation{ASI-Space Science Data Center, Via del Politecnico, I-00133, Rome, Italy}

\author[0000-0002-5259-4774]{Ana Duarte-Cabral}
\affiliation{Cardiff Hub for Astrophysics Research and Technology (CHART), School of Physics \& Astronomy, Cardiff University, The Parade, CF24 3AA Cardiff, UK}

\author[0000-0002-2199-0977]{Helena Faustino Vieira}
\affiliation{Department of Astronomy, The Oskar Klein Centre, Stockholm University, AlbaNova, SE-10691 Stockholm, Sweden}

\author[0000-0001-8608-0408]{John S. Gallagher}
\affiliation{Department of Astronomy, University of Wisconsin--Madison, 475 N. Charles Street, Madison, WI 53706--1507 USA}
\affiliation{Department of Physics and Astronomy, Macalester University, 1600 Grand Avenue, Saint Paul, MN 55105-1899 USA}

\author[0000-0002-3247-5321]{Kathryn~Grasha}
\altaffiliation{ARC DECRA Fellow}
\affiliation{Research School of Astronomy and Astrophysics, Australian National University, Canberra, ACT 2611, Australia}   
\affiliation{ARC Centre of Excellence for All Sky Astrophysics in 3 Dimensions (ASTRO 3D), Australia}   

\author[0000-0002-3871-010X]{Mark~Heyer}
\affiliation{Department of Astronomy, University of Massachusetts Amherst, 710 North Pleasant Street, Amherst, MA 01003, USA}

\author[0000-0001-9162-2371]{Leslie~K. Hunt}
\affiliation{INAF -- Osservatorio Astrofisico di Arcetri, Largo E. Fermi 5, 50125 Firenze, Italy}

\author[0000-0001-8348-2671]{Kelsey E. Johnson}
\affiliation{Department of Astronomy, University of Virginia, Charlottesville, VA, USA}

\author[0000-0002-0560-3172]{Ralf S.\ Klessen}
\affiliation{Universit\"{a}t Heidelberg, Zentrum f\"{u}r Astronomie, Institut f\"{u}r Theoretische Astrophysik, Albert-Ueberle-Str.\ 2, 69120 Heidelberg, Germany}
\affiliation{Universit\"{a}t Heidelberg, Interdisziplin\"{a}res Zentrum f\"{u}r Wissenschaftliches Rechnen, Im Neuenheimer Feld 225, 69120 Heidelberg, Germany}

\author[0000-0001-8490-6632]{Thomas S.-Y. Lai}
\affiliation{IPAC, California Institute of Technology, 1200 East California Boulevard, Pasadena, CA 91125, USA}

\author[0000-0002-7064-4309]{Desika Narayanan}
\affiliation{Department of Astronomy, University of Florida, 211 Bryant Space Science Center, Gainesville, FL 32611, USA}

\author[0000-0002-3005-1349]{G\"{o}ran \"{Os}tlin}
\affiliation{Department of Astronomy, The Oskar Klein Centre, Stockholm University, AlbaNova, SE-10691 Stockholm, Sweden}

\author[0000-0002-8222-8986]{Alex Pedrini}
\affiliation{Department of Astronomy, The Oskar Klein Centre, Stockholm University, AlbaNova, SE-10691 Stockholm, Sweden}

\author[0000-0003-2954-7643]{Elena Sabbi}
\affiliation{Gemini Observatory, NOIRLab, 950 N. Cherry Ave., Tucson, AZ 85719, USA }

\author[0000-0003-1545-5078]{John-David T. Smith}
\affiliation{Ritter Astrophysical Research Center, University of Toledo, Toledo, OH 43606, USA}

\author[0000-0002-0806-168X]{Linda J. Smith}
\affiliation{Space Telescope  Science Institute, 3700 San Martin Drive, Baltimore, MD 21218, USA}

\author[0000-0002-0986-4759]{Monica Tosi}
\affiliation{INAF - Osservatorio di Astrofisica e Scienza dello Spazio di Bologna, Via Gobetti 93/3, I-40129 Bologna, Italy }

\author[0009-0008-4009-3391]{Varun Bajaj}
\affiliation{Space Telescope Science Institute, 3700 San Martin Drive, Baltimore, MD 21218, USA}

\author[0000-0003-4850-9589]{Martha Boyer}
\affiliation{Space Telescope  Science Institute, 3700 San Martin Drive, Baltimore, MD 21218, USA}

\author[0009-0005-8923-558X]{Tony D. Weinbeck}
\affiliation{Department of Physics and Astronomy, University of Wyoming, Laramie, WY 82071, USA}

\begin{abstract}
We combine imaging data from the HST, JWST, and ground--based millimeter facilities to investigate the correlation between star formation rate (SFR) and molecular gas at the $\sim$100~pc scale of HII regions in three nearby galaxies: NGC628, NGC5194 and NGC5236. The JWST 21~$\mu$m maps of the three galaxies offer a unique insight into the dust--absorbed SFR at high resolution. We find that the relation between the surface densities of SFR and molecular gas has a slope of $\sim$1.85, in log--log scale, significantly steeper than previous results for nearby galaxies but closer to the trends found for molecular clouds in the Milky Way. The steep relation also holds on larger, $\sim$500~pc, scales, and results from the high--resolution imaging that cleanly isolates the star--forming region emission from the underlying galaxy's diffuse contribution.  The diffuse emission at 21~$\mu$m is, in fact, found to correlate with the galaxy's stellar mass. Comparisons with physical models of star formation are inconclusive; they overlap with the locus of the 100~pc data, but have difficulties in reproducing the data scatter. Possible exceptions are models that add a power law tail to the gas density probability distribution, due to the large range of free parameters allowed. We find that local HII regions, high redshift star--forming clumps, and low and high redshift starburst galaxies form a single sequence of star formation over three orders of magnitude in gas surface density.  
 \end{abstract}


\section{Introduction} \label{sec:intro}
Star formation (SF) laws
are the `standard rod' against which models and simulations of the SF  process, and the physics they implement, are tested \citep[e.g.,][]{Li+2006, Hopkins+2011, Feldmann+2011, Elmegreen+2015, Salim+2015,  Dib+2017, Krumholz+2018, Sormani+2020, Grudic+2022, Ostriker+2022, Whitworth+2022, Hassan+2024, Khullar+2024}. 

At the basic level, star formation 
results from the gravitational collapse and cooling of gas within galaxies 
modulated by magnetic fields and turbulence from accretion and stellar feedback (e.g., stellar winds, photoionization, radiation pressure and supernovae) 
which reduce the efficiency of gas conversion into stars \citep[e.g.,][]{Elmegreen+2002, Krumholz+2005, Hopkins+2014, Federrath+2015, Salim+2015, Krumholz+2019, Grudic+2019, Grudic+2022}. The feedback from stars regulates star formation either at the local level, by acting directly on the clouds to suppress future star formation \citep[e.g.,][]{Krumholz+2005, Dobbs+2011, Hopkins+2011, Dale+2011, Dib+2011b,  Federrath+2015, Krumholz+2019, Grisdale+2017, Grudic+2018}, or at the global level, by maintaining galaxies in a "quasi--steady" state, thus setting the disk scale height and the collapse rate on kpc scales \citep[e.g.,][]{Ostriker+2010, Kim+2011, Faucher+2013, Ostriker+2022}. In models and simulations, feedback is necessary to regulate star formation and keep its efficiency low, at the level of a few percent \citep{Ostriker+2010, Hopkins+2014, Orr+2018, Bending+2020, Ostriker+2022}, although it has been argued that the main role of feedback is not to regulate star formation but to halt it, while still keeping the efficiency low \citep{Ballesteros+2024}. 

Following \citet{Kennicutt+1989} and \citet{Kennicutt+1998}, the SF laws of galaxies have been expressed as:
\begin{equation}
\Sigma_{SFR} \propto \Sigma_{gas}^{N},
\label{totallaw}
\end{equation}
where $\Sigma_{SFR}$ and $\Sigma_{gas}$ are the surface densities of star formation rate (SFR) and gas, respectively, and $N\sim$1--2 is generally referred to as the `slope'.
The expression involving surface densities is an `observer--friendly' evolution of the original volumetric expression of \citet{Schmidt+1959} and \citet{Schmidt+1963}, with the two expressions being roughly identical if either $N \sim$1 or the scale--height of the star--forming disk is constant and small relative to the extent of the disk itself \citep{Kennicutt+1989}. As instruments' sensitivity and angular resolution have increased over time, studies have moved from measuring the SF laws of whole galaxies to measuring those of regions within galaxies. Within the central 5--10~kpc of galaxy disks, gas is sufficiently dense that H$_2$ surface density dominates over that of atomic gas \citep[e.g.,][]{Wong+2002, Bigiel+2008} and the expression of the SF laws in spatially--resolved studies is often expressed as:
\begin{equation}
\Sigma_{SFR} \propto \Sigma_{mol}^{n},
\label{mollaw}
\end{equation}
with $\Sigma_{mol}$ the surface density of molecular gas. In equation~\ref{mollaw} the slope is indicated with $n$ to distinguish it from $N$ in the total gas SF law of equation~\ref{totallaw}. Henceforth, we mainly review the properties of the `molecular gas' SF laws, since this paper analyzes regions located within the inner 3.5~kpc radius of three local galaxies (see below) where molecular gas dominates by a factors $>$3-10 over atomic gas \citep{Crosthwaite+2002, Bigiel+2008}.  For convenience, in what follows we refer to the molecular SF law as simply the SF law, although deviations from this convention are implemented  when needed for clarity. 

When averaged over entire disk galaxies \citep[e.g.,][]{Kennicutt+1998, Kennicutt+2012, Liu+2015, delosReyes+2019}, the SF law has a slope $n\sim 1$ across a large range of properties, which has led to the suggestion of constant depletion times for the molecular gas $\tau_{dep}$=$\Sigma_{mol}$/$\Sigma_{SFR}\approx$10$^9$~yr \citep[e.g.,][]{Leroy+2008}, albeit with significant scatter from galaxy to galaxy \citep[e.g.,][]{Shetty+2016}. Infrared--bright starburst galaxies also show a linear relation between  $\Sigma_{SFR}$ and $\Sigma_{mol}$, but with a significantly smaller value of $\tau_{dep}$, by factors $\sim$5--8, than that of star--forming disks \citep{Liu+2015, Wilson+2019, Kennicutt+2021}. The discrepancy in $\tau_{dep}$ between star--forming disks and starburst galaxies persists at high redshift, although there are uncertainties in deriving molecular gas masses from different tracers and in comparing them to one another \citep{Daddi+2010, Genzel+2010, Sharon+2013, Hodge+2015, Sharda+2018, Sharon+2019, Dessauges+2025, Accard+2025}. Taken together, galaxies show a smooth anti--correlation between the $\tau_{dep}$ and both SFR and specific SFR (sSFR=SFR/M$_{stellar}$), suggesting that a single mechanism underlies the SF process scaling up from normal star--forming galaxies to starbursts \citep{Saintonge+2011, Kennicutt+2021}.  In fact, the discrepancy disappears when H$_2$ is measured using a high--density gas tracer (e.g., HCN); in this case star--forming disks and starbursts follow a single relation with slope of unity, suggesting that dense gas is more closely related to star formation than its lower density counterpart \citep{Gao+2004}.

At the level of  $\sim$kpc--sized galaxy regions, the SF law of normal star--forming galaxies has slopes in the range $n\sim 0.9-1.3$ \citep[e.g.,][]{Kennicutt+2007, Bigiel+2008, Liu+2011, Bigiel+2011, Momose+2013, Utomo+2018, Chevance+2022, Sun+2023, Leroy+2025}, with a scatter that decreases for increasing spatial scale \citep[e.g.,][]{Onodera+2010}. The near--unity value of the slope $n$  at these resolved scales supports, like in the case of whole galaxies analyses, a universal $\tau_{dep}$ for normal disks \citep[e.g.,][]{Leroy+2013}, although \citet{Shetty+2013} find mostly $n<1$ when applying Bayesian analysis to the data of \citet{Bigiel+2008}, thus evidence for non--constant $\tau_{dep}$. The efficiency of star formation per unit free--fall time is also roughly constant, $\approx$1\%, or slightly decreasing for increasing $\Sigma_{mol}$ and gas velocity dispersion \citep{Zuckerman+1974, Elmegreen+1977, Myers+1986, Lada+1987, Krumholz+2012, Shetty+2014a, Shetty+2014b, Utomo+2018, Leroy+2025, Meidt+2025}. In the molecular--gas--poor M\,33 galaxy, \citet{Corbelli+2025}, however, find that the mid--plane hydrostatic pressure, rather than $\Sigma_{mol}$, is the main driver of SFR and that the observed scatter in the relation between the two quantities is due to variations in $\tau_{dep}$. Notable is that the linearity of the resolved molecular SF law depends on the treatment of the diffuse galaxy's emission underlying the regions, and whether this diffuse emission should or should not be included in the SFR budget; when removed, the SF law steepens, yielding $n\sim$1.5--1.8 \citep{Liu+2011, Rahman+2011, Momose+2013, Morokuma+2017, Kumari+2020}. 

In the Milky Way (MW), the high spatial resolution that can be achieved enables more nuanced analyses that, however, uncover a complicated landscape. Resolved molecular cloud analyses find SF laws that have slopes $n\sim1.6 - 2$ or steeper, but with a large scatter \citep[e.g.][]{Evans+2009, Heiderman+2010, Gutermuth+2011, Lada+2013, Willis+2015, Hony+2015, Nguyen--Luong+2016, Retes--Romero+2017, Lada+2017, Pokhrel+2020, Rawat+2025}. \citet{Heiderman+2010} breaks the relation into two regimes, separated by the threshold value Log($\Sigma_{mol} [M_{\odot}\ pc^{-2}])$=2.1: a low--density regime with slope $n\sim$4.6 and a high--density regime with $n\sim1.1$. When zooming into dense clumps, \citet{Rawat+2025} and \citet{Elia+2025} find slopes $n\sim 1.1 - 1.5$. Efficiencies per free--fall time either vary by more than one order of magnitude \citep{Evans+2014, Heyer+2016, Lee+2016} or by less than factors of 2 \citep{Pokhrel+2021, Hu+2022}, depending on the clouds' sample size, the tracers of SFR and gas density used, and the  treatment of the cloud geometry and filling factor \citep[e.g.][]{Ballesteros+2024}. In addition to a large scatter, \citet{Lee+2016}, \citet{Ochsendorf+2017} and \citet{Zhou+2025} find that the efficiencies per free--fall time decrease with increasing cloud {\em mass}, in agreement with theoretical expectations \citep{Dib+2011a, Dib+2011b}. The model of \citet{Dib+2011a} further predicts an anticorrelation between the efficiency and $\Sigma_{mol}$. For extragalactic HII regions (scales $\sim$ 150~pc), \citet{Pathak+2025} show a scatter by more than one order of magnitude in the correlation between SFR (which they express as total mass of newly formed stars) and molecular gas; however, these authors use the Balmer decrement to derive dust attenuation corrections for the newly formed stars, which can result in underestimates of the SFRs and of the young star masses in regions of high dust optical depth. 

In this paper, we leverage the combination of multiwavelength observations from HST and JWST with high--angular resolution CO maps for three nearby galaxies, NGC\,628, NGC\,5194 (M\,51a), and NGC\,5236 (M\,83), to derive the SF law of star forming regions at $\sim$100~pc scale. Our objective is to investigate whether HII regions follow the same trends identified for normal star--forming galaxies at the kpc--resolved scale ($n\sim$1) or they more closely resemble MW molecular clouds ($n\sim$2). Our ultimate goal is to understand whether the 100--pc scale star--forming regions  can be considered `scaled--down' versions of the intensely star--forming clumps and starburst galaxies observed both at low and high redshift. 

We utilize a recent calibration of the SFR indicator that combines the hydrogen recombination emission line H$\alpha$ with the 21~$\mu$m dust emission at the HII region scale \citep{Calzetti+2024, Calzetti+2025}; this indicator effectively probes both the dust--obscured and unobscured star formation, via the mid--IR and optical tracers, respectively. As shown in these and previous works \citep[e.g.,][]{Buat+1996, Li+2013, Calzetti+2013, Boquien+2014, Boquien+2016}, SFR calibrations depend on the star formation history (SFH) of the stellar populations in the region or galaxy. \citet{Calzetti+2025} find that the mid--IR SFR calibration varies by up to a factor 4, with younger (tens of Myr old) populations requiring larger scaling factors to convert their dust luminosity to SFR than older (Gyr--old) populations. As we are interested in the SF properties of HII regions, we closely reproduce the approach to photometry utilized by \citet{Calzetti+2025} and remove the diffuse light of the underlying galaxy from the emission of the star--forming regions and their molecular gas. 

The paper is organized as follows: section~\ref{sec:data} presents the observational data used in this paper; sections~\ref{sec:selection} and \ref{sec:physical} describe how the sources are selected and their photometric and physical quantities derived; section~\ref{sec:analysis} presents the analysis and results; comparisons with previous results, including high--redshift galaxies, as well as with models are given in section~\ref{sec:discussion}. The conclusions from this study are in section~\ref{sec:conclusions}. The basic parameters adopted for the three galaxies in this study are listed in Table~\ref{tab:properties}. 

\begin{deluxetable*}{llrrrc}
\tablecaption{Properties of the Galaxies\label{tab:properties}}
\tablewidth{0pt}
\tablehead{
\colhead{Parameter} & \colhead{Units} & \colhead{NGC\,628$^1$} &\colhead{NGC\,5194$^1$} &\colhead{NGC\,5236} &\colhead{References$^2$}
}
\decimalcolnumbers
\startdata
Distance & Mpc & 9.3 & 7.55 & 4.50 & (a) \\
Redshift &          & 0.00219  & 0.001745   & 0.001711  &  (b)  \\
Inclination & degrees & 9 & 22 & 24 & (c) \\
R$_{25}$ & arcsec (kpc) & 314.2 (14.15) & 336.6 (12.32) & 386.4 (8.43) & (d) \\
E(B--V)$_{MW}^3$ & mag & 0.06 & 0.031 & 0.057  & (e)  \\
M$_{star}$ & M$_{\odot}$ & 9.9$\times$10$^9$ & 2.3$\times$10$^{10}$ & 2.1$\times$10$^{10}$ & (f) \\
SFR & M$_{\odot}$~yr$^{-1}$ & 3.2  & 6.7 & 5.1 & (f) \\
12+Log(O/H)$^4$ & & 8.71 & 8.75  & 8.90 & (g) \\
Gradient$^5$ & R$_{25}$ & $-$0.40 & $-$0.27 & $-$0.37& (h) \\
M$_{dust}$/M$_{star}$ &    & 4.2$\times$10$^{-3}$  & 4.5$\times$10$^{-3}$ & 2.0$\times$10$^{-3}$ & (f) \\
\enddata
$^1$ The parameters for NGC\,628 and NGC\,5194 are the same reported in \citet{Calzetti+2024} and \citet{Calzetti+2025}, respectively. We refer the reader 
to those works for the relevant references.\\
$^2$  References for the NGC\, 5236 parameters:  (a) \citet{Thim+2003}, using Cepheids;  (b) NED, the NASA Extragalactic Database. (c) \citet{Leroy+2021}; (d) \citet{deVaucouleurs+1991}; (e) \citet{Schlafly+2011}; (f) \citet{Dale+2023}; (g) \citet{Bresolin+2016}; (h) \citet{Hernandez+2018}.\\
$^3$  Foreground Milky Way (MW) extinction.\\
$^4$ Central oxygen abundance. We adopt a solar oxygen abundance of 12+Log(O/H)=8.69, \citet{Asplund+2009}.\\
$^5$ Metallicity gradient as a function of galactocentric radius in units of R$_{25}$.\\
\end{deluxetable*}

\section{Imaging Data and Processing} \label{sec:data}

The JWST and HST data retrieval and handling, as well as the derivation of the Pa$\alpha$($\lambda$1.8756~$\mu$m) and H$\alpha$($\lambda$0.6563~$\mu$m)  mosaics for NGC\,628 and NGC\,5194 have been presented in \citet{Gregg+2024}, \citet{Calzetti+2024}, \citet{Gregg+2025} and \citet{Calzetti+2025}, and we refer the readers to those papers for details. The 21~$\mu$m images of NGC\,628 and NGC\,5194 have also been presented in \citet{Calzetti+2024} and \citet{Calzetti+2025}; in particular, the MIRI 21~$\mu$m map of NGC\,5194 is from the JWST Cycle 2 Treasury program \# 3435 (The JWST Whirpool Galaxy Treasury, P.Is.: K. Sandstrom \& D. Dale). In this study, we add to those datasets the JWST/NIRCam F277W and/or F300M images, as available, to derive stellar masses. These images are from the Cycle 1 program \# 1783 (Feedback in Emerging extrAgalactic Star clusTers, JWST--FEAST, P.I.: A. Adamo), and have been processed and calibrated as described in \citet{Gregg+2024} and \citet{Calzetti+2025}. All HST and JWST imaging data used in this work have astrometry aligned to Gaia EDR3 \citep{Gaia+2021}. 

The JWST/NIRCam mosaics of NGC\,5236 are also part  of the Cycle 1 program \# 1783. For this analysis, we use the NIRCAM/F150W, F187N, F200W to derive stellar continuum subtracted images in the light of the Pa$\alpha$ and the F300M to derive stellar masses. The process for deriving stellar continuum--subtracted Pa$\alpha$ images are described in \citet{Gregg+2025} and \citet{Knutas+2025}. 

\begin{figure}
\plotone{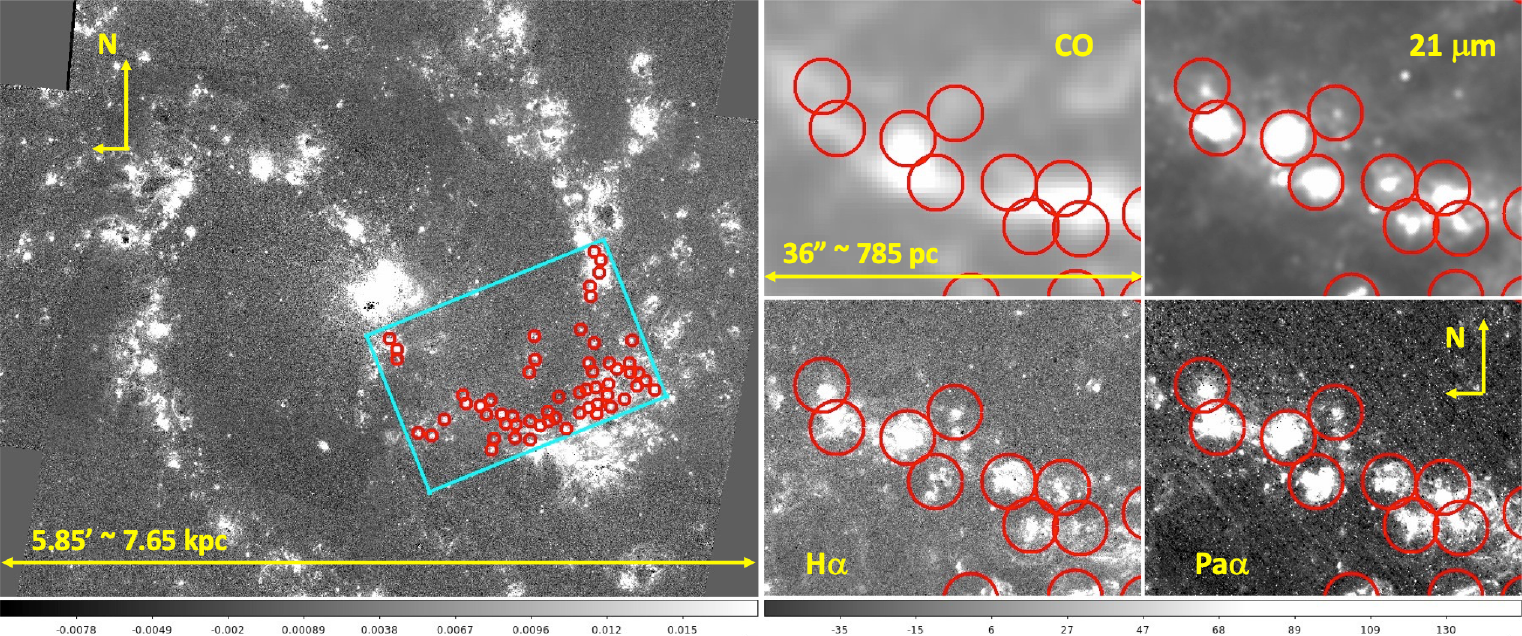}
\caption{{\bf (Left):} The stellar continuum--subtracted HST/WFC3/H$\alpha$ image of NGC\,5236, showing the footprint of the JWST/MIRI/21~$\mu$m image (cyan rectangle) with the 56 selected regions (red circles). The circles have the same radius as those used for the photometric measurements (2\farcs6).  North is up, East is left. {\bf (Right):} A detail of NGC\,5236, shown in the four bands used in this work (clockwise from top-left): CO(2--1), MIRI/21~$\mu$m, NIRCam/Pa$\alpha$ and WFC3/H$\alpha$. The location of several regions is shown with red circles, highlighting the offsets between the H$\alpha$/Pa$\alpha$/21~$\mu$m peaks and the CO peaks. 
}
 \label{fig:detail}
\end{figure}

The JWST/MIRI 21~$\mu$m images of NGC\,5236 are from the Cycle 1 program \# 2219 (Shining light on the CO-dark H2 gas in the heart of M83, P.I.: S. Hernandez). The MIRI imaging consists of a small mosaic 2$^{\prime}\times$1$^{\prime}$.3 in size to the SW of the nucleus, along the southern spiral arm (see Figure~\ref{fig:detail} for the location of the MIRI pointing on the H$\alpha$ image). A separate pointing of the sky background was obtained, which we use to remove this contribution from the science images. Both the science and background images were processed through the JWST pipeline version 1.18.1 (March 2025 release) using the CRDS context ``jwst\_1364.pmap'' \footnote{https://jwst-pipeline.readthedocs.io/en/latest/jwst/user\_documentation/reference\_files\_crds.html}. The final sky--subtracted image has pixel scale of 0\farcs11/pix and is in units of MJy/sr. 

The HST/WFC3 mosaics of NGC\,5236 in H$\alpha$ and adjacent stellar continuum filters are from two programs: \# 11360 (Star Formation in Nearby Galaxies, P.I.:  R. O'Connell) and \# 12513 (Stellar Life and Death in M83: A Hubble-Chandra Perspective, P.I.: W. Blair) and are available through the MAST Archive\footnote{MAST: Mikulski Archive for Space Telescopes at the Space Telescope Science Institute; https://archive.stsci.edu/. For the NGC\,5236 mosaics, see: https://archive.stsci.edu/prepds/m83mos/}. The details of the observations and mosaic preparation can be found in \citet{Dopita+2010} and \citet{Blair+2014}. The H$\alpha$+[NII] emission lines have been captured with the F657N filter; the bracketing stellar continuum filters are F547M/F555W and F814W to the blue and red sides of the lines, respectively. The central region of NGC\,5236 has been targeted with the F555W filter and the external regions with the F547M filter. We thus process the two 
areas separately when removing the stellar continuum from the narrow--band filter. The F555W filter contains, unlike the F547M, the [OIII]($\lambda$0.5007~$\mu$m) emission line, which however gives negligible contribution to the stellar continuum  within the broad--band filter in metal--rich environments \citep{Calzetti+2024}. 
The H$\alpha$ line emission is derived after subtraction of the [NII] contribution, using [NII]/H$\alpha$=0.53 for the sum of the two [NII] components \citep{Kennicutt+2008}. The footprint of the 21~$\mu$m image on the H$\alpha$ mosaic is sufficiently small (Figure~\ref{fig:detail}) that we can neglect changes to the [NII]/H$\alpha$ ratio due to the metallicity gradient; the flux change in H$\alpha$ from the metallicity gradient's effect on [NII]/H$\alpha$ is $\lesssim$14\%, which is within our typical uncertainty (Table~\ref{tab:properties}).  The final, flux--calibrated H$\alpha$ mosaic, corrected for foreground Milky Way extinction, covers a diameter of $\sim$6.5$^{\prime}$ ($\sim$8.5~kpc) centered on the galaxy.

Imaging in CO with resolution $\sim$1$^{\prime\prime}$--2$^{\prime\prime}$ (FWHM)  is available for all three galaxies, either in the transition $^{12}$CO(1--0) or $^{12}$CO(2--1), as listed in Table~\ref{tab:CO}. Among the products available, we will utilize moment~0 (integrated intensity) and moment~2 (velocity dispersion) maps. CO(2--1)  intensity is converted to the CO(1--0) equivalent adopting R$_{(21/10)}$=0.65 \citep{Leroy+2021}, bearing in mind that this value shows large variations, by about 50\%,  from region to region within galaxies and from galaxy to galaxy \citep{Koda+2012, Koda+2020, Koda+2025, Sandstrom+2013, Chiang+2024}. We will not carry this uncertainty in our analysis, since the majority of HII regions analyzed here are located along the spiral arms of our target galaxies, where the  R$_{(21/10)}$ values remain fairly uniform \citep{Koda+2012, Koda+2025}; thus we expect that any changes in the R$_{(21/10)}$ numerical value will be of similar magnitude for all our regions implying a rigid shift along the $\Sigma_{mol}$ axis. For NGC5194, the total intensity map is available through PdBI observations from the PAWS survey \citep{Schinnerer+2013, Colombo+2014}. For NGC628 and NGC5236 two intensity maps are available from separate ALMA observational programs \citep{Koda+2020, Leroy+2021} and processed through the PHANGS--ALMA pipeline with both "broad" and "strict" masks. The "broad" masks include all sightlines where signal is identified at any resolution, and therefore include more regions with faint emission that may appear noisier. The "strict" masks only include regions with emission at high S/N, but because of these stringent signal identification criteria, the "strict" maps typically include less of the total flux \citep{Leroy+2021a, Leroy+2021}. Therefore we utilize the "broad" masks for measuring the molecular gas, but adopt the "strict" masks when measuring the the velocity dispersions from the  moment--2 maps.

\begin{figure}
\plotone{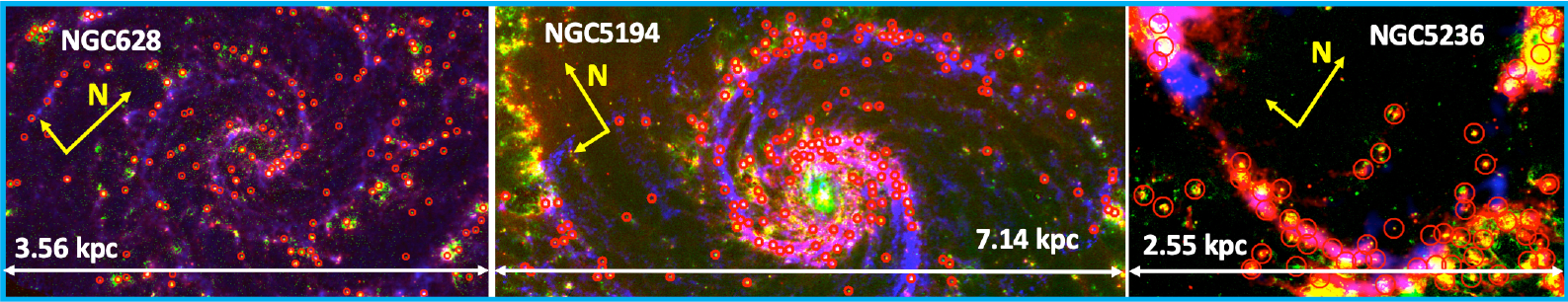}
\caption{Composites of common FoVs in CO (blue), H$\alpha$ (green), 21~$\mu$m (red) for the three galaxies: NGC\,628, NGC\,5194=M\,51a and NGC\,5236=M\,83, with the selected star forming regions shown as red circles. The NE direction is indicated in each panel.
}
 \label{fig:sources}
\end{figure}

Given the heterogeneous nature of the data, the overlap in Field--of--View (FoV) between the HST/H$\alpha$ (either WFC3 or ACS), JWST/MIRI/21~$\mu$m and CO maps is different for the different galaxies, but still large enough to ensure that we can secure a total sample of about 350 star forming regions across the three galaxies (Figure~\ref{fig:sources} and Table~\ref{tab:CO}). For NGC\,628, the overlap between the images is sufficient that we can re--use the sources selected by \citet{Calzetti+2024}. For NGC\,5194 and NGC\,5236, sources need to be selected anew. The smallest footprint pertains to NGC\,5236, due to the small MIRI coverage (Figure~\ref{fig:sources}). The FoV of the JWST/NIRCam images has excellent overlap with the MIRI footprint for NGC\,628 and NGC\,5194, but is slightly offset in NGC\,5236; for this galaxy, we can measure the Pa$\alpha$ emission and stellar mass densities for only 38 of the 56 regions. The incomplete overlap of the NIRCam images is not a limitation for this work, as we use the P$\alpha$ information only to derive general dust attenuation properties of the sources; for the missing stellar mass densities, we apply the mean value of the 38 regions with measurements to those without them (section~\ref{sec:physical}).

\begin{deluxetable*}{lccccccc}
\tablecaption{CO Imaging Data\label{tab:CO}}
\tablewidth{0pt}
\tablehead{
\colhead{Galaxy} &\colhead{CO Trans.$^1$} & \colhead{Map Size$^2$} & \colhead{FWHM(CO)$^2$} & \colhead{Reference$^3$} &\colhead{Apert. Radius$^4$} &\colhead{Fraction$^5$} &\colhead{\# Regions$^6$}\\
\colhead{} &\colhead{} & \colhead{($^{\prime}\times^{\prime}$)} & \colhead{($^{\prime\prime}$)} & \colhead{} &\colhead{($^{\prime\prime}$ (pc))} &\colhead{} &\colhead{}
}
\decimalcolnumbers
\startdata
NGC\,628 & (2--1) & 4.6$\times$3.2   &  1.12  & (1) & 1.4 (63) & 0.70 & 143\\
NGC\,5194& (1--0)  & 3.8$\times$2.8 &  1.07  & (2) & 1.6 (59) & 0.77 & 154 \\
NGC\,5236 & (2--1) &  7.5$\times$7.5  & 2.14 & (3) & 2.6 (57) & 0.70 & 56\\
\enddata
$^1$  $^{12}$CO transition for which a map is available.\\
$^2$ Size in arcmin$\times$arcmin and FWHM in arcsec of the CO map.\\
$^3$ References to the CO maps: (1) \citet{Leroy+2021}; (2) \citet{Schinnerer+2013} and \citet{Colombo+2014}; (3) \citet{Koda+2020}, reprocessed by \citet{Leroy+2021}.\\
$^4$ Radius of the photometric aperture utilized, in arcsec and, in parentheses, parsec.\\
$^5$ Fraction of the CO beam, assumed to be a round gaussian, contained within the photometric aperture.\\
$^6$ Number of regions identified in the overlapping footprint of the HST/H$\alpha$ (WFC3 or ACS), JWST/MIRI/21~$\mu$m, and CO maps.
\end{deluxetable*}

\section{Source Selection and Photometry} \label{sec:selection}

\subsection{Selection and Measurements}\label{sec:measurements}
The CO images have the lowest angular resolution among the sets of interest and, therefore, drive the size of the apertures used to perform photometry (Table~\ref{tab:CO}). We choose  aperture radii $\gtrsim$1.2$\times$FWHM$_{CO}$, ensuring that at least 70\% of the CO Point--Spread--Function (PSF) is captured by the photometric measurements. The apertures are also chosen to be as much as possible similar in physical radius, $\sim$60~pc, in the three galaxies;  these are small enough to be dominated by a single or a small cluster of HII regions, but large enough to capture most of the ionized emission from those regions. For our sources, the largest Str\"omgren radius is about 30~pc \citep{Osterbrock+2006} for uniform electron density n$_e\sim$100~cm$^{-3}$, but the sources can be more extended than this, if the ionized gas emission is from a few spatially close HII regions (Figure~\ref{fig:detail}). The goal of setting apertures close in spatial extent leads to a small mismatch between the three galaxies in the fraction of PSF$_{CO}$ captured in each case; for NGC\,5194, the photometric aperture subtends 77\% of the PSF$_{CO}$, about 10\% larger than the apertures of the other two galaxies. However, this difference is well within the typical uncertainties of CO measurements and we disregard it. We also disregard small differences in the subtended area due to differences in inclination (also $<$10\%). Our galaxies are face--on for all practical purposes (Table~\ref{tab:properties}), thus inclination corrections are not applied. 

Once the photometric radii are determined from the CO maps, we use the H$\alpha$ images to visually identify local peaks of emission and determine the location of the photometric apertures. The apertures are preferentially centered on the centroid of the strongest local peak in H$\alpha$, but deviations from this default approach are necessary when multiple adjacent
peaks are present in an area. In this case, the apertures are slightly off--centered from the local H$\alpha$ peak, also to prevent the apertures from overlapping with each other by more than 10\%; we impose this limit in the overlapping areas between adjacent apertures to avoid  `double counting' flux. Because the regions tend to be crowded, some emission peaks, usually the fainter in the area, need to be discarded altogether. These are usually regions that are faint enough to be below the limit of 3,000~M$_{\odot}$  for the mass of young stars formed; as discussed in \citet{Calzetti+2024}, we take this limit as the lowest acceptable mass for the stellar Initial Mass Function (IMF) to be sufficiently populated at all stellar masses that the ionizing photon rate varies by less than 30\% and is, thus, a reliable measure of the SFR \citep{Cervino+2002}. For a Kroupa IMF in the range 0.1--120~M$_{\odot}$ \citep{Kroupa+2001} and a 4~Myr old region, this mass limit corresponds to a minimum dust attenuation--corrected H$\alpha$ luminosity of 10$^{37.55}$~erg~s$^{-1}$. We call this limit the `stochastic sampling limit' from now on.

The choice of a Kroupa IMF for all our regions is a convenient working hypothesis; variations relative to this default have been reported in the
literature based on observations of individual star clusters \citep[see, e.g.,][]{Schneider+2018a, Hosek+2019}, though how convincing this evidence is when averaged over star cluster populations remains subject to debate  \citep[e.g.,][]{Fumagalli+2011, Weisz+2015}. Models have also been proposed that link the properties of the IMF to characteristics of the environment where the stars are formed \citep[e.g.,][]{Weidner+2006, Dib+2023}. As we cannot determine the distribution of the stellar masses within each of our regions, we report the potential impact of changing the IMF default. If the IMF of our regions were as shallow as the one reported by \citet{Schneider+2018a} for 30~Doradus in the LMC, the same condition that the ionizing photon rate varies by less than 30\% would still correspond to a minimum dust attenuation--corrected H$\alpha$ luminosity of 10$^{37.55}$~erg~s$^{-1}$, but this minimum value would be associated to a cluster mass of $\sim$1,200~M$_{\odot}$. Our conclusions, which are based on tracers of massive star luminosities, are effectively unchanged, although SFRs (see next sections) will shift to larger or smaller values depending on the IMF assumed.

After all selections are imposed, the final sample consists of 353 line emitting regions across the three galaxies (last column of Table~\ref{tab:CO}). The ionized gas emission is usually well tracked by the mid--IR emission, but both H$\alpha$ and mid--IR are generally displaced from the CO peaks and populate the `periphery' of CO--bright areas, as shown, for instance, in Figure~\ref{fig:detail}, right. Displacements between H$\alpha$ and CO peaks are routinely observed along the spiral arms of disk galaxies and have been quantified in the literature \citep[e.g.,][]{Egusa+2004, Kreckel+2018, Kruijssen+2019, Querejeta+2025}, although in the majority of cases the displacement tends to be along the downstream direction of the spiral pattern's rotation. This is consistent with a picture, also observed in the MW, in which HII regions stream away from the inner part of the cloud ("blister model"), i.e., away from the dense area into a less dense area, as expected in the presence of a density or pressure gradient from the cloud core to the inter--cloud medium. The impact of this displacement on our results is discussed in section~\ref{subsec:summary}.

In order to verify whether our sample, selected by eye, carries biases, we compare the summed--up luminosity of the selected regions against the total luminosity for different bands, within the common footprints of the H$\alpha$--21~$\mu$m--CO images of each galaxy. 
The sources include $\sim$50\% of the dust attenuation--corrected H$\alpha$ of NGC\,628 and NGC\,5194 and $\sim$75\% in NGC\,5236, after extrapolating  the summed--up emission along the mass function of star--forming regions down to the lowest expected mass, $\sim$100~M$_{\odot}$ \citep[see, e.g.,][for the methodology]{Calzetti+2025}. We, thus, find that about half of the ionized gas emission in both NGC\,628 and NGC\,5194 is in HII regions, as expected in star--forming disks where about half of the H$\alpha$ emission is diffuse \citep{Oey+2007, Pellegrini+2012}. The larger fraction recovered for NGC\,5236 is consistent with the MIRI footprint being centered on a spiral arm (Figure~\ref{fig:detail}, left). The fraction of recovered CO emission is, conversely, highly uncertain as it depends on the adopted extrapolation to low masses of M$_{mol}$ at the fixed aperture size of 60~pc. Adopting the steep trend between $\Sigma_{SFR}$ and $\Sigma_{mol}$ derived in this work, we find that the CO emission included within the star--forming regions approaches 100\%, but it can be as low as 50\% for less steep trends. In summary, our sample includes the vast majority of the HII regions above the 3,000~M$_{\odot}$ cut located within the common imaging footprints of the three galaxies, relieving concerns of bias from our selection procedure.

Photometry is performed in each circular region, and the diffuse emission from the galaxy is removed from each measure to isolate the flux of the star--forming region. As already mentioned above, we do not correct our measurements for inclination, as all three galaxies are practically face--on and such corrections are $<$10\%. The removal of the diffuse emission is performed differently for H$\alpha$/Pa$\alpha$/21~$\mu$m and CO; this is because the photometric aperture 
radii are only 20\%--50\% larger than the PSF$_{CO}$, but are $>$2$\times$ and $>$17$\times$ larger than the 21~$\mu$m (PSF$_{21}$=0\farcs674) and H$\alpha$/Pa$\alpha$  (PSF$_{H\alpha}$=0\farcs08, PSF$_{Pa\alpha}$=0\farcs06) PSFs, respectively. As shown in \citet{Gregg+2025}, local background subtraction applied to photometry in apertures of size comparable to the PSF leads to biased results, thus requiring here a different approach for the H$\alpha$/Pa$\alpha$/21~$\mu$m photometry and the CO photometry.  

For H$\alpha$, Pa$\alpha$ and 21~$\mu$m photometry, the diffuse emission is measured in an annulus around the aperture, extending from $\sim$60~pc to 100~pc in radius. The size of the annulus is almost twice the radius of the photometric aperture to exclude the majority of the emission in the wings of the 21~$\mu$m PSF. Conversely, the H$\alpha$/Pa$\alpha$ PSF is sufficiently smaller than the aperture size that photometry biases due to power in the PSF wings are negligible. The diffuse emission is determined from the mode of the pixel value distribution after iterative $\sigma$ clipping to remove any source present in the annulus. The nature of this diffuse emission is discussed in section~\ref{sec:diffuse}. Aperture corrections are applied to the 21~$\mu$m photometry to account for the flux in the wings of the PSF: 17\%, 15\%, and 6\% for NGC\,628, NGC\,5194 and NGC\,5236, respectively. No aperture corrections are applied to H$\alpha$ and Pa$\alpha$ measurements, but they are corrected for foreground Milky Way extinction (Table~\ref{tab:properties}). 

The above  photometric strategy aims at closely matching the procedure adopted in \citet{Calzetti+2025}, whose SFR calibration we use. However, to test the sensitivity of our photometry to the choice of background for the subtraction, we increase the outer radius of the annulus by a factor two, resulting in a factor $\sim$six increase in the area of the background region. The SFRs derived from the new background subtraction are, on average, $\sim$0.03~dex higher ($<$10\%) than our default ones (section~\ref{subsec:sfr}) and, at low values, their scatter increases by $\sim$0.03~dex. These are small variations, well within our uncertainties. We thus conclude that our approach to background subtraction is relatively robust. 

The presence of diffuse CO emission in galaxies is debated, and \citet{RomanDuval+2016}  determine that it is no more than 25\% on average in the Milky Way, reaching $\sim$30\% in the central region, but generally remaining lower than that within the inner 6~kpc radius \citep[see, also,][]{Duarte+2021}. Adopting an average 30\% of total emission 
as our absolute maximum for the diffuse CO component, we remove this component from the photometric measurements on the moment 0 maps, and determine that its impact is small on the resulting intensity I$_{CO}$ (in units of K~km~s$^{-1}$), as our sources are generally bright in CO. For consistency with the other measurements, however, we adopt CO photometry with the diffuse emission removed. We do not correct CO fluxes for the portion of the PSF outside the photometric aperture; it is unclear what such correction should be, as most sources are located in the outskirts of bright CO regions, and the wings of the brighter areas may contaminate our measurements. The lack of PSF wings correction will cause all our $\Sigma_{mol}$ values to be systematically underestimated by $\sim$30\%--40\% at most, but with galaxy--to--galaxy variations of $\lesssim$10\% (Table~\ref{tab:CO}, last column). 

Photometry in the same apertures as the other bands is also performed on: the CO moment 2 maps (km~s$^{-1}$) of all three galaxies, and the peak intensity maps (K)  and equivalent width maps (km~s$^{-1}$) of NGC628 and NGC5236. The available moment 2 maps for NGC628 and NGC5236 are intensity--weighted from the `strict' maps; about 10\% of sources in NGC\,628 do not have valid data in the strict maps as they have too low signal--to--noise (S/N) values and we exclude these sources when analyzing trends in velocity dispersion. For uniformity, we create an intensity--weighted map for NGC5194, and perform photometry on this map. The equivalent width is a noise-- and profile--insensitive version of the velocity dispersion \citep{Heyer+2001, Leroy+2016}.

\subsection{The Nature of the Diffuse Emission}\label{sec:diffuse}

The local background subtraction performed on photometry at H$\alpha$/Pa$\alpha$/21~$\mu$m has the common goal of removing the diffuse emission that is not related to the star--forming region of interest, but for slightly different reasons in different bands. For the ionized gas (H$\alpha$ and Pa$\alpha$), the intent is to remove the contribution of photon leakage and scattered light coming from neighboring regions. Ionized photon leakage can reach large distances, up to $\sim$1~kpc from the source, contributing to a diffuse ionized gas (DIG) component that comprises 50\% or more of the total line emission \citep[e.g.,][]{Reynolds+1984, Reynolds+1990, Ferguson+1996, Hoopes+1996, Hoopes+2003, Voges+2006, Oey+2007, Zhang+2017, Watkins+2024}. In the next section, we discuss the impact of ionizing photon leakage {\em out of our regions} on the SFR indicators, while the impact of the diffuse ionized gas emission on photometry is quantified in Appendix~\ref{sec:appendixA}.

\begin{figure}
\plottwo{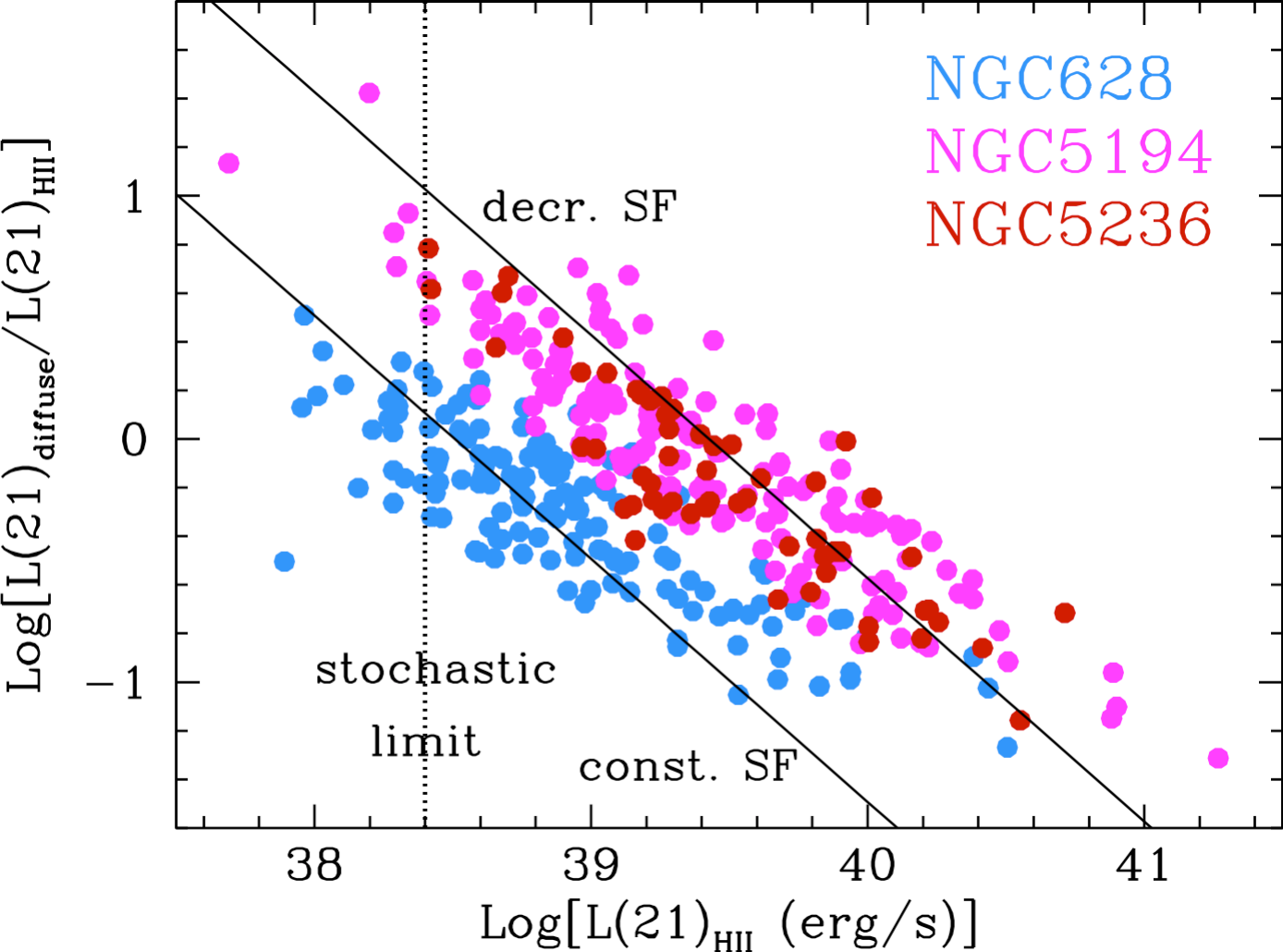}{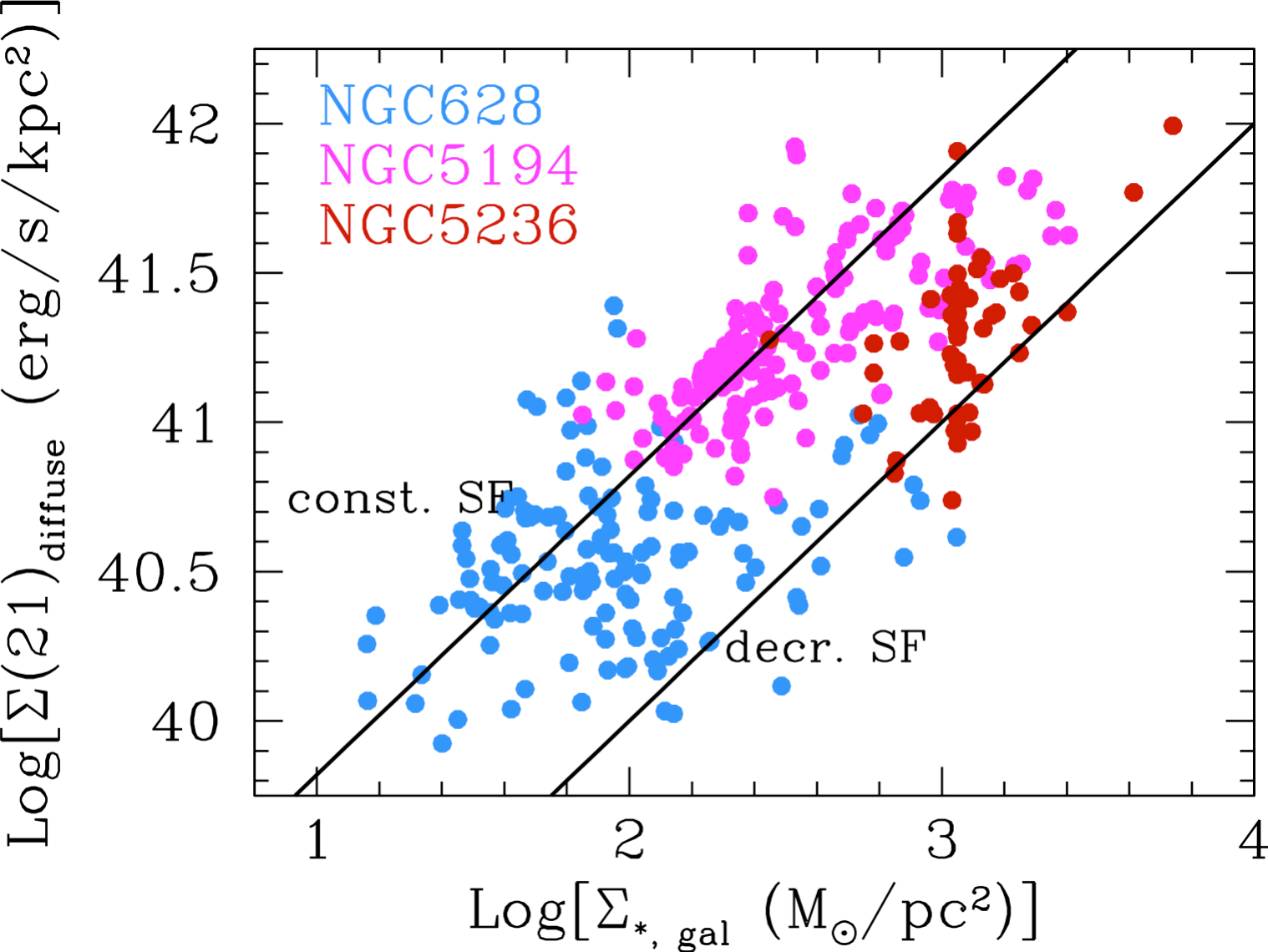}
\caption{{\bf (Left):} The ratio of diffuse--to--HII region emission at 21~$\mu$m for the star--forming regions in the three galaxies, shown in different color symbols 
for the different galaxies (teal=NGC\,628; magenta=NGC\,5194; dark red=NGC\,5236). The stochastic sampling limit (Stochastic limit in the figure) is marked as a vertical dotted black line.  
The solid black lines mark the trends expected by models, in the case of  (lower line) constant SF and (upper line) decreasing SF over the past few Gyr, the latter to mimic the SFH of NGC\,5194 \citep{Martinez+2018}; see text and Appendix~\ref{sec:appendixB} for more details on the models. The y$\sim -$x trend in log--log scale is expected for roughly constant diffuse emission independent of HII region luminosity. 
{\bf (Right):} The diffuse 21~$\mu$m surface density as a function of the stellar mass surface density of the underlying galaxy for all three galaxies. The same models as those in the left panel are shown (black solid lines).   
}
 \label{fig:diffuse}
\end{figure}

The largest contamination to our SFR indicators comes from the diffuse emission at 21~$\mu$m, which originates from dust heated by the general galaxy's stellar population as predicted by models \citep{DraineLi2007, Galliano+2018} and measured to be about 30\%--80\% of the total mid--IR emission in nearby star--forming galaxies \citep{Calzetti+2005, Calapa+2014, Boquien+2016, Leroy+2023}. At the level of individual star--forming regions, it has a large, luminosity--dependent effect on photometry, as shown in Figure~\ref{fig:diffuse}, left, where the diffuse emission represents less than 10\% of the bright regions'  21~$\mu$m luminosity, but increases by almost two orders of magnitude at the faint end. The solid lines in the left panel of Figure~\ref{fig:diffuse} show the possible range of ratios permitted by a range of star formation histories using the models in \citet[][briefly described in Appendix~\ref{sec:appendixB}]{Calzetti+2025}. The models are for constant star formation and exponentially decreasing star formation over the past few~Gyr, mimicking the SFH of NGC\,5194 \citep{Martinez+2018}. Despite the simplifying assumption that the same SFH works for all regions in a galaxy, the models can explain, within a factor $\sim$2, the observed trend and its dispersion; the former is a simple 1/x trend (in linear scale) as expected if the diffuse emission luminosity is roughly independent of HII region luminosity and the latter depends on the SFH. NGC\,5194 experienced a peak in SFR about 6 times higher than current in the past 2~Gyr, remaining high until recently and then sharply decreasing to the present day \citep{Martinez+2018}. This may explain why its diffuse L(21) emission is significantly higher than the one in NGC\,628. This emission is due to a combination of thermal and non-thermal dust heated by both non--ionizing UV and optical photons \citep{DraineLi2007, Galliano+2018, Draine+2021}. Optical photons are present at all ages and  non--ionizing UV photons persist for $\gtrsim$100~Myr, over 10 times longer than ionizing photons; population build--up from decreasing star formation contributes to the excess UV$+$optical luminosity over the ionizing one \citep{Calzetti+2021}. Both the 1/x trend and the scatter strongly indicate that  L(21)$_{diffuse}$ cannot be part of the star--forming regions' emission, i.e., of the SFR budget. If the latter were the case, we would expect L(21)$_{diffuse}$/L(21)$_{HII}\sim$constant, because bright regions would remain bright also in their `wings', which is not what is  observed in Figure~\ref{fig:diffuse}; furthermore, there should be little or no scatter in  L(21)$_{diffuse}$/L(21)$_{HII}$ at constant L(21)$_{HII}$, while a large, galaxy--dependent (i.e., SFH--dependent) scatter is observed.

The diffuse 21~$\mu$m emission correlates with the stellar mass surface density of the galaxy underlying the star--forming regions, as shown in Figure~\ref{fig:diffuse}, right. Stellar mass surface densities $\Sigma_{*, gal}$ are derived from the JWST/NIRCam F277W and/or F300M images, as described in Appendix~\ref{sec:appendixB}. The relation, which follows closely a 1--to--1 trend in log--log scale, has a large scatter. The scatter, in both $\Sigma_{*, gal}$ and $\Sigma(21)$, is due to differences in star formation histories as well as differences in dust color excess of the different stellar populations. The models briefly presented in Appendix~\ref{sec:appendixB} bracket much of the observed scatter between the diffuse $\Sigma(21)$ and $\Sigma_{*, gal}$, but an even broader range can be achieved by, say, including exponentially increasing star formation histories (smaller $\Sigma_{*, gal}$ at constant  $\Sigma(21)_{diffuse}$) and dust color excess E(B--V)$<$0.1~mag for the oldest stellar population (larger $\Sigma_{*, gal}$ at constant  $\Sigma(21)_{diffuse}$). 

In addition to global galaxy--to--galaxy differences, there are also local differences in the recent SFHs, to which L(21)$_{diffuse}$ is sensitive, while the stellar mass is less so, as shown in Appendix~\ref{sec:appendixB}. For instance, the data for NGC\,5194 may seem inconsistent in the SFH they follow between the left and right panels of Figure~\ref{fig:diffuse}: they are explained by decreasing SF in the left panel, and by constant SF in the right panel. The two panels, however, represent different timescales: in the left hand--side panel, L(21)$_{diffuse}$ is compared with the 21~$\mu$m luminosity of HII regions ($\lesssim$10~Myr), while in the right hand--side panel L(21)$_{diffuse}$ is compared with stellar mass ($\approx$10~Gyr). Thus, the behavior of the data for NGC\,5194 can be explained if the SFH of this galaxy was roughly constant for most of its lifetime, only drastically decreasing in SFR in recent ($<$100~Myr) times, consistent with the inferred SFH \citep{Martinez+2018}. 

In summary, subtraction of the local diffuse component from the star--forming regions's emissions is a necessary step to ensure that the light contribution of the underlying galaxy is removed before deriving SFRs. 

\section{Physical Quantities}\label{sec:physical}

The list of sources and derived physical quantities is given in three separate Tables, one per galaxy (see Tables~\ref{tab:NGC628}, \ref{tab:NGC5194}, \ref{tab:NGC5236}).

\subsection{SFR Surface Density and the Surface Density of Young Stars}\label{subsec:sfr}

The presence of H$\alpha$ emission ensures that the regions we measure  host star clusters younger than $\sim$6--7~Myr and with median age $\sim$3~Myr \citep{Pedrini+2025, Knutas+2025}.  While HII regions are often considered to be powered by instantaneous-burst populations, thus questioning an attribution of a SFR to such regions, large sample analyses show that the mean properties of HII regions are consistent with those of constant star formation over 3-6~Myr timescales \citep[e.g.,][]{Calzetti+2025}, also confirmed by detailed analyses of nearby HII regions \citep[e.g.,][]{Schneider+2018}. 

\begin{figure}
\plottwo{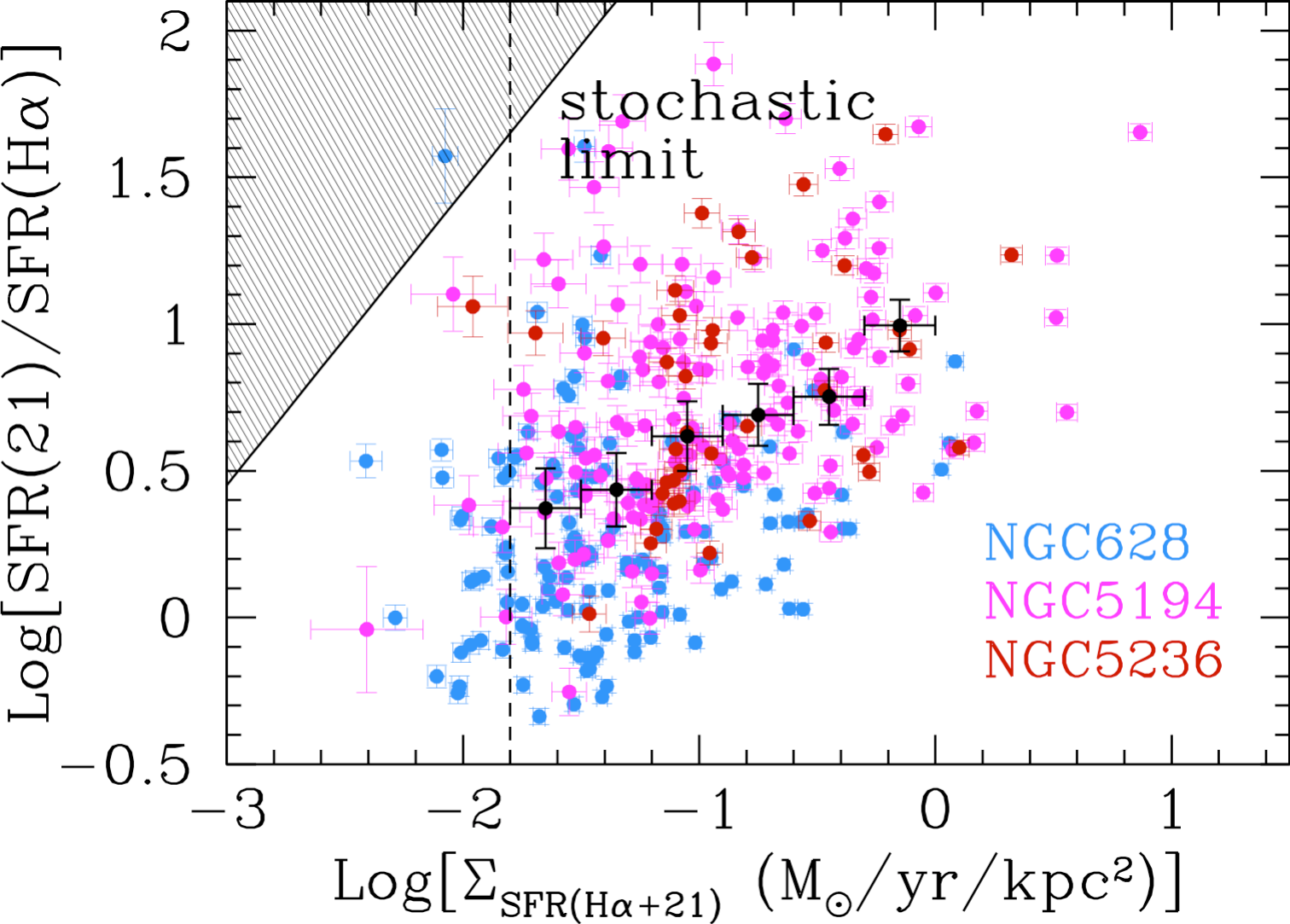}{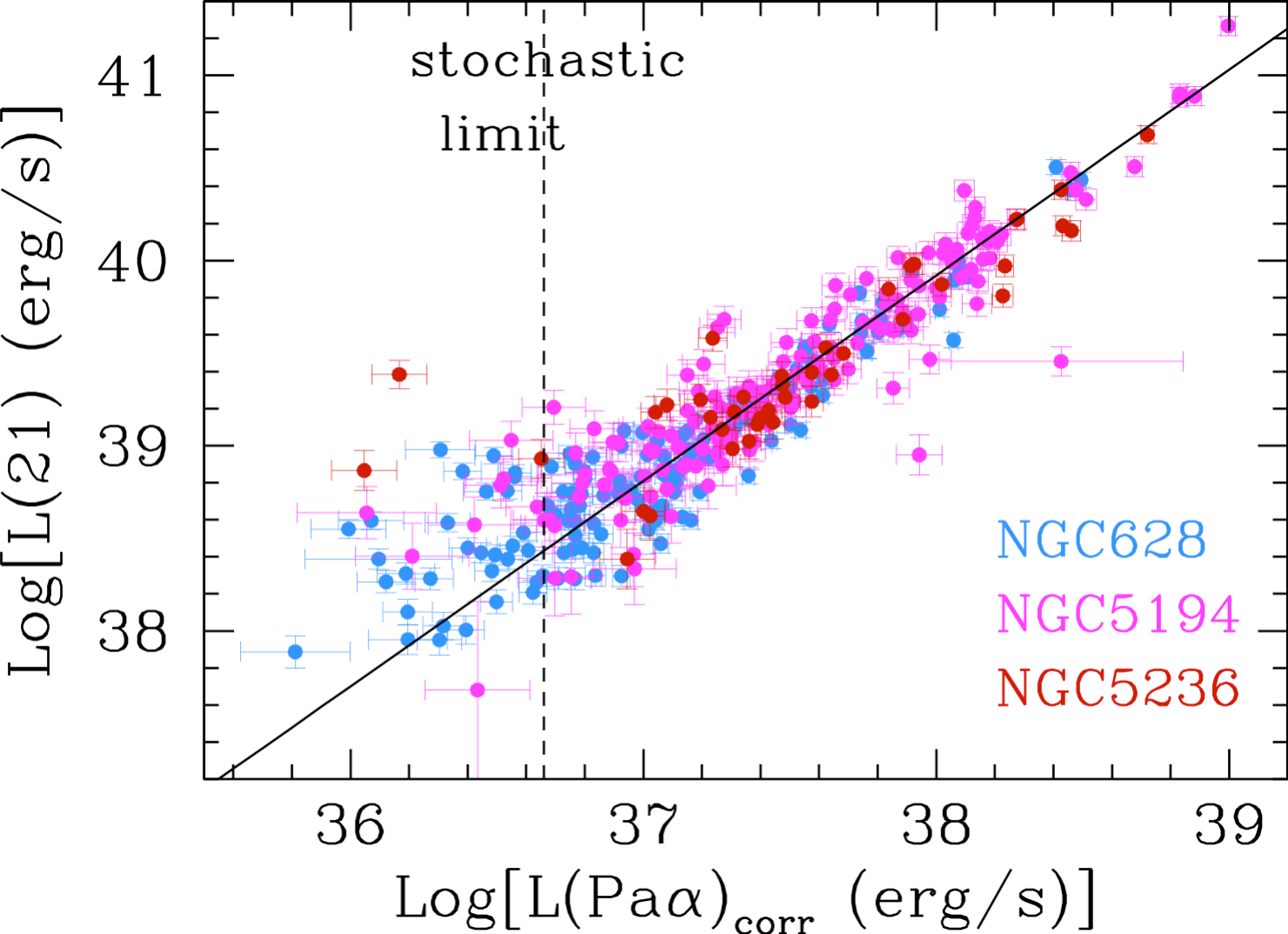}
\caption{{\bf (Left):} The 21~$\mu$m/H$\alpha$ SFR ratio as a function of the  SFR(21$+$H$\alpha$) surface density for the star forming regions in the three galaxies (teal=NGC\,628, magenta=NGC\,5194, dark red=NGC\,5236) with 1$\sigma$ uncertainties. The stochastic sampling limit in $\Sigma_{SFR}$ 
is marked with a dashed vertical line. The shaded area marks the 3$\sigma$ detection limit based on the average depth of the images for the three galaxies. The binned averages for data above the stochastic sampling limit are shown as black filled circles with uncertainties calculated as root--mean--squares of the uncertainties of the data in each bin. 
{\bf (Right):} The 21~$\mu$m luminosity of the star forming regions as a function of the dust attenuation--corrected Pa$\alpha$ luminosity, with the best fit for data above the stochastic sampling limit shown as a solid back line. 
}
 \label{fig:sfr}
\end{figure}

Dust attenuation, as measured via the color--excess E(B--V), increases for increasing SFRs in HII regions, as shown in both NGC\,628 and NGC\,5194 by \citet{Calzetti+2024} and \citet{Calzetti+2025}. This is illustrated here in Figure~\ref{fig:sfr}, left, where the ratio of SFRs captured by the 21~$\mu$m and H$\alpha$ luminosities, respectively, is plotted as a function of the SFR surface density of each region in the three galaxies. We use this ratio as a proxy for E(B--V), since SFR(21) captures the dust--obscured SF, while SFR(H$\alpha$) tracks the light from SF directly emerging in the optical. The ratio of the two SFRs shows a trend of increasing 21~$\mu$m emission relative to the H$\alpha$ one for increasing $\Sigma_{SFR}$, particularly evident in the lower envelope of the relation. This lower envelope is not due to detection limits as all regions are detected well above the limits, by about an order of magnitude, at each wavelength. The same figure reports the binned averages, as black filled circles with uncertainties\footnote{The uncertainties for the binned averages are calculated as root--mean--squares of the uncertainties of the measurements contained in each bin. The standard deviation of the distribution of the data in the bins, $\sim$0.30~dex, is not shown, because it has a specific physical meaning: it measures the amount of dust in the galaxies combined with our detection limits. Dustier galaxies will display larger standard deviations in this plot \citep{Calzetti+2025}. Thus, the standard deviation is a measure of the range of dust attenuations probed in each bin, not the intrinsic scatter of each measurement relative to the mean trend.},
of the data above the stochastic sampling limit; this limit, for region sizes of 60~pc radius, corresponds to Log($\Sigma_{SFR})\sim -1.8$, in units of M$_{\odot}$~yr$^{-1}$~kpc$^{-2}$. Even at the lowest value of $\Sigma_{SFR}$, SFR(21) is about 2.5 higher than SFR(H$\alpha$), i.e., the dust obscured SFR that emerges in the mid--IR dominates over the unobscured one that emerges in the optical. If about 50\% of ionizing photons are lost to leakage out of the regions, $\Sigma_{SFR}$ will be underestimated by $\sim$17\% on average at the lowest end of the range and by less than a few \% at high values. Thus, leakage of ionizing photons out of regions has a small impact on our SFR determinations, since the 21~$\mu$m emission is dominated by dust heated by non--ionizing UV photons \citep{DraineLi2007, Galliano+2018}. 

The 21~$\mu$m luminosity is tightly correlated with the Pa$\alpha$ luminosity, shown in Figure~\ref{fig:sfr}, right \citep[see, also,][]{Calzetti+2025}. The best fit through the data above the IMF stochastic sampling limit, performed using LINMIX \citep{Kelly+2007}, is:
\begin{equation}
Log[L(21)] = (1.11\pm0.02) Log[L(Pa\alpha)] - (2.26\pm 0.58),
\end{equation}
with a scatter=0.16 and with all quantities in units of erg~s$^{-1}$. The Pa$\alpha$ luminosity has been corrected for dust attenuation using the H$\alpha$/Pa$\alpha$ ratio. The tight correlation between L(21) and the ionizing gas emission mitigates any concerns that the 21~$\mu$m emission may be contaminated by emission from sources unrelated to the young star formation. The best fit slope is within 1$\sigma$ of the slope reported by \citet{Calzetti+2025}, suggesting that the relation between the two luminosities is super--linear, i.e., the L(21) luminosity increases faster than L(Pa$\alpha$) for increasing values. Multiple reasons can account for this behavior, such as, for increasing amounts of dust, the ionizing photons are directly absorbed by dust and/or the regions become progressively more opaque in Pa$\alpha$ \citep{Inoue+2001, Dopita+2003, Krumholz+2009, Draine+2011}. Both effects make the Pa$\alpha$ luminosity unreliable as a SFR indicator at high dust opacities. About 10\% of the star forming regions in the three galaxies have A$_{Pa\alpha}\gtrsim$1~mag. Because of this, we elect to use the combination of H$\alpha$ and 21~$\mu$m as an unbiased SFR indicator for our star forming regions, relying on the mir--IR emission to probe into the dust. We adopt the calibration by \citet{Calzetti+2025} for this work:
\begin{equation}
SFR(H\alpha + 21) (M_{\odot} yr^{-1})  = 5.45\times 10^{-42} [L(H\alpha) + (0.077\pm 0.022)  L(21)],
\label{equa:mixSFR21}
\end{equation}
calibrated for a Kroupa IMF in the range 0.1--120~M$_{\odot}$ \citep{Kroupa+2001} and with the luminosities in units of erg~s$^{-1}$. As a reminder, changing the assumption on the IMF will change the calibration constant that links the luminosities to the SFR, as already discussed in Section~\ref{sec:measurements}.

A quantity that is related to $\Sigma_{SFR}$ is $\Sigma_{young\ stars}$, i.e., the mass surface density in stars produced in the star forming region integrated over the SF event.  This is simply $\Sigma_{young\ stars}$=$\Sigma_{SFR} \times \tau_{SF}$, where $\tau_{SF}$ is the characteristic duration of star formation in the region.  $\Sigma_{young\ stars}$ is a convenient measure for comparing data with some models that predict the  net efficiency of star formation \citep[e.g.][]{Kim+2018}, a common measure in Milky Way studies \citep[][ and references therein]{Heiderman+2010, Kennicutt+2012}.

\subsection{Molecular Gas Surface Densities and Velocity Dispersions}\label{subsec:mol}

Molecular gas surface densities, $\Sigma_{mol}$ are derived for each source using the relation  $\Sigma_{mol}$(M$_{\odot}$~pc$^{-2}$)=$\alpha_{CO}$ I$_{CO}$(K km s$^{-1}$) \citep{Bolatto+2013, Leroy+2021}, where we disregard the small inclination corrections and have already converted CO(2-1) to the equivalent values of CO(1-0). For $\alpha_{CO}$, we adopt the expression from \citet{Bolatto+2013}:
\begin{equation}
\alpha_{CO}= 2.9 ~  \rm{exp}\Bigl({0.4\over Z}\Bigr) \biggl[{\Sigma_{tot} \over (100~M_{\odot}~pc^{-2})}\biggr]^{-\gamma},
\label{alphaco}
\end{equation}
in units of M$_{\odot}$~pc$^{-2}$~(K km s$^{-1}$)$^{-1}$, where $\gamma$=0.5 for $\Sigma_{tot} >$100 M$_{\odot}$~pc$^{-2}$ and $\gamma$=0 otherwise. Z is the metallicity of the region, which we approximate with the central value of our galaxies (Table~\ref{tab:properties}). Including the metallicity gradients in the calculation of $\alpha_{CO}$ for the different regions only adds a 0.04~dex scatter to the results, well below our typical uncertainty.  $\Sigma_{tot} = \Sigma_{*, gal}+\Sigma_{mol, tot}$, where $\Sigma_{*, gal}$ is the stellar mass surface density of the underlying galaxy (Figure~\ref{fig:diffuse}, right, and Appendix~\ref{sec:appendixB}) and $\Sigma_{mol, tot}$ is the total molecular gas surface density integrated along the entire line of sight. This value differs slightly from $\Sigma_{mol}$ as the latter is derived from I$_{CO}$ after removal of the galaxy's diffuse emission. Because the final $\Sigma_{mol}$ depends on itself via the expression for $\alpha_{CO}$ (equation~\ref{alphaco}), its derivation is performed iteratively; we verify that a single iteration is sufficient to converge to $\Sigma_{mol}$ values that differ from the previous ones by less than a few percent. The dependency on $\Sigma_{tot}$ in equation~\ref{alphaco} was proposed by \citet{Bolatto+2013} as a `starburst' correction to the standard value of $\alpha_{CO}\sim$4.35 in the Milky Way, expanding on the original concepts developed by \citet{Narayanan+2012}. Subsequently, \citet{Sandstrom+2013} and \citet{Chiang+2024} have shown that a strong dependency on both $\Sigma_{*, gal}$ and $I_{CO}$  is present also in normal star--forming disks at the $\sim$kpc scale. For the samples analyzed by these authors $\Sigma_{*, gal} > \Sigma_{mol, tot}$, which is true also for most, but not all, of our sources. Thus, we carry out the calculation of $\alpha_{CO}$ using the expression in equation~\ref{alphaco} with $\Sigma_{tot}$. We are also assuming that equation~\ref{alphaco}, derived for whole starbursts and $\sim$kpc--sized regions, holds for our high--spatial resolution measurements (120~pc diameter). Our HII regions are well described as young bursts of star formation, which makes equation~\ref{alphaco} an acceptable approximation. Later in this paper we evaluate the impact of assuming equation~\ref{alphaco} against other recipes, including constant $\alpha_{CO}$ for all sources. The 1$\sigma$ detection limits of the CO images correspond to molecular mass surface densities $\Sigma_{mol}\sim$3--7~M$_{\odot}$~pc$^{-2}$. For our fits, we elect to exclude all data with $\Sigma_{mol}\le$3~M$_{\odot}$~pc$^{-2}$; choosing the higher limit excludes an additional $\sim$1\% of the points and does not affect the results of the fits.  For reference, the surface brightness limit is consistent with the surface density above which HI transitions to molecular hydrogen \citep{Martin+2001, Park+2023, Schneider+2025}. 

Velocity dispersions $\sigma_v$ (km~s$^{-1}$) are derived from the moment~2 maps, after deconvolution of the maps' intrinsic channel width, 2.12~km~s$^{-1}$ for NGC\,5194 and 2.5~km~s$^{-1}$ for NGC\,628 and NGC\,5236 \citep{Rosolowsky+2006, Colombo+2014, Leroy+2021}. Unlike other authors, we do not extrapolate the values of $\sigma_v$ to their 0~K limit, since we make limited use of velocity dispersions in this work. When possible, we compare the results from our $\sigma_v$ with those from the equivalent widths, to check for robustness. 

\section{Analysis and Results} \label{sec:analysis}

\subsection{The Molecular Star Formation Law of HII Regions}\label{subsec:SKlaw}

\begin{figure}
\plottwo{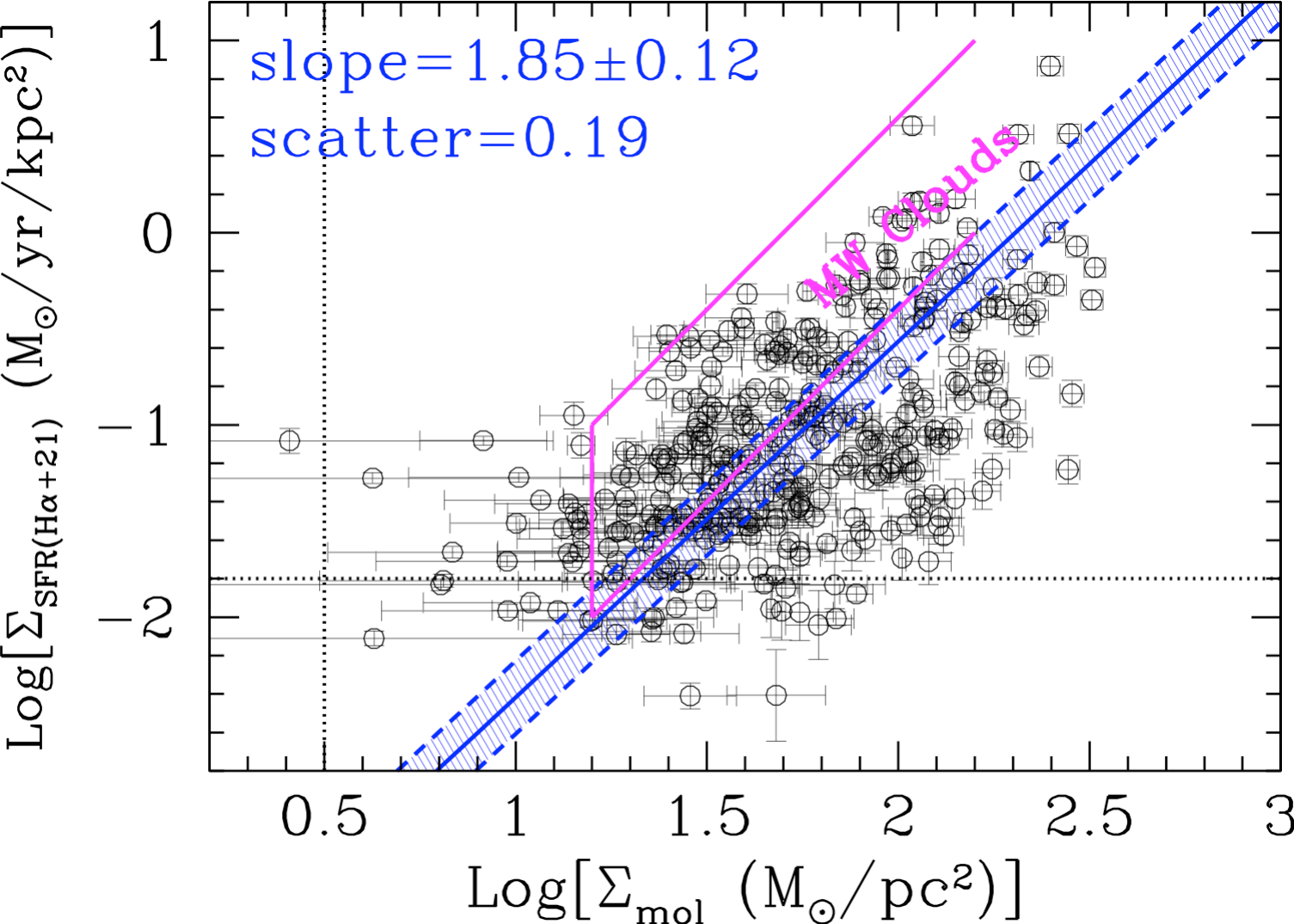}{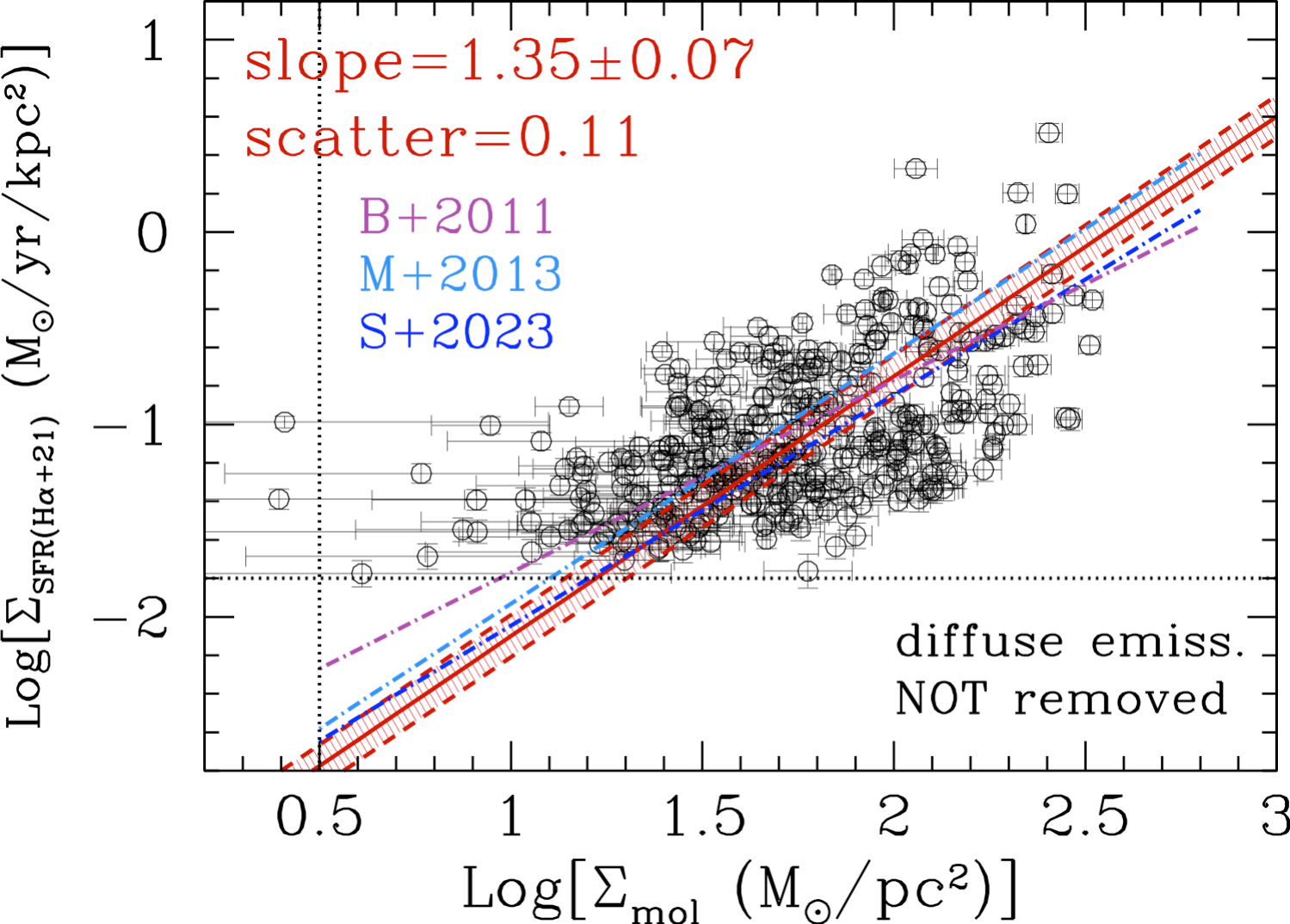}
\caption{{\bf (Left):} The SFR surface density as a function of the molecular gas surface density for the 353 regions in our sample (black circles with 1$\sigma$ uncertainties) together with the best fit through the data returned by the LINMIX algorithm \citep[blue lines with shaded area showing the scatter][]{Kelly+2007}. Only regions to the top--right of the two dotted lines are used in the fits. The approximate locus occupied by the molecular clouds in the Milky Way \citep[e.g.][]{Evans+2009, Heiderman+2010, Gutermuth+2011, Lada+2013, Willis+2015, Nguyen--Luong+2016, Retes--Romero+2017, Lada+2017, Pokhrel+2020} is also shown as a magenta outline for reference. 
{\bf (Right):} The same as the left panel, but {\em without removing the underlying diffuse emission} from both the SFR and the molecular gas tracers. The best fit, also from the LINMIX algorithm, is shown as dark--red lines with shaded area. The fits through $\sim$0.8--1.5~kpc regions in samples of nearby galaxies derived by \citet[][B+2011, purple line]{Bigiel+2011}, \citet[][M+2013, teal line]{Momose+2013}, and \citet[][S+2023, blue line]{Sun+2023} without removing the diffuse emission are shown for comparison.
}
 \label{fig:sklaw}
\end{figure}

The distribution of surface brightnesses in both SFR(H$\alpha$+21) and molecular gas is shown as a scatter plot for the 353 star forming regions from all three galaxies in Figure~\ref{fig:sklaw}, left. The locus occupied by Milky Way molecular clouds is also shown, as a magenta outline; resolved cloud analyses find the SF law to have slope $n\sim1.6 - 2$ or larger in our Galaxy and the SMC, all showing that the normalization for the clouds is higher than the one for our HII regions \citep[e.g.][]{Evans+2009, Heiderman+2010, Gutermuth+2011, Lada+2013, Willis+2015, Hony+2015, Nguyen--Luong+2016, Retes--Romero+2017, Lada+2017, Pokhrel+2020}. 

Given the large scatter in our data, we perform linear fits in log--log space using three different approaches, listed in Table~\ref{tab:fits}. We use the LINMIX package\footnote{The python version of LINMIX used in this work can be found at: https://github.com/jmeyers314/linmix}, which applies a hierarchical Bayesian approach to linear regression \citep{Kelly+2007}. In addition, we use the bi--regression algorithm FITEXY \citep{Press+1992} extended by \citet{Tremaine+2002} to include a scatter component in the fit. \citet{Tremaine+2002} also extend the bivariate correlated error and intrinsic scatter (BCES) algorithm of  \citet{Akritas+1996} to be symmetric in x and y, and we use this estimator as well. As our data are censored in both x and y (Log($\Sigma_{SFR}$)$> -$1.8 and Log($\Sigma_{mol}$)$>$0.5), all three algorithms return different fits when performed in x--vs.--y or y--vs.--x; we, thus, take the average of the fits along the two directions for each algorithm. The limits in the two surface brightnesses exclude 34 data points from the fits; of these, only two have Log($\Sigma_{mol}$)$\le$0.5, while the remaining 32 have Log($\Sigma_{SFR}$)$\le -$1.8. Table~\ref{tab:fits} reports the fits results, which give a consistent picture, with all algorithms returning slopes $n\sim$1.85--1.92; however, the scatter differs by about a factor of 1.5, with LINMIX returning the smallest value. The LINMIX fit results are shown in Figure~\ref{fig:sklaw}, left. Appendix~\ref{sec:appendixC} discusses the residuals around the best fit line.

The impacts of different choices for the size of the regions used to derive the SF law and for the formulation of $\alpha_{CO}$ are shown in Appendix~\ref{sec:appendixD} and Appendix~\ref{sec:appendixE}, respectively. 

\begin{deluxetable*}{lrrrl}
\tablecaption{Fit Results of Log($\Sigma_{SFR}$) versus Log($\Sigma_{mol}$)$^1$\label{tab:fits}}
\tablewidth{0pt}
\tablehead{
\colhead{Algorithm} &\colhead{n (Slope)} & \colhead{A (Intercept)} & \colhead{scatter} & \colhead{Comments} }
\decimalcolnumbers
\startdata
LINMIX$^2$                   & 1.85$\pm$0.12 & -4.27$\pm$0.26 & 0.18 &\\
Extended Bi--Regr$^3$ & 1.85$\pm$0.14 & -4.25$\pm$0.29 & 0.30 &\\
Symmetric BCES$^4$   &  1.92$\pm$0.08 & -4.35$\pm$0.16 & 0.27 & gaussian errors\\
                                       &         $\pm$0.12 &          $\pm$0.20 &        & uniform errors\\
\enddata
$^1$  Fits are expressed in the form: Log($\Sigma_{SFR}$)= n Log($\Sigma_{mol}$) + A. Fits are performed for Log($\Sigma_{SFR}$)$> -$1.8 and Log($\Sigma_{mol}$)$>$0.5 (see test).\\
$^2$ Python version of LINMIX, based on the algorithm of \citet{Kelly+2007}.\\
$^3$ Bi-Regression algorithm \citep{Press+1992}, extended to include scatter \citep{Tremaine+2002}.\\
$^4$ Bivariate correlated error and intrinsic scatter (BCES) algorithm \citep{Akritas+1996}, extended by \citet{Tremaine+2002} to be symmetric in x and y.
\end{deluxetable*}

\subsection{The Effect of the Diffuse Emission on the Molecular Star Formation Law}\label{subsec:nobck_removal}

We also investigate the effect of {\em not removing} the diffuse emission from both the SFR tracer and the molecular gas tracer. The results are shown in Figure~\ref{fig:sklaw}, right. Here the SFR surface density is derived using the H$\alpha$+21~$\mu$m relation derived by \citet{Belfiore+2023}, where the proportionality constant between the two tracers is found to be 0.031, rather than 0.077. This relation has been used in the literature and we adopt it to enable comparisons between our HII regions and previous results. We separate below the different effects on our results.

A fit through the data, using the same censoring as above, yields the relation: Log($\Sigma_{SFR}$)= (1.35$\pm$0.07) Log($\Sigma_{mol}$) -(3.45$\pm$0.16) with a scatter of 0.11, with the LINMIX algorithm. For comparison, the Extended Bi-Regression  and the Symmetric BCES algorithms yield slopes $n=1.33\pm 0.07$ (scatter=0.28) and $n=1.16\pm 0.06$ (scatter=0.23), respectively. This is shallower, by 3.6$\sigma$, than the relation found for the case where the diffuse emission is removed from the tracers. For comparison, we also show in Figure~\ref{fig:sklaw}, right, the relations found by \citet{Bigiel+2011}, \citet{Momose+2013} and \citet{Sun+2023} for 0.8--1.5~kpc regions in samples of nearby galaxies. The intercepts in the published relations are scaled up by $\sim$0.19~dex to account for our smaller regions (120~pc size versus those authors' 0.8-1.5~kpc), which are comparable or smaller than the typical disk thickness. Stellar disk scale heights are 0.2--0.4~kpc  \citep{Ranaivoharimina+2024} and molecular gas disks have generally heights that are roughly half of that \citep{Patra+2019, Elmegreen+2025, Corbelli+2025}. For our scaling we adopt a gas scale--height of 100~pc. An additional correction is implemented to account for our photometric apertures capturing only about 70\%--77\% of the CO PSF. We report the \citet{Sun+2023}'s relation derived using the CO--to--H2 conversion of \citet{Bolatto+2013}, which is consistent with our derivation. The relations from the literature are slightly shallower (but insignificantly so when accounting for the uncertainties) than our results, as expected given the factor $\sim$10 increase in the region sizes analyzed. The light contribution from the underlying galaxy is minimized in small regions centered on the star forming clusters, accounting for our slightly steeper slope. Overall, however, we find that when the light contribution from the underlying galaxy is not removed, the SFR--gas relation of HII regions is shallower by $\Delta(n)=0.5\pm 0.14$.

Disaggregating the different contributions, we find that using the \citet{Belfiore+2023} recipe for the SFR tracer only contributes 25\% of the difference in  $\Delta(n)$. The remaining 75\% of the difference is due to removing/not--removing the underlying diffuse emission from the galaxy. Of this, removing the diffuse CO emission causes the slope to become shallower (as opposed to steeper) by $<$15\% relative to not removing it, while removing the diffuse H$\alpha$ and 21~$\mu$m emissions steepens $n$ by $\Delta(n)\sim 0.45$.

Heuristically, keeping the diffuse emission in the data results in a shallower slope because the underlying galaxy's luminosity has a larger effect on the $\Sigma_{SFR}$ values of faint sources relative to brighter ones. At the bright end for our sources, the contribution to the total luminosity from the galaxy's diffuse emission is generally a minor effect, $<$10\%, while it can dominate over the source's luminosity at the faint end of our luminosity range (Figures~\ref{fig:diffuse}, left, and \ref{fig:diffuseHa}). This effect is significantly more pronounced in the IR than in H$\alpha$ or CO. The overall effect is to push the $\Sigma_{SFR}$ of faint sources to proportionally higher values than those of the bright sources, thus flattening the overall $\Sigma_{SFR}$--$\Sigma_{mol}$ trend. This is particularly problematic since the IR luminosity from the underlying galaxy is not related to current star formation, but is correlated with the cumulative stellar mass surface density of the galaxy itself (Figure~\ref{fig:diffuse}). For this reason, we adopt the results from the previous section (i.e., with the contribution of the diffuse emission removed) as our default for the rest of the analysis and discussion. 

\subsection{The Efficiency per Free--Fall Time and the Molecular Depletion Timescale}\label{subsec:efficiency}

A common expression for the SF laws is \citep{Kennicutt+1989, Elmegreen+1994, Kennicutt+1998, Elmegreen+2002, Krumholz+2005, Krumholz+2012}:
\begin{equation}
\Sigma_{SFR} = {\epsilon_{ff}\over \tau_{ff}} \Sigma_{mol},
\label{eff1}
\end{equation}
where we have specialized the expression to use the molecular gas surface density, since it is the dominant gas component for our regions. In this expression, $\epsilon_{ff}$ is the efficiency per free-fall time and $\tau_{ff}$ is the free--fall time. The timescale appearing in equation~\ref{eff1} is more generally a characteristic timescale of the gas \citep[e.g.,][]{Ballesteros+2024}, 
which different authors have adapted to different cases depending on the application. For galaxy disks, the orbital timescale $\tau_{orb}$ is often used instead of $\tau_{ff}$ \citep{Wyse+1989, Elmegreen+1997, Silk+1997, Kennicutt+1998, Krumholz+2012}. An alternative expression for equation~\ref{eff1} is given in \citet{Dib+2011a}, where $\tau_{ff}$ is replaced by $\tau_{exp}$, i.e., the time it takes for the gas to be expelled from the protocluster region, with the efficiency referring to that timescale. This alternative expression can be made equivalent to the one in equation~\ref{eff1} by deriving the conversion between $\tau_{exp}$ and $\tau_{ff}$ and between $\epsilon_{ff}$ and $\epsilon_{exp}$, as done in \citet{Dib+2011a}.

Since our regions' sizes are comparable to the sizes of HII regions, we will use $\tau_{ff}$ for our analysis. Under the assumption that the cloud is undergoing gravitational collapse, $\tau_{ff}\propto \rho^{-0.5} \sim (R/\Sigma_{mol})^{0.5}$, where $\rho$ is the volumetric density of the collapsing region and $R$ is the region's radius, for spherical approximation. This expression of $\tau_{ff}$ is appropriate when the region's size is comparable or smaller than the gas scale--height; for larger region's sizes, $R$ is commonly replaced with the scale--height itself.  Molecular scale--heights are in the range $\sim$50--200~pc within the central $\sim$4~kpc radius in disk galaxies \citep{Patra+2019}, thus we can assume that the photometric radii of our regions are comparable to or smaller than the gas scale--height. As an additional check, using the size--line~width relation \citep{Larson+1981} as revised by \citet{Wong+2019}, the velocity dispersions $\sigma_v$ of our sources yield an average R=(5$^{+5}_{-3}$)$\times$10~pc, implying that the line--of--sight radii are consistent with the photometric radii, supporting the previous statement. We, therefore, adopt the photometric radius, R$\sim$60~pc, to calculate  $\tau_{ff}$ in all three galaxies. It should be stated again that calculating $\tau_{ff}$ of a region requires the actual physical size of that region, which we do not have access to because of the resolution limit of the CO maps; thus, ours is an `average' $\tau_{ff}$, and all quantities derived from it will carry the same limitation.

The use of spherical density is, in addition, a rough approximation for calculating $\tau_{ff}$;  molecular gas is distributed in long filamentary structures within galaxies, as shown by both dust and CO maps \citep{Molinari+2010, Andre+2010, Li+2016, Mattern+2018, Su+2019, He+2026}, with star formation occurring  in correspondence of the densest regions and occupying a very small fraction of the molecular gas' volume \citep{Ballesteros+2007, Vazquez+2009, Ballesteros+2011, Padoan+2014, Dib+2020, Pokhrel+2021, Schneider+2022}. However, \citet{Hu+2022} have shown that a more careful derivation of $\rho$ has minimal impact on the resulting $\epsilon_{ff}$; the value of the latter decreases by 0.13~dex and the scatter is only reduced by 0.02~dex relative to the spherical case; these small corrections will be neglected in this analysis. 

\begin{figure}
\plottwo{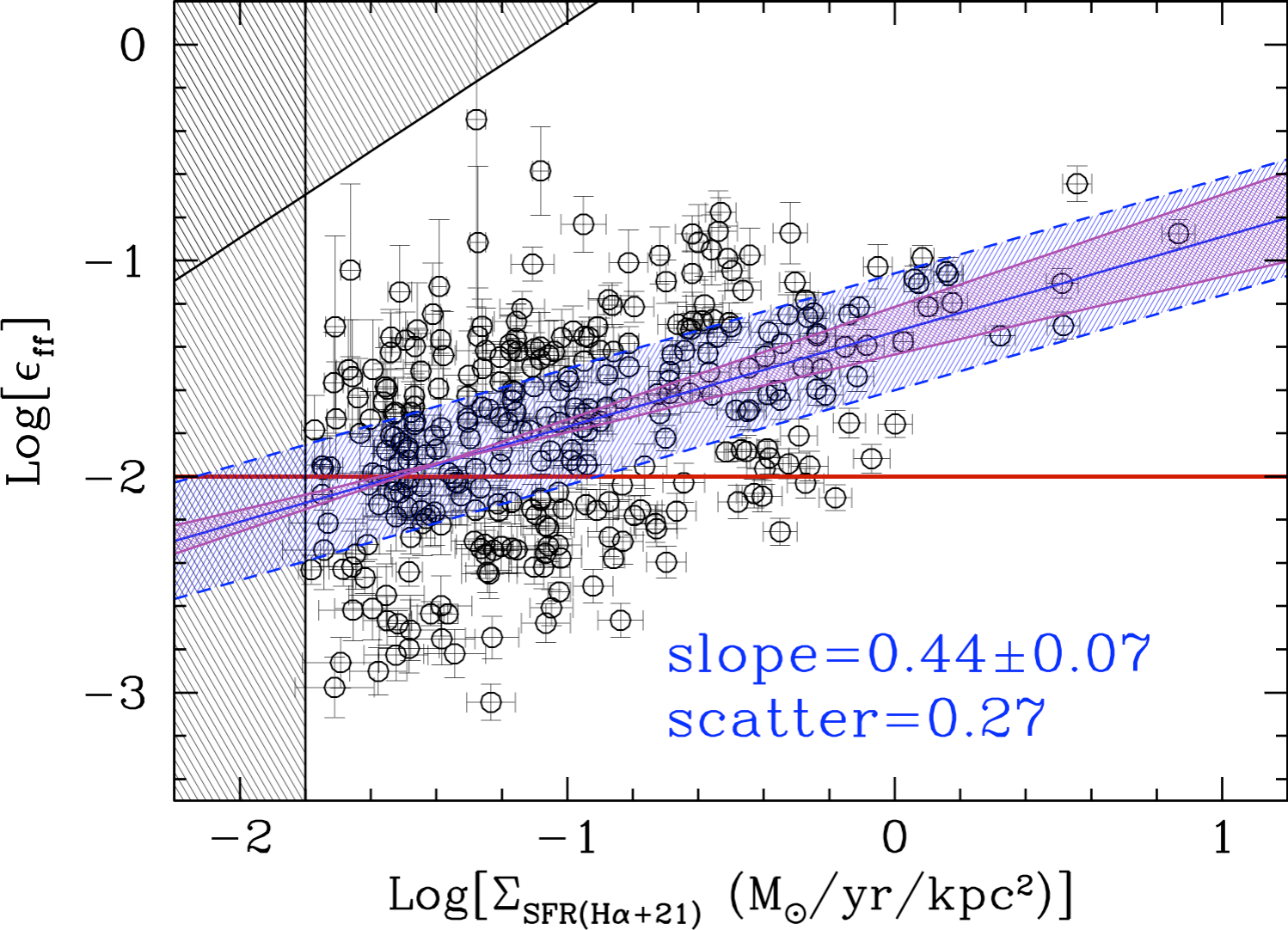}{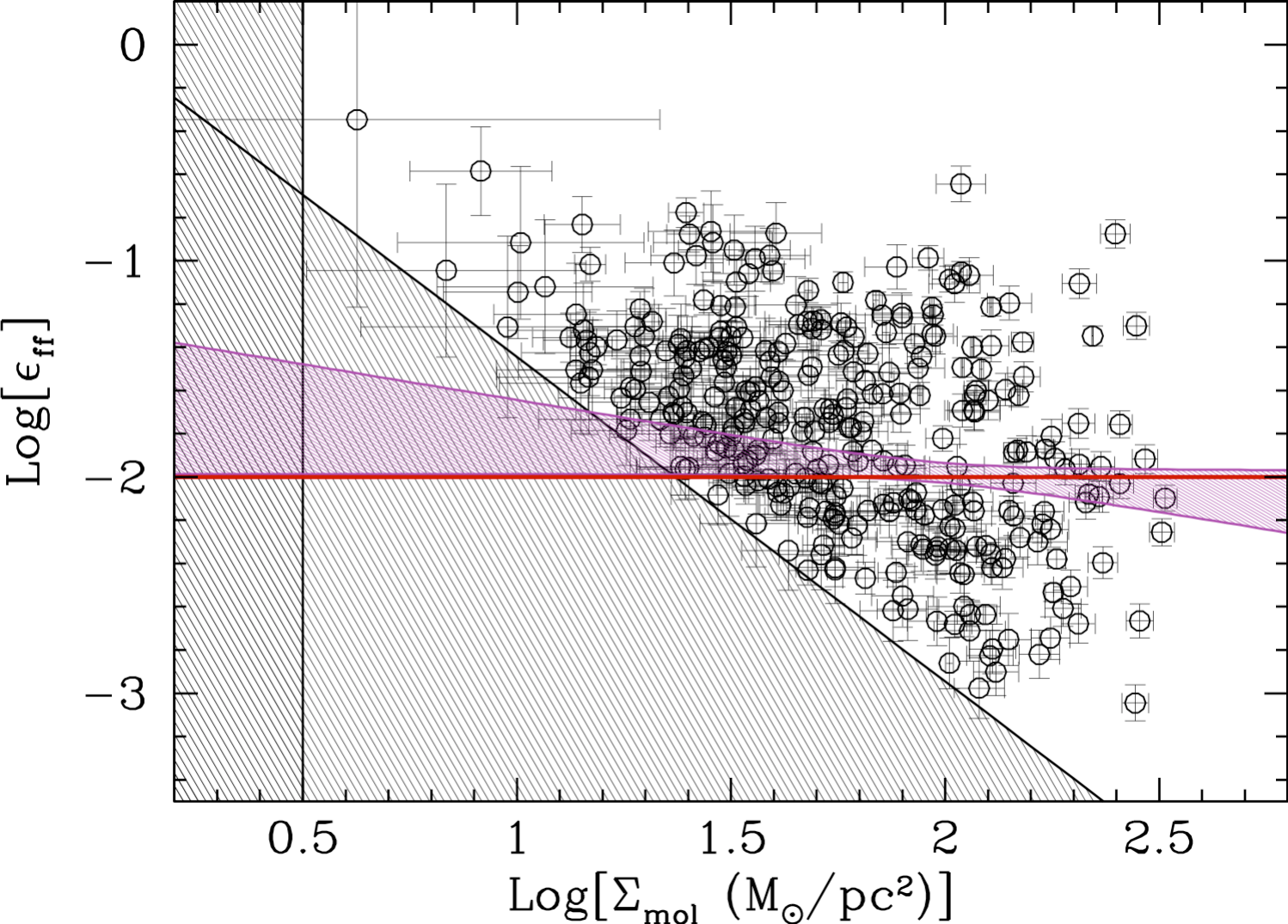}
\plottwo{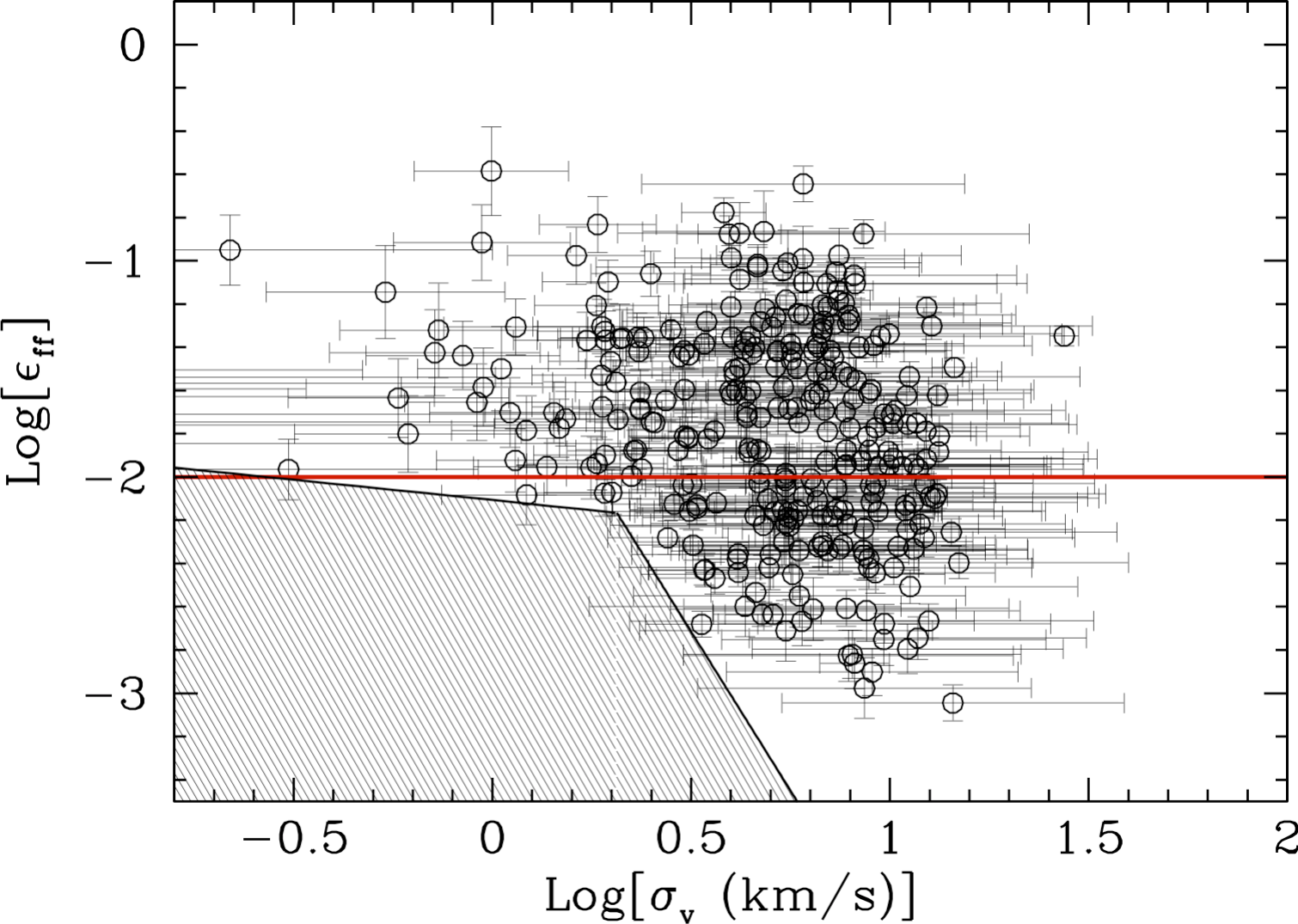}{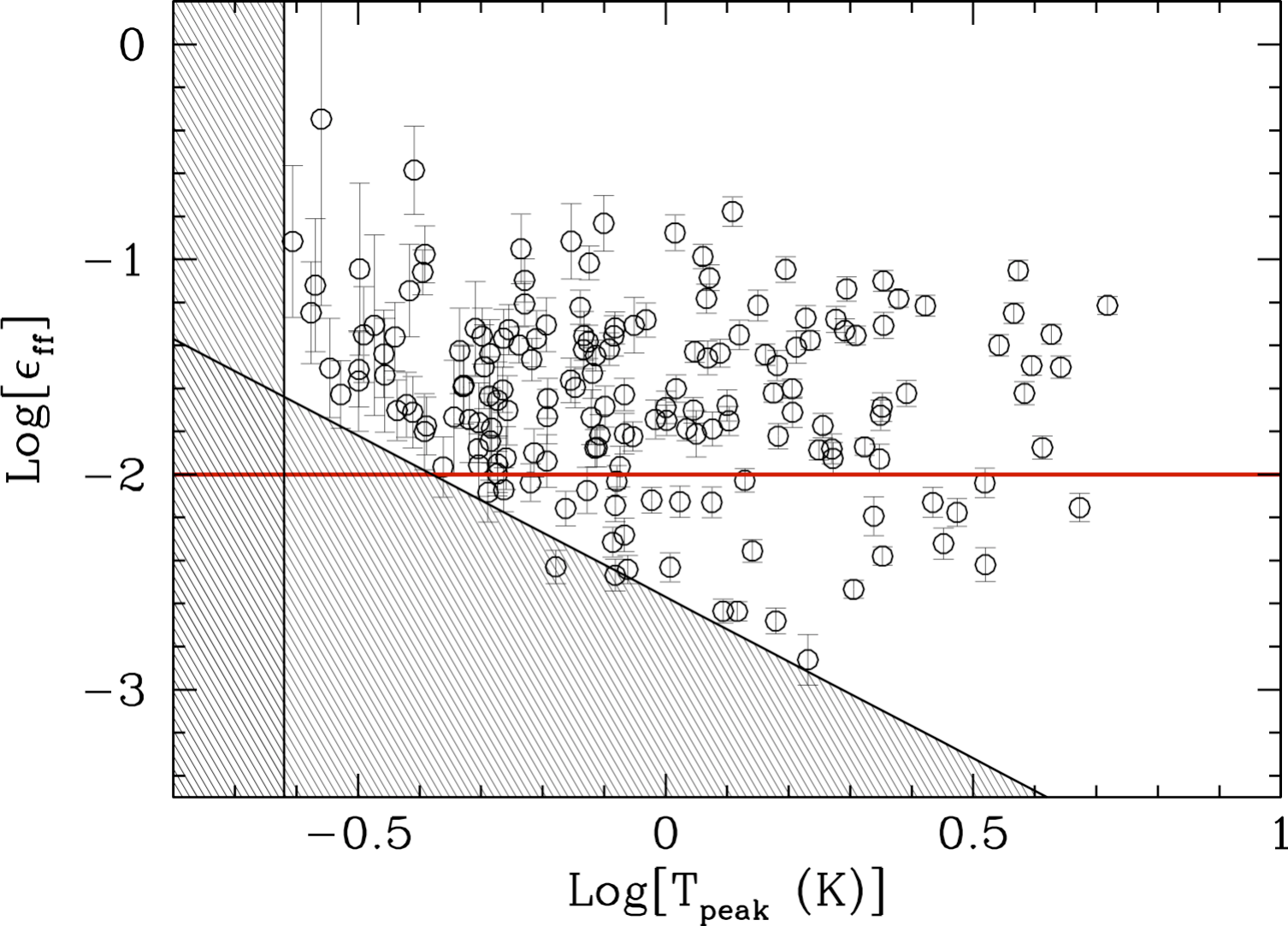}
\caption{The efficiency per free--fall time, $\epsilon_{ff}$, for the regions in our sample as a function of: $\Sigma_{SFR}$ (top--left); $\Sigma_{mol}$ (top--right); the gas velocity dispersion $\sigma_v$ (bottom--left); and the gas peak temperature T$_{peak}$ (bottom--right). The areas excluded by the limits Log($\Sigma_{SFR}$)$\ge -$1.8 and Log($\Sigma_{mol}$)$\ge$0.5 are marked as black lines with grey shaded regions.  The horizontal dark--red line shows the location of constant $\epsilon_{ff}$=0.01. The purple shaded areas in the two top panels show the 1$\sigma$ range of possible slopes returned by a forward modeling approach to the data distribution (see text). For the $\epsilon_{ff}$--versus--$\Sigma_{SFR}$ plot, the best fit line (blue solid) with its scatter (shaded blue) from the LINMIX algorithm are also shown.
} 
\label{fig:eff}
\end{figure}

With the above caveats, the expression for the efficiency per free-fall time becomes:
\begin{equation}
\epsilon_{ff} \propto{\Sigma_{SFR}\over \Sigma_{mol}^{1.5}} \sim {\Sigma_{SFR}\over (\sigma_v T_{peak})^{1.5}} ,
\label{eff2}
\end{equation}
since, in first approximation, $ \Sigma_{mol}\sim \sigma_v T_{peak}$ \citep{Leroy+2016}, and we use a constant value R=60~pc for all regions. 
Figure~\ref{fig:eff} shows the trends of $\epsilon_{ff}$ as a function of $\Sigma_{SFR}$, $\Sigma_{mol}$, the dispersion velocity $\sigma_v$ and (for NGC\,628 and NGC\,5236) the CO emission's peak temperature T$_{peak}$. The shaded areas mark the regions excluded by our limits in $\Sigma_{SFR}$ and $\Sigma_{mol}$. For $\sigma_v$ and $T_{peak}$ the excluded regions are calculated after fitting each variable as a function of $\Sigma_{mol}$. The observed trends do not change if the EW is used instead of $\sigma_v$. Any trend between $\epsilon_{ff}$ and $\Sigma_{mol}$, $\sigma_v$ and T$_{peak}$ is heavily affected by the selection limits, specifically the limit Log($\Sigma_{SFR}$)$\ge -$1.8, as shown by the shaded areas in Figure~\ref{fig:eff}. 

The trend between $\epsilon_{ff}$ and $\Sigma_{SFR}$ shows a marked positive correlation, as would be expected from the results of Table~\ref{tab:fits}. A formal fit yields:
\begin{equation}
Log(\epsilon_{ff}) = (0.44\pm0.07) Log(\Sigma_{SFR}) - (1.33\pm0.12),
\label{fit_epsilon}
\end{equation}
with scatter=0.27. We only consider the y--versus--x direction of the fit, since this relation, as well as any relation between $\epsilon_{ff}$ and $\Sigma_{mol}$, $\sigma_v$ and T$_{peak}$, is a polynomial combination of these variables, which imparts significant covariance in any fit of $\epsilon_{ff}$ relative to these observables. The only truly independent quantities in our analysis are $\Sigma_{SFR}$ and $\Sigma_{mol}$ as they come from measurements of separate datasets. A direct application of the fit results from Table~\ref{tab:fits}, i.e., $\Sigma_{SFR}\propto \Sigma_{mol}^{1.85}$ would have yielded 
$\epsilon_{ff}\propto \Sigma_{SFR}^{0.19\pm0.05}$, which is about 2.5$\sigma$ lower than the formal fit's result. 

When plotting $\epsilon_{ff}$ as a function of $\Sigma_{mol}$, $\sigma_v$ and T$_{peak}$ (Figure~\ref{fig:eff}, top--right and bottom panels), the data are  heavily censored by the limit Log($\Sigma_{SFR})\ge -1.8$ and show large scatter. For instance, the scatter plot between Log($\epsilon_{ff}$) and Log($\Sigma_{mol}$) returns $\sigma\sim$0.57. To mitigate the effects of censoring we use  the forward modeling approach described in Appendix~\ref{sec:appendixG}, recovering a 1$\sigma$ range [-0.33, 0.01] for the possible values of the slope between Log($\epsilon_{ff}$) and Log($\Sigma_{mol}$) (Figure~\ref{fig:eff}, top--right). For consistency, we use the same approach to check the distribution of slopes for Log($\epsilon_{ff}$)--versus--Log($\Sigma_{SFR}$), finding a 1$\sigma$ range [0.36, 0.52] (Figure~\ref{fig:eff}, top--left); this is in agreement with the results from the direct fit (equation~\ref{fit_epsilon}), as expected for data that are minimally affected by censoring. 
From the SF law (Table~\ref{tab:fits}), the expectation $\epsilon_{ff} \propto \Sigma_{mol}^{0.35}$ is, like in the previous case with $\Sigma_{SFR}$,  discrepant from the observed trend.

Appendix~\ref{sec:appendixF} addresses the discrepancy between the observed and the expected exponents for $\epsilon_{ff}$ with $\Sigma_{SFR}$ and $\Sigma_{mol}$, showing that it is an effect of the covariance between $\epsilon_{ff}$ and the two variables and originates from the scatter of the data about the best fit line (Appendix~\ref{sec:appendixC}). In other words, covariance accounts for the majority of the discrepancy between the slopes expected from the fits in Table~\ref{tab:fits} and the slopes measured from the data using forward modeling. With this in mind, we  conclude that the formal fits to the data (Figure~\ref{fig:eff}, top panels) yield artificial results and the underlying trends of $\epsilon_{ff}$ with both $\Sigma_{SFR}$ and $\Sigma_{mol}$ are consistent with what is inferred directly from the coefficients in Table~\ref{tab:fits}. 
Thus, the data are consistent with a steady increase by about a factor 3.5 in $\epsilon_{ff}$ across a factor $\sim$500 in $\Sigma_{SFR}$ and $\sim$30 in $\Sigma_{mol}$, although steeper increases with $\Sigma_{SFR}$ and shallower changes with $\Sigma_{mol}$ cannot be excluded given the level of the uncertainties.

Using the cloud radius determined with the relation of \citet{Wong+2019} to derive $\tau_{ff}$ and, therefore, $\epsilon_{ff}$, while tempting, is actually circular. The radius is R$\sim \sigma_v^{1.54}$ \citep{Wong+2019} and we find  $\sigma_v\sim \Sigma_{mol}^{0.6}$ for  $\Sigma_{mol}\gtrsim$10~M$_{\odot}$~pc$^{-2}$ in our regions. This implies R$\sim \Sigma_{mol}^{0.9}$, which nearly cancels out any dependence of $\tau_{ff}$ on $\Sigma_{mol}$, leaving a dependency on $\Sigma_{SFR}$ to the power of $\sim$0.5 and on $\Sigma_{mol}$ to the power of $\sim$0.8, through the correlation between the two quantities.

\begin{figure}
\plottwo{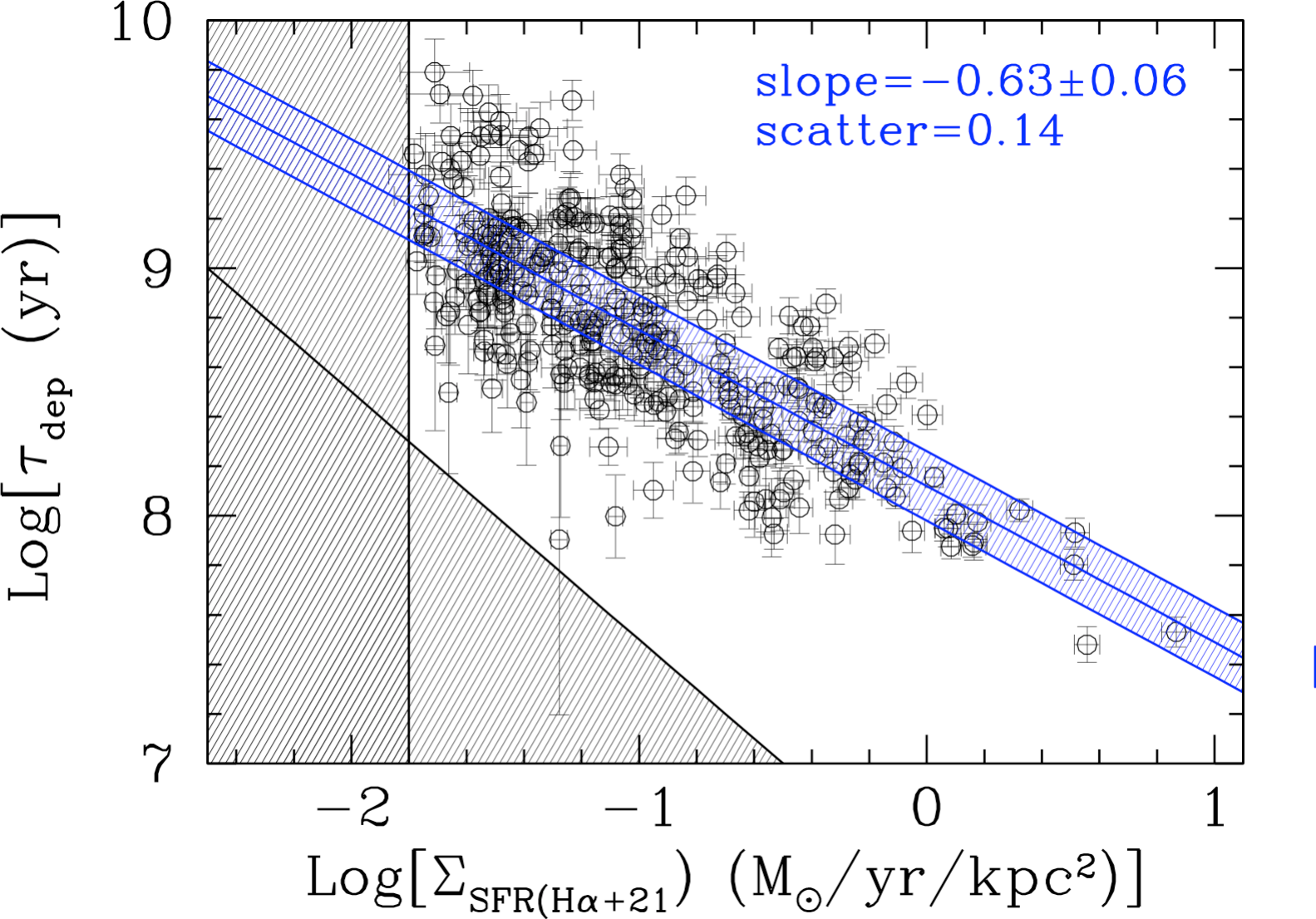}{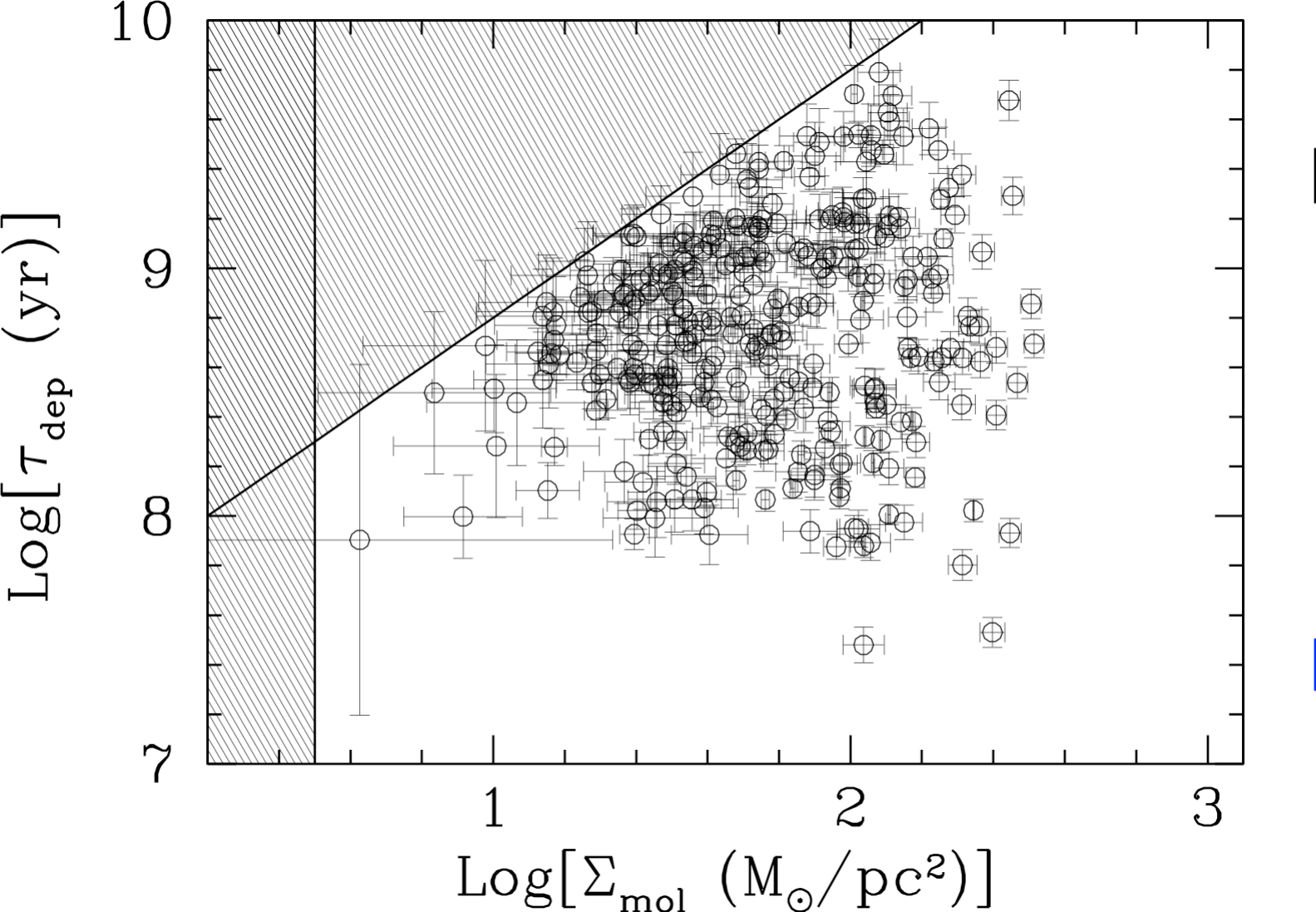}
\caption{The molecular gas depletion timescale, $\tau_{dep}$, for the regions in our sample as a function of: $\Sigma_{SFR}$ (left) and $\Sigma_{mol}$ (right). The areas excluded by the limits Log($\Sigma_{SFR}$)$\ge -$1.8 and Log($\Sigma_{mol}$)$\ge$0.5 are marked as black lines with grey shaded regions.  For the $\tau_{dep}$--versus--$\Sigma_{SFR}$ plot, the best fit line (blue solid) with its scatter (shaded blue) are also shown.
} 
\label{fig:taumol}
\end{figure}

A more direct relation between our independent variables is given by the depletion timescale \citep{Elmegreen+1997, Krumholz+2005}: 
\begin{equation}
\tau_{dep} = {\Sigma_{mol}\over \Sigma_{SFR}} = {\tau_{ff}\over \epsilon_{ff}}
\label{tau_mol_1}
\end{equation}
shown as a function of both $\Sigma_{SFR}$ and $\Sigma_{mol}$, respectively, in Figure~\ref{fig:taumol}.  As for the case of $\epsilon_{ff}$, the only notable correlation is between $\tau_{dep}$ and $\Sigma_{SFR}$, with a formal fit:
\begin{equation}
Log(\tau_{dep}) = (-0.63\pm0.06) Log(\Sigma_{SFR}) + (8.12\pm0.40).
\label{fit_taumol}
\end{equation}
and scatter=0.14. The slope $-0.63\pm0.06$ is, like in the case of $\epsilon_{ff}$,  $\sim$2.5$\sigma$ steeper than the slope of $-0.46\pm0.04$ expected from a direct application of the results in Table~\ref{tab:fits}. Again, the large scatter about the best fit of  $\Sigma_{SFR}$--vs.--$\Sigma_{mol}$ plays a role in the measured relation, and the underlying relation is likely to be $\tau_{dep}\propto \Sigma_{SFR}^{-0.5}$. For the same reason, the intrinsic relation between $\tau_{dep}$ and $\Sigma_{mol}$ is likely to have exponent $-0.85$. 

\section{Discussion}\label{sec:discussion}

\subsection{Summary of Results}\label{subsec:summary}

The sample of $\sim$350 120--pc--size star--forming regions in three nearby galaxies analyzed in this study shows that the relation between SFR and molecular gas surface densities has a slope $n\simeq$1.85 (in Log--Log scale, Figure~\ref{fig:sklaw}, left). The steepness of the SF law needs to be critically evaluated, especially in light of the fact that, as already stated above, peaks of H$\alpha$+21~$\mu$m emission do not often coincide with peaks in CO emission (Figure~\ref{fig:detail}) and that we use a fixed aperture radius of $\sim$60~pc for all our measurements. However, a re--analysis of the data using larger, 500--pc~size, regions yields the same steep value of $n$ (Appendix~\ref{sec:appendixD}), supporting the findings at HII region scales and indicating that the observed displacement between H$\alpha$+21~$\mu$m emission peaks and CO emission peaks does not influence the results.

\begin{figure}
\plottwo{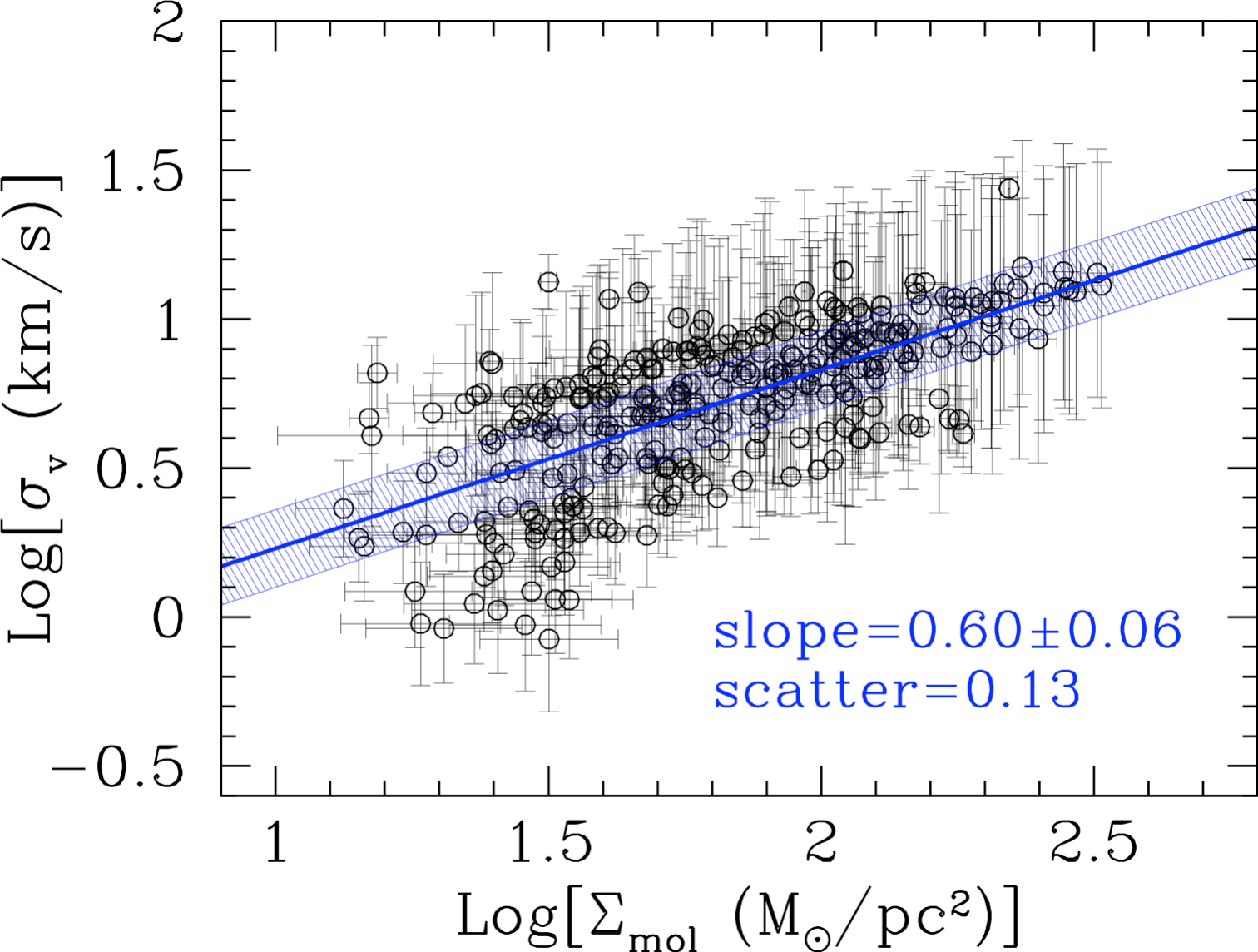}{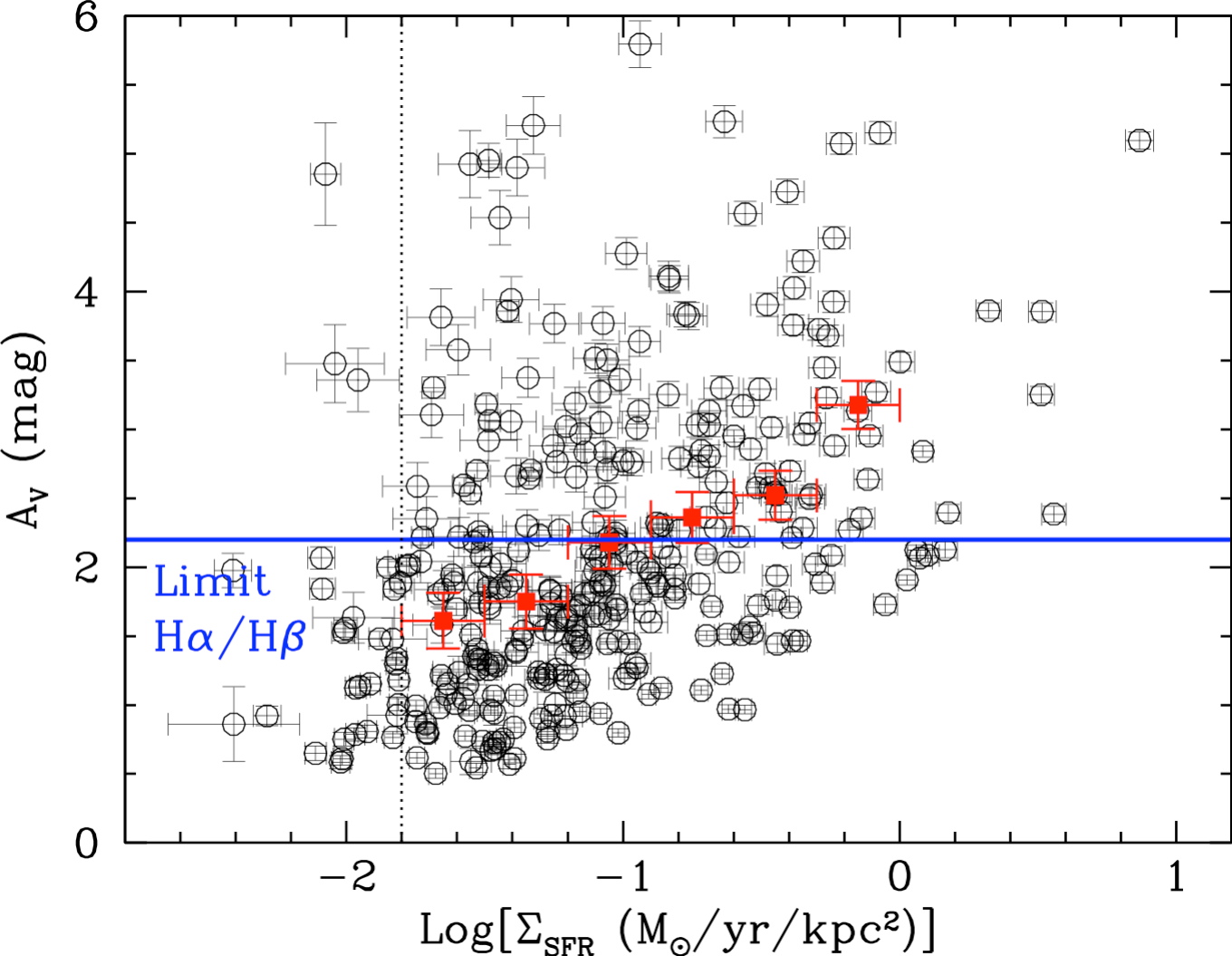}
\caption{{\bf (Left):} The distribution of dispersion velocities for the HII regions with $\Sigma_{mol}\ge$10~M$_{\odot}$~pc$^{-2}$ in the three galaxies, with the best fit shown together with its scatter (blue line and blue shaded region). The values of the best fit slope and scatter are given in the panel.  {\bf (Right):} The V--band attenuation A$_V$, calculated from the 21~$\mu$m/H$\alpha$ ratio, as a function of $\Sigma_{SFR}$, for all sources in our sample (black circles) with 1$\sigma$ uncertainties. The binned averages for data above the stochastic sampling limit (vertical dotted line) are shown as red filled squares with uncertainties. The value of A$_V$ above which the H$\beta$ recombination line is suppressed by more than one order of magnitude is marked by a blue horizontal line.} 
\label{fig:vel}
\end{figure}

As determined in section~\ref{subsec:efficiency}, the radii of the parent clouds of our sources are in the range $\sim$20--100~pc; thus, they are reasonably well sampled by our fixed aperture of 60~pc radius. This radius translates to $\Sigma_{mol}$=10$^{2.1-2.2}$~M$_{\odot}$~pc$^{-2}$ for the largest cloud fully sampled by our aperture. Removing all data above this limit (a total of 31 points), we get a slope $n=2.14\pm0.25$, which is slightly steeper, by less than 1$\sigma$, than the slope derived from the full sample. The regions are also consistent with being virialized, since we find that $\sigma_v\sim \Sigma_{mol}^{0.6\pm0.06}$ for  $\Sigma_{mol}\ge$10~M$_{\odot}$~pc$^{-2}$ (Figure~\ref{fig:vel}, left), in good agreement with expectations from the Larson's relations \citep{Larson+1981, Heyer+2009, Wong+2019}.

We investigate whether we are detecting all the star formation in our regions. We could have underestimated $\Sigma_{SFR}$, for instance, for 
geometrical reasons, e.g., the star formation is deeply embedded in dust or is emerging on the opposite side of the cloud. We exclude this possibility, for two reasons. First, from a visual inspection of our images, 
  we do not find 21~$\mu$m peaks that are completely undetected in H$\alpha$, unless they are removed from areas that contain gas -- such isolated 21~$\mu$m peaks are likely to be older stars \citep[e.g., AGB stars and dust--enveloped proto--planetary nebulae][]{Volk+2011, Preston+2025}. Absence of {\em both} H$\alpha$ and 21~$\mu$m emission would require that the star formation is opaque even at 21~$\mu$m \citep[A$_V>$70~mag,][]{Gordon+2023}. This value of the dust attenuation is significantly larger than the amount inferred from the clouds' surface densities, A$_V\sim$0.3--30~mag \citep{Bohlin+1978, Zhu+2017}, when adopting $\mu$=1.36 for the mean nucleon's weight and R$_V$=5.5~mag for dense gas \citep{Cardelli+1989, Ascenso+2013} and neglecting the contribution of HI. Second, even in the case that star formation occurs close to the edge of gas clouds and most of the cloud is optically thick, the random orientations of the clouds coupled with the higher attenuations for the larger values of $\Sigma_{mol}$ would produce a flatter SF Law than the true one, implying that $n=1.85$ would be a lower limit to the actual value. 
  
\begin{figure}
\plotone{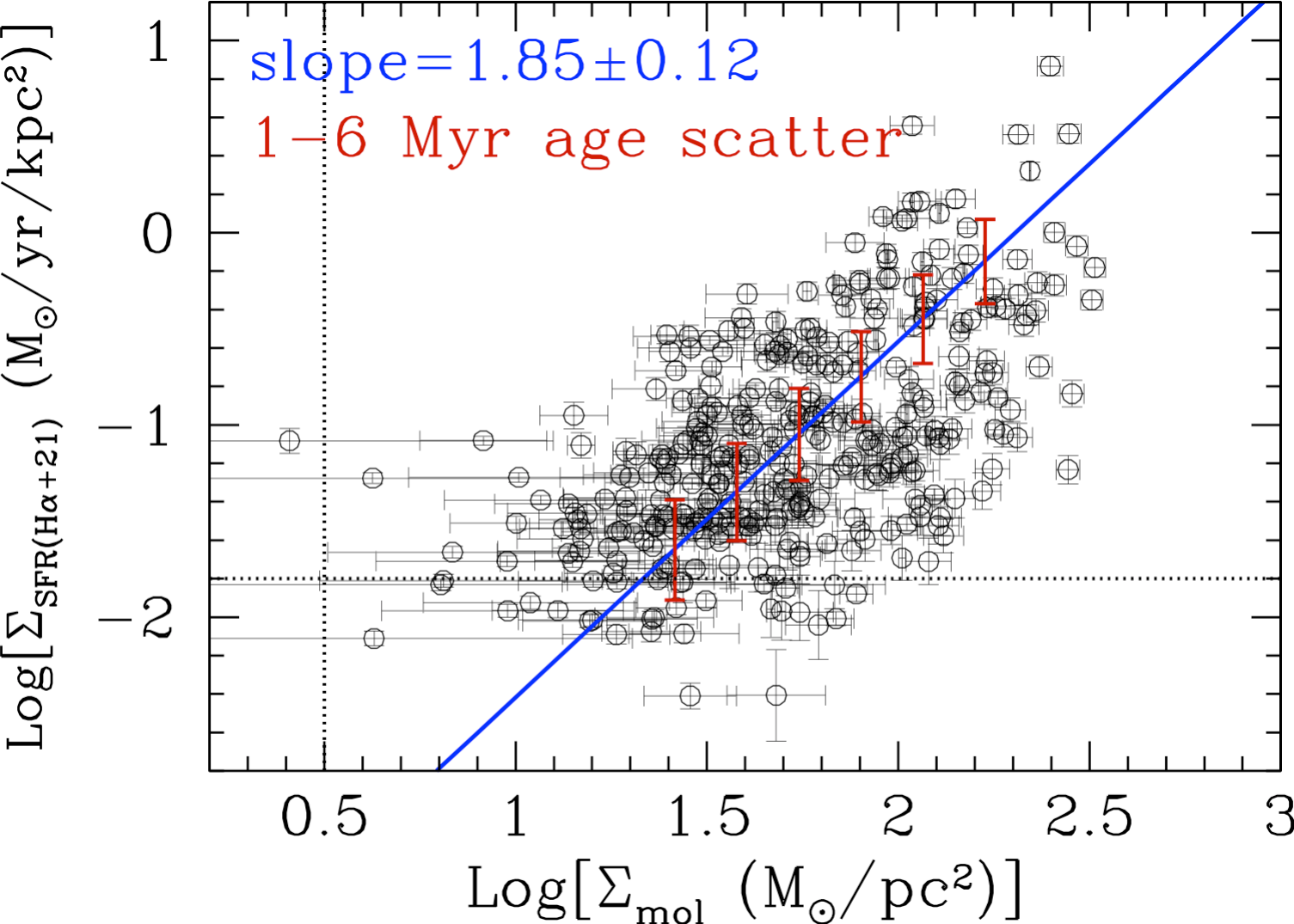}
\caption{The effects of age variations, between 1~Myr and 6~Myr (vertical dark--red bars), on $\Sigma_{SFR}$ for our regions, shown as a function of  $\Sigma_{SFR}$ for the same bins as the left--hand--side Figure; the vertical bars are centered on the mean $\Sigma_{SFR}$--$\Sigma_{mol}$ relation (blue line). The variations in $\Sigma_{SFR}$ range from 0.52~dex at the low end to 0.44~dex at the high end, mirroring the increase in the contribution of the 21~$\mu$m emission to the SFR budget from low--to--high $\Sigma_{SFR}$. } 
\label{fig:age}
\end{figure}

Although we treat our regions as having uniform ages, small age differences can be expected \citep{Calzetti+2024, Calzetti+2025}, since H$\alpha$ remains bright for the first 5--6~Myr, decreasing sharply  afterwards; at 6~Myr of age, H$\alpha$ is about a factor 30 fainter than at its peak \citep{Leitherer+1999}. The general effect on the plot in Figure~\ref{fig:sklaw} is to impart scatter along the vertical direction. We estimate the scatter induced by age variations by calculating the expected $\Sigma_{SFR}$ for instantaneous burst populations in the range 1--6~Myr,  using Starburst99 models with solar metallicity \citep{Leitherer+1999}; we adopt the average dust attenuation A$_V$ calculated in each bin of $\Sigma_{SFR}$ from the SFR(21)/SFR(H$\alpha$) of Figure~\ref{fig:sfr} and reported in Figure~\ref{fig:vel}, right. The spread in $\Sigma_{SFR}$ caused by age differences in the HII regions is shown in Figure~\ref{fig:age}, as vertical dark--red bars, centered on the mean $\Sigma_{SFR}$--$\Sigma_{mol}$ relation (blue line), for each $\Sigma_{SFR}$ bin. The age--induced scatter is $\lesssim$0.5~dex, top--to--bottom, smaller than the observed scatter but still large enough to explain part of it. The calculated scatter, 0.5~dex,  is smaller than the scatter expected in H$\alpha$ alone, $\sim$1.5~dex, for the same age variations. The reason for the smaller scatter is because SFR(21) dominates over SFR(H$\alpha$) in all bins of  $\Sigma_{SFR}$ (Figure~\ref{fig:sfr}, left), and SFR(21) is mostly due to the dust heated by the non--ionizing UV and optical emission of the regions, which varies less with age than the ionizing photon flux.  Adopting a model of star formation that increased the mass of young stars over 3--4~Myr, as opposed to an instantaneous burst, produces a similar scatter of 0.5~dex \citep{Corbelli+2025}.  The conclusion from this test is that age variations in the regions do not account for the spread in the data, which is likely intrinsic and possibly due to variations in the efficiency of star formation from region to region. 
  
The use of an expression for $\alpha_{CO}$ that depends on the weight of the galaxy's disk (intended here as the stellar+molecular gas mass surface density), while consistent with recent findings \citep{Sandstrom+2013, Chiang+2024}, has not been common in the literature. A re-analysis of the HII regions' data using a constant, MW--like $\alpha_{CO}$ value finds an overall shallower relation for $\Sigma_{SFR}$--$\Sigma_{mol}$, with slope $n\sim$1.3 (Appendix~\ref{sec:appendixE}). This shallower slope is, however, the effect of a shift in CO luminosity between NGC\,628 and the two other galaxies, NGC\,5194 and NGC\,5236; these two galaxies have 
larger average $\Sigma_{*, gal}$ values than NGC\,628 (Figure~\ref{fig:diffuse}, right), suggesting larger disk weights and, as a result, larger CO luminosities relative to those of NGC\,628. When fitting the HII regions' $\Sigma_{SFR}$--$\Sigma_{mol}$ within each individual galaxy, we get $n\sim 1.75-1.90$ for constant $\alpha_{CO}$,  consistent with the overall slope of 1.85. The basic conclusion is that equation~\ref{alphaco} does not steepen existing $\Sigma_{SFR}$--$\Sigma_{mol}$ relations, but simply removes offsets in CO luminosity due to differences in the weight of galactic disks. 
  
The efficiency per free-fall time shows the following trends: $\epsilon_{ff}\propto\Sigma_{SFR}^{a}$ with a=[0.36,0.52] and $\epsilon_{ff}\propto\Sigma_{mol}^{b}$ with b=[-0.33,0.01], see section~\ref{subsec:efficiency}. These are in apparent contradiction with expectations from  the best fit between $\Sigma_{SFR}$ and $\Sigma_{mol}$, which predict: $\epsilon_{ff}\propto\Sigma_{SFR}^{0.19}$ and $\epsilon_{ff}\propto\Sigma_{mol}^{0.35}$. Monte Carlo sampling of simulated data that follow the best fit and scatter in $\Sigma_{SFR}$--$\Sigma_{mol}$ as the real data demonstrate that the measured exponents for $\epsilon_{ff}$ are not the intrinsic ones (Appendix~\ref{sec:appendixF}). The measured exponent is more negative than the intrinsic one for $\epsilon_{ff}$--$\Sigma_{mol}$ and more positive for $\epsilon_{ff}$--$\Sigma_{SFR}$. Furthermore, the differences between intrinsic and measured slopes increase for increasing scatter:  the exponent of $\Sigma_{mol}$ becomes progressively more negative and the one of $\Sigma_{SFR}$ more positive as the scatter in the data grows. The covariance between the independent and dependent variables, coupled with the spread in the data, masks the intrinsic trends, which are likely close to expectations. 
 
A key assumption for deriving efficiencies is that the timescales probed by the SFR tracer are comparable with or smaller than the free--fall timescales \citep[e.g.,][]{Ballesteros+2024}; star formation timescales longer than the typical free--fall time would measure the time--integrated SFR in the region over multiple free--fall timescales. The values of $\tau_{ff}$ for our sources peak at 10~Myr, with  standard deviation $\pm$5~Myr, and full range $\sim$4--20~Myr, which are consistent with or slightly longer than the timescales probed by our mixed H$\alpha$+mid--IR SFR indicator. These free--fall timescales are calculated for a fixed radius of $\sim$60~pc, which is the minimum spatial scale we can use given the angular resolution of the CO maps.

We can also exclude major effects of molecular gas processing or destruction within our regions. Gas processing/destruction could be a concern, if present, as it could alter the overall measurements, for instance by increasing scatter and/or producing artificial trends. There are two lines of reasoning that argue against such effects. First, the re--analysis of the regions at 500~pc scale yields the same results as the 120~pc scale (Appendix~\ref{sec:appendixD}). Second, when the diffuse emission is not removed from the data, the relation between $\Sigma_{SFR}$ and $\Sigma_{mol}$ for the star forming regions is very similar to the relations measured for 0.8--1.5~kpc regions by many authors (section~\ref{subsec:nobck_removal} and Figure~\ref{fig:sklaw}, right) 
and the scatter=0.11 of our data about the best fit relation  is significantly smaller than the scatter ($\sim$0.3) found by, e.g., \citet{Sun+2023}. We compare our results with those from these authors, as they utilize the same fitting algorithm as we do. Both our 500~pc regions and \citet{Sun+2023}'s 1.5~kpc regions  are sufficiently large to average out effects of gas destruction and formation, indicating that also our measurements are mostly free from biases from such effects. 
Furthermore, none of our regions has extremely high values of $\epsilon_{ff}$ (Figure~\ref{fig:eff}, top--left), being typically around 1\%, with excursions to values as high as $\sim$10\% and as low as $\sim$0.1\%. If the regions with $\epsilon_{ff}$=10\% were to have  `true'  $\epsilon_{ff}$=1\%, this would require that about 10 times more molecular gas than currently present would have been around them originally; such an assumption would push some regions to values of $\Sigma_{mol}\gtrsim$10$^{3.5 - 3.8}$~M$_{\odot}$~pc$^{-2}$. In the MW, gas surface densities $\sim$10$^{3.2}$~M$_{\odot}$~pc$^{-2}$ are only observed within the inner 2--3~pc region of molecular clouds, while clouds in nearby galaxies imaged in $^{12}$CO rarely go above $\sim$10$^{3}$~M$_{\odot}$~pc$^{-2}$  \citep{Rosolowsky+2021, Pathak+2025}.

\subsection{Comparisons with Previous Results}\label{subsec:previous}

Studies that remove the underlying galaxy emission from the SFR tracers of resolved regions within galaxies have been conducted at the lower resolution of the Spitzer/MIPS~24$\mu$m observations (FWHM$\sim$6\farcs5), which has limited the spatial resolution probed. \citet{Liu+2011} pushed the spatial resolution to $\sim$250~pc in two galaxies, NGC\,5194 and NGC\,3521, finding slopes $n\sim$1.4--1.9. Similarly, \citet{Rahman+2011} found n$\sim$1.5 in one galaxy, NGC\,4254. \citet{Momose+2013} analyzed a sample of 10 nearby galaxies down to 500~pc resolution, finding $n\sim1.8$. Using a smaller sample of three galaxies (NGC\,3627, NGC\,5055 and NGC\, 5236), \citet{Morokuma+2017} derived $n\sim$1.7 for 600--900~pc regions. \citet{Kumari+2020} analyzed nine nearby galaxies down to $\sim$500~pc; taking only the seven galaxies with inclination $<$50$^o$, they derive an average slope n$\sim$1.5. 
All these results are obtained assuming a constant value of $\alpha_{CO}$ for the galaxies in each sample. Furthermore, we only report the slopes of the molecular gas--SFR relations derived by each author, while galaxies are often dominated by HI gas outside of the central regions. When taking into account these differences relative to our treatment, which involves only H2--dominated regions and uses $\alpha_{CO}$ from equation~\ref{alphaco}, those previous results are consistent with our finding that $n\sim1.88$ in 500~pc regions 
(Appendix~\ref{sec:appendixD}). Interestingly, \citet{Liu+2011} find that $n$ decreases for increasing region's size, from $\approx$1.4--1.9 at 250~pc scale down to $n\sim$1.2 at 1~kpc scale, possibly owing to increasing difficulty in identifying the diffuse emission at larger spatial scales. 

The works listed above use different approaches to identifying and removing the diffuse emission of the underlying galaxy from the sources. \citet{Liu+2011},  \citet{Momose+2013} and  \citet{Morokuma+2017} use the package HIIPhot \citep{Thilker+2000} to identify local regions of emission and then model the diffuse emission.  \citet{Rahman+2011} employ unsharp masking, but also remarks that the size of the kernel used in the algorithm determines the fraction of diffuse emission recovered and, ultimately, the slope of the SF relation. \citet{Kumari+2020} use the Nebulosity Filter software\footnote{http://casu.ast.cam.ac.uk/publications/nebulosity-filter}, which is based on a iteratively--clipped, two--dimensional median filtering method to separate diffuse emission from the more compact emission (HII regions). In the present work, the diffuse emission is calculated as the mode of pixels values within annuli surrounding the sources, after iteratively sigma--clipping compact emission. Despite the different approaches adopted, all authors obtain relatively steep relations, $n \sim$1.5--1.9, for $\Sigma_{SFR}$--vs.--$\Sigma_{mol}$. This lends additional support, beyond the arguments presented in section~\ref{sec:diffuse}, to the methodology adopted in this work for the removal of the underlying galaxy emission.

Recently, \citet{Pathak+2025} investigated the properties of 18,000 HII regions at 150~pc scale in 19 nearby galaxies, finding a shallow positive correlation between the mass of young stars and $\Sigma_{mol}$. For reference, their results can be related to ours by recalling that $\Sigma_{SFR}$=10$^{-1.8}$~M$_{\odot}$~yr$^{-1}$~kpc$^{-2}$ corresponds to a stellar mass $\sim$3,000~M$_{\odot}$ for a 4~Myr old region with 60~pc radius. Thus, the conversion from $\Sigma_{SFR}$ to M$_{young\ stars}$ is a rigid shift along the vertical axis. As a result, those authors' trend is shallower than the one we find. However, while \citet{Pathak+2025}'s work is, in terms of physical sizes probed, the closest to the present analysis, it also differs in a few aspects. \citet{Pathak+2025} uses the Balmer decrement (H$\alpha$/H$\beta$) to obtain extinction--corrected H$\alpha$ luminosities, from which SFRs and the total masses in young stars are derived. The blue and weak H$\beta$ line ($\lambda$0.4861~$\mu$m) is heavily affected by dust attenuation, and its luminosity is suppressed by an order of magnitude already for A$_V\simeq$2.2~mag. Our sample includes a significant number of sources with higher A$_V$ values (Figure~\ref{fig:vel}, right). Furthermore, as the average A$_V$ increases with increasing $\Sigma_{SFR}$, the luminosity of higher $\Sigma_{SFR}$ regions is proportionally more affected by dust attenuation. For instance, at Log($\Sigma_{SFR}$)$> -$0.2, the average A$_V$=3.2~mag and the H$\beta$ flux is suppressed by a factor of 30. A second difference between our work and \citet{Pathak+2025}'s is in the lowest mass limit  considered: we only use regions with a cumulative mass  $\gtrsim$3,000~M$_{\odot}$ in young stars to ensure a reliable conversion between H$\alpha$ luminosity and SFR, while \citet{Pathak+2025} include regions as low as a few 10's~M$_{\odot}$. Because of these significant differences, comparing the two works is challenging. 

Local samples of starburst galaxies tend to mostly include infrared--bright sources (U/LIRGs), which we use to compare with our results for HII regions. As the starburst generally dominates the galaxy's luminosity output at all wavelengths, we do not remove the underlying galaxy emission from the starburst's light. \citet{Kennicutt+2021} collected and investigated a sample of 112 nearby circumnuclear starbursting regions and starbursts; of this sample, we only use the 89 starbursts and discard the 23 circumnuclear regions, owing to the presence of the underlying galaxy's diffuse  contribution in the these regions that we cannot easily remove. These authors utilize the IR emission as a SFR indicator, which makes an appropriate term of comparison with our hybrid SFR determinations as the optical emission is sub--dominant in U/LIRGs. \citet{Wilson+2019} observed a total of 53 $\sim$400--500~pc regions in five U/LIRGs, using free--free emission as a SFR indicator. We use the three galaxies in common between the two samples to bring the free--free SFR calibration into agreement with the IR SFR one, requiring the former to be increased by $\sim$45\%. For both samples, we use equation~\ref{alphaco} to recalculate $\Sigma_{mol}$, adopting $\Sigma_{tot}\sim \Sigma_{mol}$; this choice is justified because the average $\Sigma_{*, gal}\sim 100$~M$_{\odot}$~pc$^{-2}$ \citep{Shangguan+2019}, which is much smaller than $\Sigma_{mol}$ for the vast majority of the U/LIRGs in the two samples. The 89 U/LIRGs and 53 regions from five U/LIRGs are shown in a $\Sigma_{SFR}$--vs.--$\Sigma_{mol}$ plot in Figure~\ref{fig:highz}, left, with magenta and dark--red symbols, respectively. Both samples represent reasonably well a high--$\Sigma_{mol}$ extension of the trend from our sample of HII regions, providing a coherent picture: when star formation dominates the luminous output of a region or galaxy, the SF relation is described by a relatively tight relation with slope n$\sim$1.85 over 3 orders of magnitude in molecular gas surface density. 

The absence of, or slightly negative, trend between $\epsilon_{ff}$ and $\Sigma_{mol}$  in our data (Figure~\ref{fig:eff}, top--right) reflects similar findings by other authors \citep[e.g.,][]{Utomo+2018, Leroy+2025, Meidt+2025}. We show that the measured trend does not reflect the true trend, since it is heavily affected by the scatter of the data about the best fit between $\Sigma_{SFR}$ and $\Sigma_{mol}$, coupled with the covariance between $\epsilon_{ff}$ and $\Sigma_{mol}$ (Appendix~\ref{sec:appendixF}). The larger the scatter, the more negative the measured exponent of $\epsilon_{ff}$--$\Sigma_{mol}$. For our HII regions, the intrinsic trend between the two variables is consistent with being a power law with a slightly positive exponent, as expected from the fits in Table~\ref{tab:fits}, implying a mean efficiency that increases by a factor $\sim$3.5 between $\Sigma_{mol}$=10~M$_{\odot}$~pc$^{-2}$ and $\Sigma_{mol}$=300~M$_{\odot}$~pc$^{-2}$. Although the increase in $\epsilon_{ff}$ we observe is modest, the trend is in the direction of supporting models that require large efficiencies at high gas masses and densities to form massive bound star clusters \citep[e.g.,][]{Polak+2024}. We caution that, due to the large uncertainties in the measured trend between $\epsilon_{ff}$ and $\Sigma_{mol}$ these conclusions are not definitive, and a more extensive study that includes a larger dynamical range for $\Sigma_{mol}$ while minimizing the effects of covariance is required to obtain a firmer result. \citet{Lee+2016} and \citet{Ochsendorf+2017} find a negative trend between $\epsilon_{ff}$ and cloud mass, in agreement with the model of \citet{Dib+2011a}. Other authors have suggested that a correlation between $\epsilon_{ff}$ and cloud mass may be spurious, caused by the destruction of star-forming clouds by feedback on small scales \citep{Feldmann+2011a, Kruijssen+2014, Krumholz+2019}. Since we do not have access to cloud masses, we cannot perform a direct comparison between those and our results.

The average efficiency is still low, $\epsilon_{ff}\sim$1\%, but with a large measured scatter, about a factor of 4,  as inferred from Figure~\ref{fig:eff}. 
That the conversion of gas to stars proceeds with low efficiency in the Milky Way has been known for over five decades \citep{Zuckerman+1974, Elmegreen+1977, Myers+1986, Lada+1987}; it has been repeatedly confirmed by modern observations  to be around 1\%--4\% for both galactic molecular clouds and extragalactic $\sim$kpc--sized regions 
\citep[e.g.][]{Leroy+2008, Evans+2009, Gutermuth+2011, Krumholz+2012, Lee+2016, Utomo+2018, Pokhrel+2021, Hu+2022, Leroy+2025, Meidt+2025}. Estimates of the scatter in the efficiency in Milky Way clouds vary either by $\sim$one order of magnitude \citep{Evans+2014, Heyer+2016, Lee+2016} or by less than a factor of 2  \citep{Evans+2009, Pokhrel+2021, Hu+2022}. However, these differences are mostly driven by the approach to evaluating $\epsilon_{ff}$; \citet{Lee+2016} analyzed a sample of $\sim$200 MW molecular clouds deriving cloud--to--cloud variations, while \citet{Pokhrel+2021} analyzed only a dozen clouds, deriving the free--fall time (and the efficiency) from decreasing volumes of each cloud to capture increasing densities and SFRs. 
The latter approach is different from the approach employed in extragalactic studies, which are resolution--limited and generally only utilize a single region size and compile large samples of independent regions. With this caveat, the scatter in the efficiency for our extragalactic star forming regions is 2.5$\times$ larger than what determined by \citet{Pokhrel+2021} and \citet{Hu+2022}, but 2$\times$ smaller  than what found by \citet{Lee+2016} for Milky Way clouds. 

The decreasing $\tau_{dep}$ with increasing $\Sigma_{SFR}$, $\tau_{dep}\propto \Sigma_{SFR}^{-0.6}$ (or $\tau_{dep}\propto \Sigma_{SFR}^{-0.5}$ if taking covariance into account), is a direct consequence of the SF law we find for our sample of HII regions. The decrease in molecular depletion timescale for increasing activity, as expressed by the SFR, is an established result for galaxies. For nearby galaxies, \citet{Saintonge+2011} found that  $\tau_{dep}\propto sSFR^{-0.5}$, although a shallower dependency, with slope (in log--log scale) of $-0.24$, with $\Sigma_{SFR}$; similarly, \citet{Kennicutt+2021} find a slope $\sim -0.5$ for the relation between $\tau_{dep}$ and SFR. \citet{Genzel+2015} and \citet{Tacconi+2018} express this trend as a function of the deviation from the Main Sequence of Star Formation \citep[SFR--stellar mass relation, e.g.,][]{Speagle+2014} at constant stellar mass, finding also slope of $\sim -0.5$. Finally, \citet{Sun+2025} determines that $\sim$kpc--sized regions in nearby galaxies show a tight relation between $\tau_{dep}$ and sSFR, with slope $\sim -0.9$. Our results add a datapoint at the level of $\approx$100~pc region sizes consistent with results at larger scales.

\subsection{Connection to the High Redshift}\label{subsec:highz}

The available spatially--resolved observations of galaxies at redshift z$\gtrsim$1 show that the clouds in those galaxies tend to have larger gas and SFR surface densities and larger velocity dispersions than the gas clouds in the local universe \citep[e.g.,][]{Dessauges+2023, Accard+2025}. This is the result of the combination of multiple factors, including that high--redshift galaxies have higher molecular gas masses and higher SFRs overall than local galaxies, a more turbulent ISM and molecular depletion timescales that decrease with increasing redshift \citep{Genzel+2015, Tacconi+2018, Tacconi+2020}. 

\begin{figure}
\plottwo{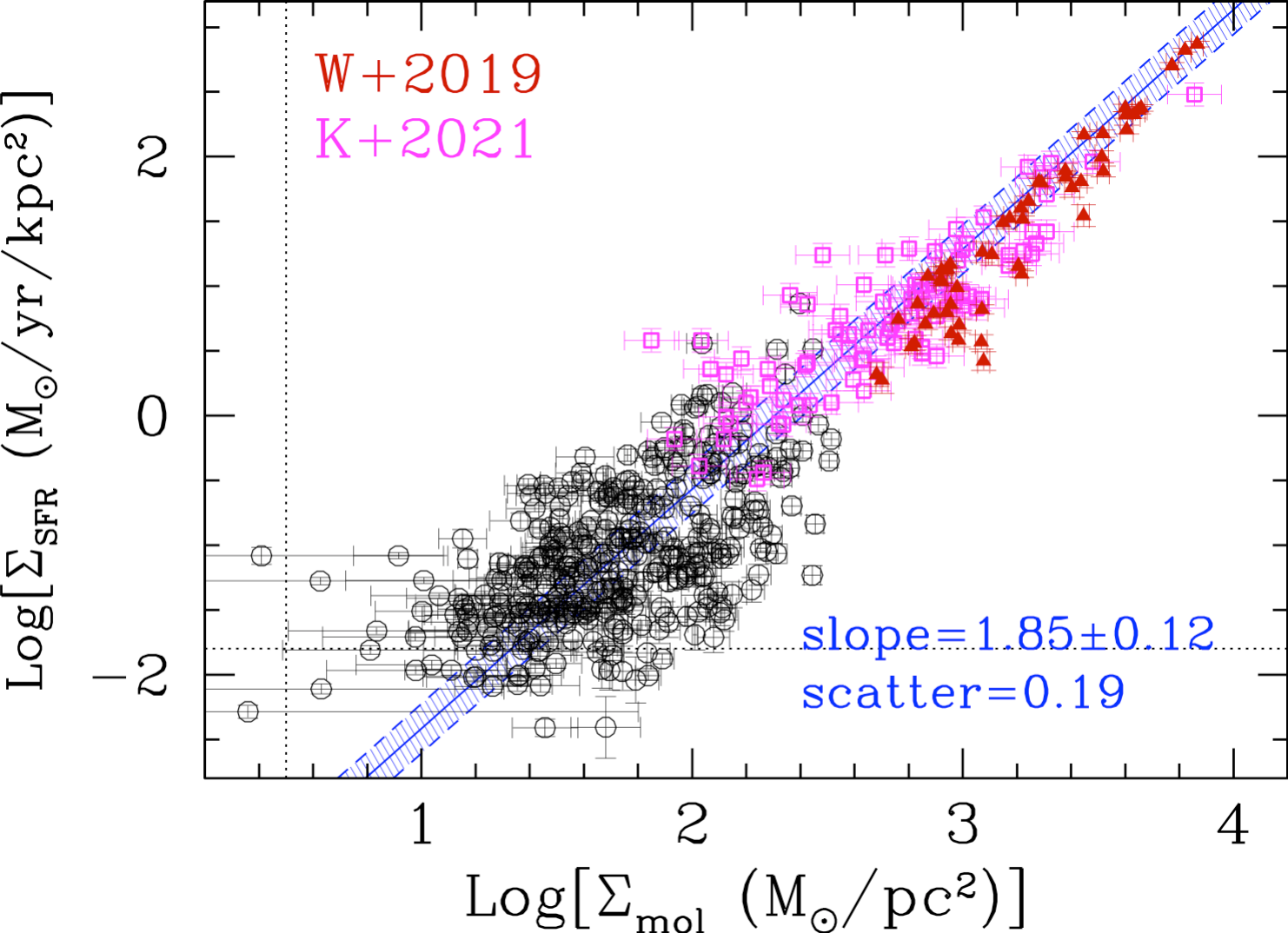}{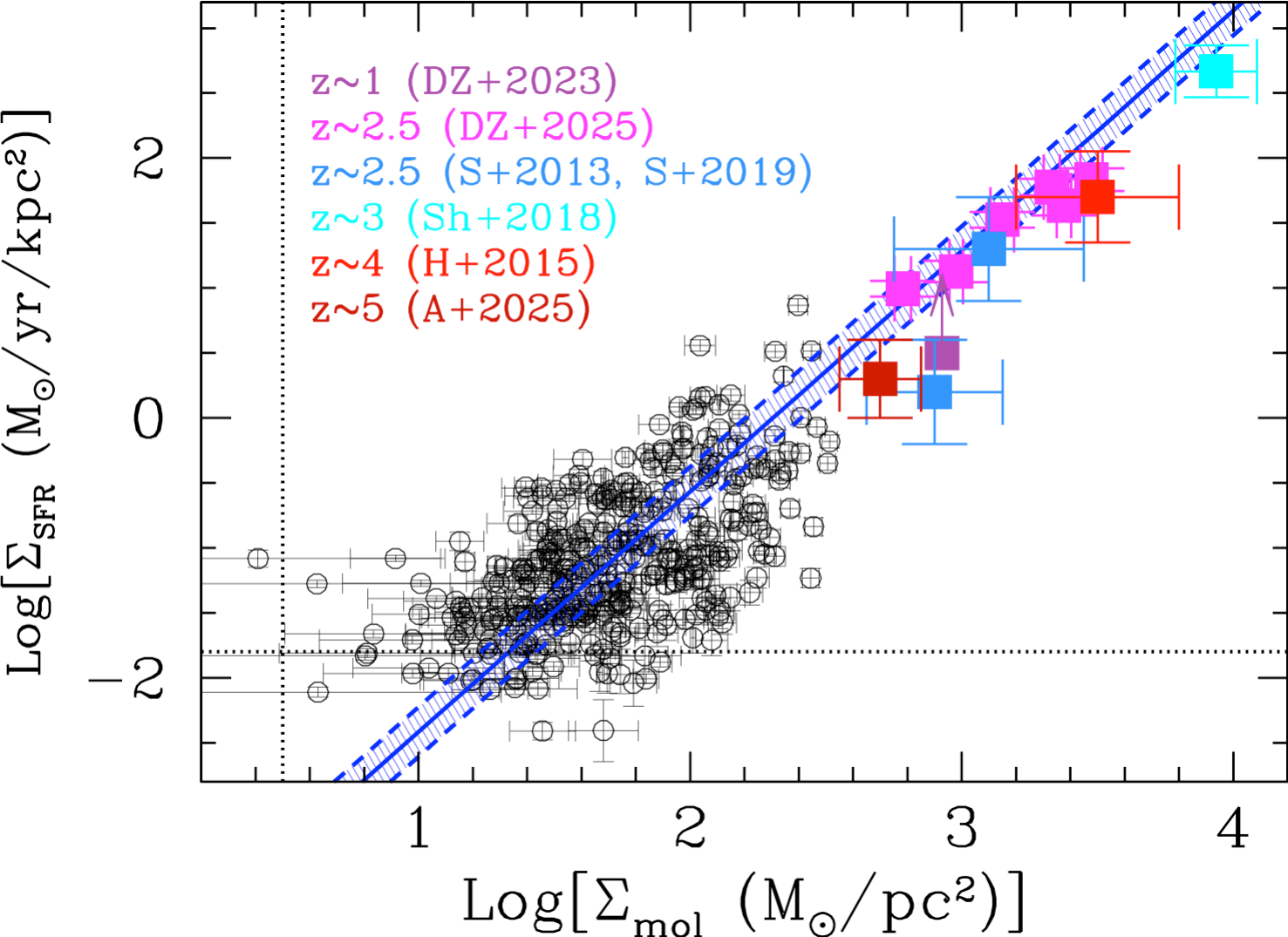}
\caption{{\bf (Left):}  The molecular SF law of the local HII regions in comparison with that of local infrared--bright starbursts \citep[U/LIRGs, magenta squares,][]{Kennicutt+2021} and $\sim$400--500~pc regions in a few local U/LIRGs \citep[dark--red triangles,][]{Wilson+2019}. The blue line and shaded region are the best fit for the HII regions (see Figure~\ref{fig:sklaw}, left).  {\bf (Right):} The molecular SF law of the local HII regions in comparison with that of the average values of resolved regions at redshifts: z$\sim$1 \citep[purple square]{Dessauges+2023, Messa+2022}, z$\sim$3 \citep[cyan square,]{Sharda+2018}, z$\sim$4 \citep[red square,]{Hodge+2015}  and z$\sim$5 \citep[dark--red square][]{Accard+2025} and that of a few starburst galaxies at z$\sim$2.5 \citep[blue and magenta squares with uncertainties][]{Sharon+2013, Sharon+2019, Dessauges+2025}. The z$\sim$1 point is shown as a lower limit as some star--forming regions are unresolved \citep{Messa+2022}. The blue lines and shaded area are the best fit from Figure~\ref{fig:sklaw}, left. 
} 
\label{fig:highz}
\end{figure}

Figure~\ref{fig:highz}, right, shows the data for a few high--z measurements in comparison with the SF law we derive from the HII regions in our sample. For this comparison, we assume that at high--z  the subtraction of the underlying galaxy's diffuse light from the star forming regions' emission is not necessary, since the star formation is sufficiently luminous to overshine the host galaxy. Furthermore, as the gas measurements are heterogeneous, we do not implement a disk--weight correction for $\alpha_{CO}$, but simply adopt each author's measurements. This may lead to overestimated $\Sigma_{mol}$ for the high redshift sample relative to the local HII regions.

We combine the resolved regions' data of the z$\sim$1 lensed system A521-sys1 into a single average point in $\Sigma_{SFR}$ and $\Sigma_{mol}$ because the regions detected in CO are spatially distinct from those showing star formation activity and some level of homogeneity in the star formation activity needs to be assumed \citep{Dessauges+2023, Messa+2022}. Some of the star--forming regions in this system are also unresolved, hence the data shown as a lower limit in Figure~\ref{fig:highz}, right. The highest $\Sigma_{SFR}$ point in Figure~\ref{fig:highz}, right, is from the central $\sim$150~pc starburst clump in the z$\sim$3 SDP 81 galaxy \citep{Sharda+2018}, for which we recalculate the SFR from the IR luminosity using the conversion factor in \citet{Kennicutt+2012}. \citet{Hodge+2015} resolves the z$\sim$4 starburst galaxy in 15 separate kpc--sized clumps, for which we report the median  value in both $\Sigma_{SFR}$ and $\Sigma_{mol}$, since the data for the individual clumps are not tabulated. The $\sim$1.5~kpc regions in galaxies at z=5 \citep{Accard+2025} are also combined into a single data point, for the same reason; we utilize the mean value of the SF law obtained by \citet{Accard+2025} using a surface--density--dependent [CII]--to-gas conversion factor. The uncertainty in both the z$\sim$4 and z$\sim$5 median values represents scatter in the data. \citet{Dessauges+2025} recently reported SFRs, molecular gas masses (from CO) and sizes for a small sample of starburst galaxies at z$\sim$2.5; of this sample, we use only the six galaxies with detections in both SFR and molecular gas. Finally, the two z$\sim$2.5 starbursts from \citet{Sharon+2013} and \citet{Sharon+2019} are added, as these are used by \citet{Accard+2025} for comparison with their sample. Overall, the high--z star forming clouds and starburst galaxies follow our SF law reasonably well within the uncertainties. \citet{Dessauges+2023} calculate for their molecular clouds a range for $\epsilon_{ff}\sim$4\%--11\%, which is consistent with our results: for those author's mean $\Sigma_{SFR}\sim$3~M$_{\odot}$~yr$^{-1}$~kpc$^{-2}$, we find $\epsilon_{ff}\sim$3\%--18\% .

We will not push these comparisons further, as there are significant differences in the SFR and gas tracers used  between our low--z sample of star forming regions and the high--z samples, as well as uncertainties in the sizes of the high--z regions/galaxies. The goals of this brief (and incomplete) comparison is to highlight that high--z star--forming clumps and starbursts are likely a higher~gas--higher~SFR extension of local HII regions.

\subsection{Comparisons with Models}\label{subsec:models}

Models of star formation are basically divided into two main groups: `local' and `global'. In local models the balance is set between the local gravity and turbulence with star formation following from turbulent fragmentation \citep{Elmegreen+2002, MacLow+2004, Krumholz+2005, Hopkins+2011, Padoan+2011, Federrath+2012}, while in global models the molecular gas is confined by the mid--plane pressure of the disk \citep{Kim+2007,Ostriker+2010, Kim+2011, Faucher+2013}. The probability distribution function (PDF) of the gas density resulting from the presence of turbulence is generally described with a log--normal distribution, with a width that depends on the sonic Mach number of turbulence, a turbulence forcing parameter and, in some cases, an additional parameter capturing magnetic pressure \citep[see reviews in][]{Padoan+2014, Klessen+2016}. Once collapse sets in, a power law tail develops in the high density region of the PDF; star formation occurs above a critical density in the tail, which different authors define in different ways \citep[see review in][]{Federrath+2012}. Models are further divided between those that select a density--independent mean free--fall time for each collapsing cloud \citep[single--freefall,][]{Krumholz+2005, Padoan+2011} and those that allow the free--fall time to depend on the local density of each fluctuation \citep[multi--freefall,][]{Hennebelle+2011, Federrath+2012, Hennebelle+2013}. Modifications of the positive tail of the density PDF to add a flexible power--law extension attempt to reproduce the hierarchical structure of the star forming portion of the clouds \citep{Ballesteros+2007, Vazquez+2009, Gomez+2014, Vazquez+2017}, in agreement with observations \citep{Kainulainen+2009, Lombardi+2010a,Schneider+2012, Lombardi+2015, Schneider+2016, Dib+2020, Schneider+2022}, and can help avoid extreme values in the turbulence to explain the large range of star formation behaviors in galaxies \citep{Burkhart+2018, Burkhart+2019}. \citet{Dib+2011a} and \citet{Dib+2011b} proposed a variation to the local turbulent--driven model by implementing stellar feedback as the mechanism  for ending star formation; in this model, the critical timescale is not $\tau_{ff}$, but $\tau_{exp}$, the timescale for expelling gas from the star forming region.

We do not expect to do justice to the rich literature on the topic of star formation, but propose a few selected comparisons to provide a general view of how models perform in the case of extragalactic star forming regions. In all cases, the original SF laws provided by the authors are reported, without any attempt at modifying those models to fit our data. This is an important caveat: models aim at reproducing either SF in MW clouds, which, as shown in Figure~\ref{fig:sklaw}, have on average higher $\Sigma_{SFR}$ for a given $\Sigma_{mol}$ than our regions, or at fitting results for kpc--sized regions. These distinctions will be highlighted for each model. As before, the comparisons will be performed by keeping $\Sigma_{SFR}$ separate from $\Sigma_{mol}$, as these are the only two independent variables in the data. Several models report their results in terms of net efficiency (also called Star Formation Efficiency or SFE), which is the ratio of the total mass in stars produced by a given mass of molecular gas integrated over the entire star formation event, and expressed as: $\epsilon_{\star} = M_{star}/(M_{mol}+M_{star})$, where M$_{mol}$ is the current (remaining) mass in gas \citep{Evans+2009}. For our applications, we will use the same expression per unit area:
\begin{equation}
\Sigma_{young\ stars} = {\epsilon_{\star} \over 1-\epsilon_{\star} } \Sigma_{mol}.
\label{neteff}
\end{equation}
with $\Sigma_{young\ stars}$ defined in section~\ref{subsec:sfr}. This formulation assumes that molecular gas has not been destroyed or processed by the star formation, which, as shown in section~\ref{subsec:summary}, is a reasonable assumption for our regions. In order to explicitly calculate $\Sigma_{young\ stars}$, the star formation duration $\tau_{SF}$ needs to be defined; for convenience, we choose $\tau_{SF}\sim$6~Myr. This value is approximately the maximum age traced by the H$\alpha$ emission \citep{Leitherer+1999} and is also the mean stellar age traced  by the 21~$\mu$m emission in regions dominated by young populations \citep{Kennicutt+2012}. An important caveat is that individual star clusters can present a large range of mean durations for the star formation, from as short as a few tenths of Myr to $\sim6$~Myr \citep{DaRio+2010, Kudryavtseva+2012, Schneider+2018}; however, our 120~pc diameter regions often include several star clusters each, thus smoothing out age spreads and differences \citep{Calzetti+2025}.

\begin{figure}
\plottwo{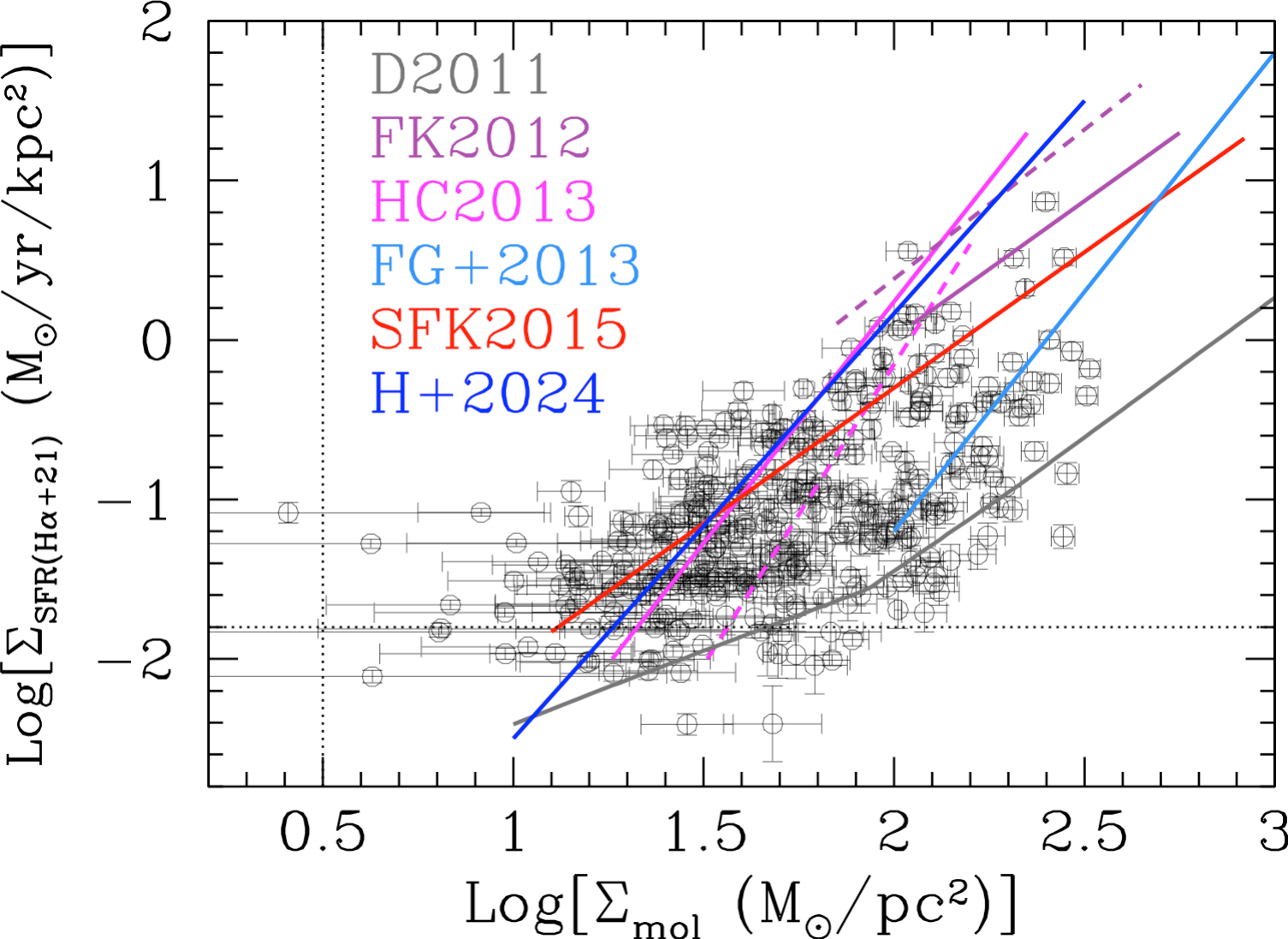}{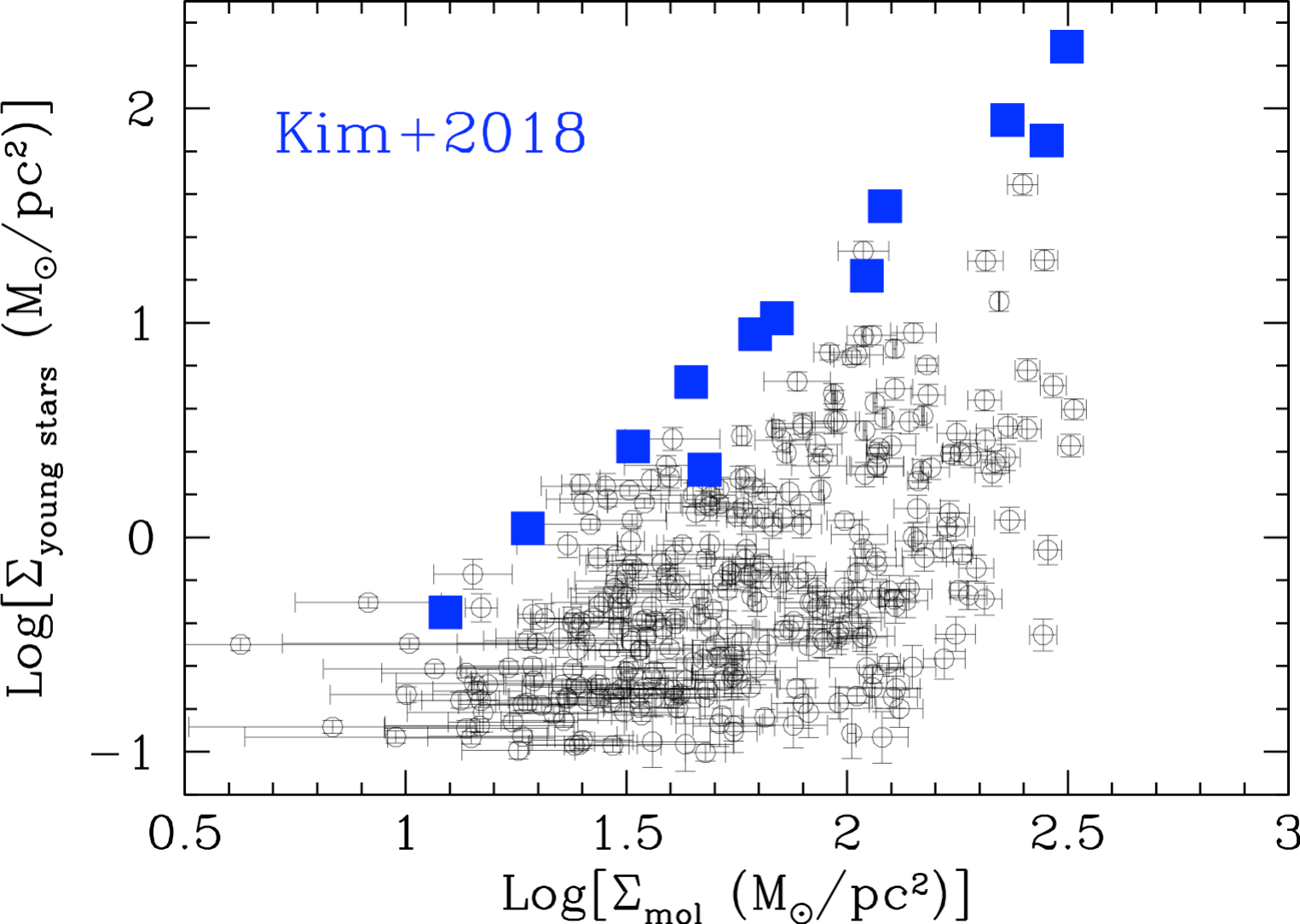}
\caption{{\bf (Left):} The SFR surface density as a function of the molecular gas surface density for our star forming regions (black circles with 1$\sigma$ uncertainties), compared with models. `Local', multi--freefall models shown include: the central ridge of the multi--freefall model by \citep[][FK2012]{Federrath+2012} for net efficiency $\epsilon_{\star}$=0.01 (solid purple line) and  0.10 (dash purple line); approximate linear tracks for the multi--freefall model of \citep[][HC2013]{Hennebelle+2013}, using the authors' examples for clump sizes of 5~pc and maximum fraction of cloud size for deriving SFR y$_{out}$=0.25: isothermal model (solid magenta line) and non--isothermal model (dash magenta line); and the sonic Mach number--based model by \citet[][SFK2015]{Salim+2015} (red solid line). The `local', feedback--regulated model of \citet{Dib+2011a}, published for kpc--sized regions, is shown for the solar metallicity case (grey line). `Global' models shown include: the disk--pressure regulated model by \citet[][FG+2013]{Faucher+2013}, which the authors only derive for $\Sigma_{mol}\ge 100$~M$_{\odot}$~pc$^{-2}$ (teal solid line);  and the SF law derived by \citet[][H+2024]{Hassan+2024} (blue solid line), based on the pressure--regulated feedback--modulated model of \citet{Ostriker+2022}. {\bf (Right):} The surface density of young stars as a function of the molecular gas surface density for our sources (black circles with 1$\sigma$ uncertainties), shown in comparison with the predictions of the model by \citet{Kim+2018} for molecular clouds, which include the effects of dispersal by UV radiation feedback. 
} 
\label{fig:model1}
\end{figure}

Figure~\ref{fig:model1} (left) shows the SF law for our sources in comparison with several models. Three `local' multi--freefall models are considered, from \citet{Federrath+2012}, \citet{Hennebelle+2013} and \citet{Salim+2015}, all three calibrated on MW clouds, with the exception of  \citet{Salim+2015}, which also include observations in the Small Magellanic Cloud. The `local' model of \citet{Dib+2011a} is shown for the solar metallicity case. The two `global' models are from \citet{Faucher+2013} and \citet{Hassan+2024} and aim at reproducing larger scale results in galaxies. None of the models/simulations reproduces the observed trends in detail, although, given the large scatter, all pass through the locus of the data. It should be added that the models have  some parametric freedom that could improve the agreement with the data, but which we do not use here. Of the three multi--fall models, the one proposed by \cite{Salim+2015} is perhaps the closest to the trend shown by the data. Of the two global models, the one by \citet{Hassan+2024} is the closest to the multi--freefall models while the one by \citet{Faucher+2013} is closer to the lower bound of the scatter in the data. The approximate value of the slope(s) of each model are listed in Table~\ref{tab:slopes}.
Figure~\ref{fig:model1} (right) compares our data with the predictions of molecular clouds evolution by \citet{Kim+2018}, which include the effects of both UV photoionization and radiation pressure in the simulations, and produces the net efficiency $\epsilon_{\star}$ as a function of the initial gas mass surface density. Our data can be compared with these simulations, as our lowest mass region corresponds to M$_{mol}\sim$3$\times$10$^5$~M$_{\odot}$ (for $\approx$1\% efficiency) and \citet{Kim+2018} simulates clouds with mass $>$10$^4$~M$_{\odot}$. We use equation~\ref{neteff} to convert the model's predictions to quantities that we can use with our data. While the general trend shown by the simulations agrees with the upper envelope of the data, the predicted net efficiencies are generally too high. 

\begin{deluxetable*}{llcll}
\tablecaption{Model Slopes$^1$\label{tab:slopes}}
\tablewidth{0pt}
\tablehead{
\colhead{Label$^2$ } & \colhead{Type$^3$} &\colhead{Slope} & \colhead{Comments} & \colhead{Reference} }
\decimalcolnumbers
\startdata
D2011             &local      & 0.91 & solar metall., Log($\Sigma_{mol}$)$<$1.93 & \citet{Dib+2011a}\\
 ...                   &local       & 1.71 & solar metall., Log($\Sigma_{mol}$)$>$1.93 &      \\
 FK2012          &local      &  1.71 & $\epsilon_{\star}$=0.01 (solid line)               &   \citet{Federrath+2012} \\
 ...                    &local      &  1.88 & $\epsilon_{\star}$=0.1  (dash line)               &   \\
 HC2013          &local     & 3.03  & isothermal model (solid line)                         & \citet{Hennebelle+2013}\\
 ...                    &local      & 3.76 & non--isothermal model (dash line)                &         \\
 FG+2013        &global     &  3.00  &                                                                      & \citet{Faucher+2013}\\
 SFK2015        &local      &  1.70  &                                                                     & \citet{Salim+2015}\\
 H+2024          &global     &  2.67   &                                                                     & \citet{Hassan+2024}\\
\enddata
$^1$ The approximate slopes of the models shown in Figure~\ref{fig:model1}, left, within the validity range of our data. The values reported are approximate representations of often--complex models.\\
$^2$ Model Label in Figure~\ref{fig:model1}, left.\\
$^3$ Type of model: local or global.\\
\end{deluxetable*}

\begin{figure}
\plottwo{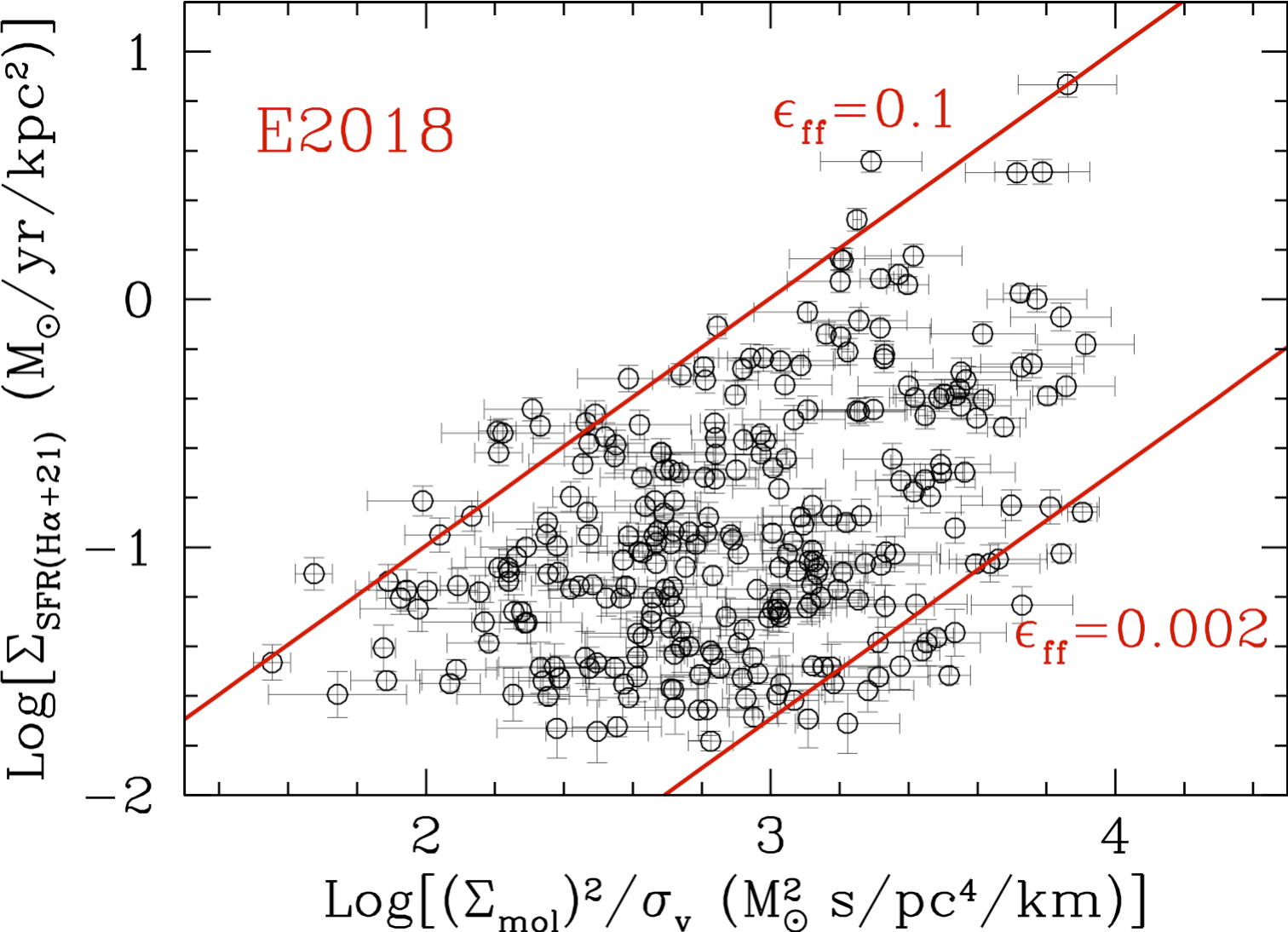}{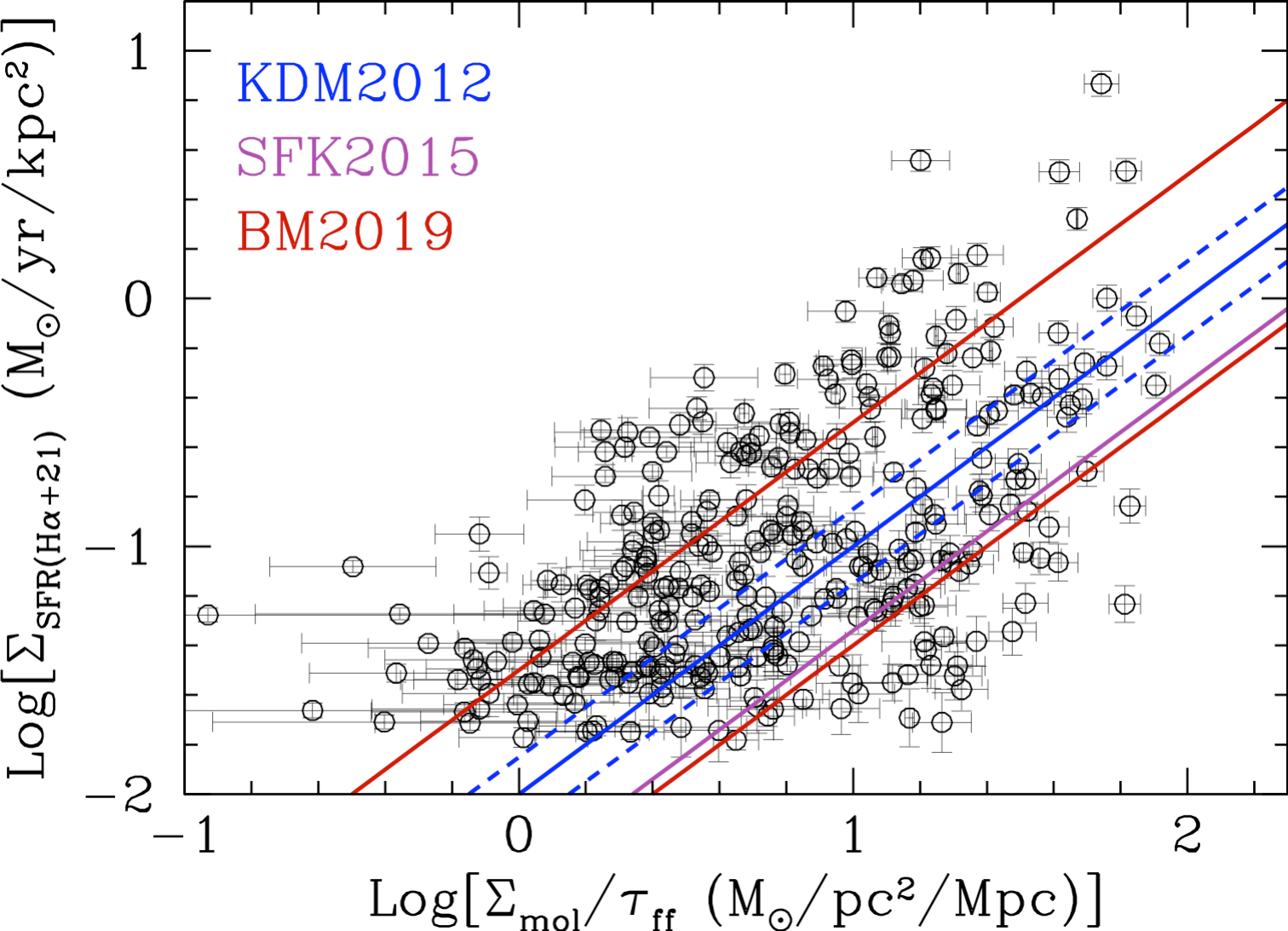}
\caption{The SFR surface density as a function of the molecular gas surface density per free--fall time for our star forming regions (black circles with 1$\sigma$ uncertainties) is compared with additional models from the literature. By construction, the models shown in both panels have slope=1.  {\bf (Left):} The `local' model of \citet{Elmegreen+2018} (dark--red solid lines), which leaves the efficiency $\epsilon_{ff}$ as a free parameter, adjusted to encompass the data as shown in the Figure.  {\bf (Right):}  The `local' single--freefall model of \citet[][KDM+2012]{Krumholz+2012} is shown for $\epsilon_{ff}$=0.01 (blue solid line) together with a scatter of $\pm$0.15~dex (blue dash lines) about the mean $\epsilon_{ff}$ value, as predicted by \citet{Krumholz+2020}.  The `local' model of \citet{Burkhart+2019}, which adds a power--law tail to the density PDF of the gas, is shown for two choices of the sonic Mach number M$_s$ and the turbulence forcing parameter $b$ (dark--red solid lines), for exponent $\alpha$=2 of the power law: top--left line for M$_s$=20 and $b$=0.3 and bottom--right line for M$_s$=50 and $b$=0.7. The prediction from the model of \citep[][SFK2015]{Salim+2015}, already introduced in Figure~\ref{fig:model1}, is also shown (purple solid line). 
} 
\label{fig:model2}
\end{figure}

Models and simulations are sometimes reported as $\Sigma_{SFR}$ versus $\Sigma_{mol}$/$\tau_{ff}$ or versus $\Sigma_{mol}^2$/$\sigma_v$; these two versions of the horizontal axis are equivalent for practical purposes \citep{Elmegreen+2018}, but we keep them separate in our comparisons. The vertical and horizontal  axes are still observationally independent, since $\tau_{ff}$ and $\sigma_v$ only require information from the CO maps. Figure~\ref{fig:model2} (left) shows $\Sigma_{SFR}$ versus $\Sigma_{mol}^2$/$\sigma_{v}$ for our regions compared with the trend discussed in \citet{Elmegreen+2018}; this author argues that the different observed values of $n$ in different environments are the result of combining different degrees of self--gravity with selection biases in the gas tracers employed. In \citet{Elmegreen+2018}'s  `local' model, the efficiency $\epsilon_{ff}$ is a free parameter, which we adjust in Figure~\ref{fig:model2} (left) to encompass most of the HII region data. We find that $\epsilon_{ff}$ needs to cover the range $\sim$0.002--0.1, similar to what already found in section~\ref{subsec:efficiency}. Figure~\ref{fig:model2} (right) reports the expectations of the single--freefall model of \citet{Krumholz+2012}, the power--law--tail model of \citet{Burkhart+2019} and the multi--freefall model of \citet{Salim+2015}, the latter already shown in Figure~\ref{fig:model1} (left). For the \citet{Krumholz+2012}'s model, we show the expected dispersion ($\pm$0.15~dex) as derived in \citet{Krumholz+2020}. For the \citet{Burkhart+2019}'s model, two separate selections of the sonic Mach number and the turbulence forcing parameter are used, as reported in those authors' paper. While no single model can explain the dispersion in the data for the parameters shown, the trends appear qualitatively consistent with the data; in addition, a large range of $\epsilon_{ff}$ from \citet{Elmegreen+2018}'s model and/or a broader choices for the parameters in \citet{Burkhart+2019} could potentially account for the observed dispersion.

In summary, the models we considered do not perfectly fit our data, but none was designed to match extragalactic star forming regions measured at fixed spatial scale. Importantly, the models struggle to reproduce the scatter observed in the data. However, most models contain adjustable parameters which could help bring them  into better agreement with the observations. For instance, the model of \citet{Burkhart+2018} and \citet{Burkhart+2019}, with two different choices of the sonic Mach number and the turbulence forcing parameter, is able to bracket most of the observed spread as well as the trend in the data; a similar agreement is obtained using a wide range of $\epsilon_{ff}$ in the model of \citet{Elmegreen+2018}, and could potentially be explained also by the effect of existing stars on the SFR \citep{Dib+2017}. 

\section{Summary and Conclusions} \label{sec:conclusions}
 
The combination of high--angular resolution data in the optical, infrared, and millimeter from HST, JWST and ALMA/PdBI has enabled the investigation of the star formation law at the 100~pc scale of HII regions in three nearby galaxies, yielding a relation between $\Sigma_{SFR}$ and $\Sigma_{mol}$ with a slope (in Log--Log scale) $n=1.85 \pm 0.12$. This slope is significantly steeper than those found for $\sim$kpc sized extragalactic regions \citep[e.g.][]{Kennicutt+2007, Bigiel+2008, Bigiel+2011, Utomo+2018, Chevance+2022, Sun+2023, Leroy+2025} and whole galaxies \citep{Kennicutt+1998, Kennicutt+2012, Liu+2015, delosReyes+2019, Kennicutt+2021}, but close to the trends found for MW molecular clouds \citep{Evans+2009, Heiderman+2010, Gutermuth+2011}. The steep slope translates into an efficiency per free--fall time, $\epsilon_{ff}$, that increases by a factor of $\sim$3.5 over a factor 500 in SFR surface density and 30 in molecular gas surface sensity, with a mean value of 1\%, once the effects of covariance are taken into account. Published tracks from several models and simulations of star formation provided limited agreement with the observations, but none of those tracks were produced for extragalactic star forming regions; additional experimentations with such models may provide better matches for the data, specifically for the large observed scatter. 

The steep relation between $\Sigma_{SFR}$ and $\Sigma_{mol}$ relies on a careful treatment of the contamination of the SFR and gas tracers from spurious effects, and remains steep even when larger areas than those of HII regions are analyzed. For a sample of $\sim$500~pc diameter regions, the slope is found $n\sim$1.88 (Appendix~\ref{sec:appendixD}), in agreement with previous findings at lower spatial resolution. We have analyzed in detail the effect of the contribution from the galaxy's diffuse stellar population to the heating of the dust emitting at 21~$\mu$m. This diffuse population is unrelated to the observed star forming regions but can provide a significant contribution, over an order of magnitude (Figure~\ref{fig:diffuse}, left) to $\Sigma_{SFR}$ if not removed. The net result of not removing contamination in the tracers is a `flattening' of the $\Sigma_{SFR}$--$\Sigma_{mol}$ relations, bringing it close to the findings for kpc--sized regions. This contribution likely becomes more difficult to remove when the regions of star formation are not as well defined as in our case, with a mostly centrally--concentrated active area, which may explain also why the slope of the $\Sigma_{SFR}$--$\Sigma_{mol}$  relation tends to decrease with increasing region size \citep{Liu+2011}. However, retaining the underlying galaxy's emission in the measurements of star--forming regions is equivalent to isolating regions with different SFHs, depending on the relative brightness of the star--formation relative to that of the diffuse emission. In this case, the use of a single SFR indicator is problematic, especially when employing infrared--based tracers, because  SFR calibrations are SFH--dependent \citep[][and reference therein]{Kennicutt+1998a, Calzetti+2025}.

In this work, the expression used for $\alpha_{CO}$ depends on the weight of each disk (equation~\ref{alphaco}). However, the steep relation between $\Sigma_{SFR}$ and $\Sigma_{mol}$ of local star forming regions is not an effect of this choice. We show that selecting a single, constant value of $\alpha_{CO}$ for all galaxies 
leaves galaxy--dependent offsets in CO luminosity uncorrected. As shown in Appendix~\ref{sec:appendixE}, choosing a fixed $\alpha_{CO}$=4.35 for all regions in our sample yields slopes for individual galaxies that are steep,  $n\sim1.75-1.90$, but an overall slope for the combined sample that is flattened to $n\sim$1.3. A visual inspection of the result shows that the CO luminosity offsets from galaxy to galaxy are the reason for this flattening. 

The molecular SF law of HII regions is found to be an extension to low $\Sigma_{mol}$ values of the molecular SF law of low--redshift starbursts and high redshift star forming clumps and starburst galaxies, which have significantly higher SFRs than those of local galaxies \citep[e.g.,][]{delosReyes+2019, Kennicutt+2021}. Thus, local HII regions and low and high redshift starbursts follow a single SF law, with a slope $n\approx$1.8. Future investigations will need to expand this study along two directions: (1) by including more regions at high $\Sigma_{SFR}$ to confirm the observed trends and understand whether the increase in $\epsilon_{ff}$ with increasing $\Sigma_{SFR}$  and $\Sigma_{mol}$ can be confirmed; and (2) by extending the range of explored metallicities to the low values relevant for investigations in the high redshift regime.
 
\begin{acknowledgments}
The authors thank the referee for their comments and suggestions which have helped improve this manuscript.

This work is based in part on observations made with the NASA/ESA/CSA James Webb Space Telescope. The dara were obtained from the Mikulski Archive for Space Telescopes at the Space Telescope Science Institute, which is operated by the Association of Universities for Research in Astronomy, Inc., under NASA contract NAS 5-03127 for JWST. These observations are associated with GO programs \# 1783 and  \#3435. Support for US investigators in programs GO \# 1783 and  GO \#3435 was provided by NASA through grants from the Space Telescope Science Institute, which is operated by the Association of Universities for Research in Astronomy, Inc., under NASA contract NAS 5-03127. 

The work also made use of archival data from the NASA/ESA Hubble Space Telescope and from the NASA/ESA/CSA James Webb Space Telescope, obtained from the Space Telescope Science Institute, which is operated by the Association of Universities for Research in Astronomy, Inc., under NASA contracts NAS 5-26555 and NAS 5-03127, respectively. HST and JWST data were retrieved from the Mikulski Archive for Space Telescopes at the Space Telescope Science Institute. 

The specific observations analyzed can be accessed via DOIs: \dataset[10.17909/q3rh-mk45]\ , \dataset[10.17909/jp5q-s259]\ , and \dataset[10.17909/wtq6-f859] .

The CO datasets of NGC\,5194 used in this work are from the PAWS survey (P.I.: Schinnerer) at the Plateau de Bure Interferometer; the processed maps were retrieved from: https://www2.mpia-hd.mpg.de/PAWS/PAWS/Data.html. The CO datasets of NGC\,628 and NGC\,5236 are from the ALMA programs: ADS/JAO.ALMA\#2013.1.01161.S  (PI Sakamoto) for NGC 5236 (M83) and 
ADS/JAO.ALMA\#2012.1.00650.S (PI Schinnerer) for NGC 0628 (M74). The processed maps for NGC\,628 and NGC\,5236 were retrieved from: https://www.canfar.net/storage/list/phangs/RELEASES/PHANGS-ALMA/

This research has made use of the NASA/IPAC Extragalactic Database (NED, \dataset[10.26132/NED1]\ ) which is operated by the Jet
Propulsion Laboratory, California Institute of Technology, under contract with the National Aeronautics and Space
Administration.

ADC acknowledges the support from a Royal Society University Research Fellowship (URF/R1/19160 and URF/R/241028). RSK acknowledges financial support from the ERC via Synergy Grant ``ECOGAL'' (project ID 855130) and from the German Excellence Strategy via the Heidelberg Cluster ``STRUCTURES'' (EXC 2181 - 390900948). In addition RSK is grateful for funding from the German BMWE in project ``MAINN'' (funding ID 50OO2206), and from DFG and ANR for project ``STARCLUSTERS'' (funding ID KL 1358/22-1). MRK acknowledges support from the Australian Research Council through Laureate Fellowship FL220100020.

\end{acknowledgments}

\vspace{5mm}
\facilities{James Webb Space Telescope (NIRCam, MIRI)}
\software{JWST Calibration Pipeline \citep[][]{Bushouse+2022, Greenfield+2016}, Drizzlepac \citep[][and the STSCI Development Team]{Gonzaga+2012}, IRAF \citep{Tody1986, Tody1993}, SAOImage DS9 \citep{Joye+2003}, GFortran (https://gcc.gnu.org/fortran/).}

\appendix

\section{The Diffuse Ionized Gas Component}\label{sec:appendixA}

\begin{figure}
\plotone{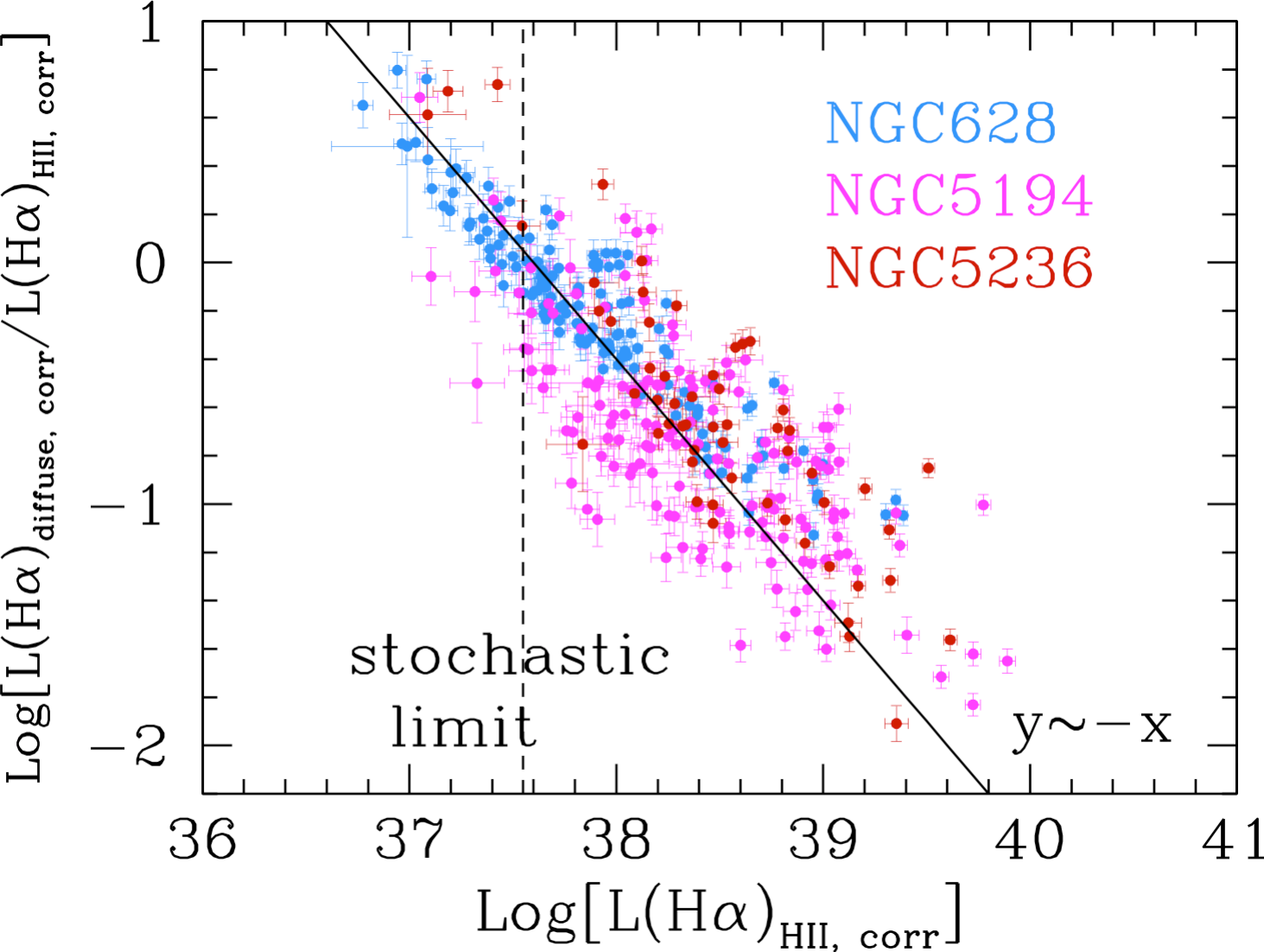}
\caption{The ratio of diffuse--to--HII region H$\alpha$ emission for the sources in our sample, shown in different color symbols 
for the different galaxies (teal=NGC\,628; magenta=NGC\,5194; dark red=NGC\,5236).  Both the diffuse and HII~region H$\alpha$ have been corrected for dust attenuation, using the H$\alpha$/Pa$\alpha$ recombination line ratio. The stochastic sampling limit is marked by the vertical dashed black line. 
The solid black line marks the $y\propto -x$ trend, in log--log scale, expected if the diffuse emission is roughly constant; this line is not a fit to the data.}
 \label{fig:diffuseHa}
\end{figure}

Figure~\ref{fig:diffuseHa} shows the contribution of the diffuse ionized gas component to the measurements in our 60~pc radius photometric apertures. Both the HII~regions and diffuse H$\alpha$ emission are corrected for the effects of dust attenuation, using the H$\alpha$/Pa$\alpha$ recombination line ratio; an intrinsic value L(H$\alpha$)/L(Pa$\alpha$)=7.82 is adopted, as appropriate for our metal--rich sources \citep{Osterbrock+2006}. While the use of L(21)/L(H$\alpha$) would be preferable to derive dust attenuation corrections, this ratio cannot be used for the diffuse emission, due to the different nature of L(21) and L(H$\alpha$) in this component: while the 21~$\mu$m emission is due to in--situ dust heating by evolved stellar populations, the diffuse H$\alpha$ originates from ionizing photon leakage from sources as distant as $\sim$1~kpc \citep[][and references therein]{Levy+2019}. The average color excess of the diffuse emission, derived as described in \citet{Calzetti+2025}, is E(B--V)$\sim$0.15--0.30~mag, depending on the galaxy. 

The diffuse--to--HII~regions H$\alpha$ emission ratio follows the $y\propto -x$ trend  as a function of the HII region luminosity (in log--log scale) expected for a roughly constant diffuse component. It has a smaller spread than the diffuse 21~$\mu$m emission at constant HII region luminosity, as expected if the H$\alpha$ is powered by a much shorter timescale component of the stellar population than the 21~$\mu$m dust emission.  The diffuse H$\alpha$ contributes only a few \% of the total inside the photometric aperture for the brightest regions, but is comparable or larger than the HII~regions at the lowest luminosity values; yet it remains overall a lower contamination factor, by factors of a few, than the diffuse 21~$\mu$m.

\citet{Elmegreen+2025} find that the diffuse emission in NGC\,5194 is only  $\sim$1\% of the average emission measure, while \citet{Belfiore+2023} conclude that the diffuse H$\alpha$ emission contamination in their HII region measurements is overall small. In order to reconcile our findings with these authors', we need to consider the different approaches adopted in evaluating the contribution of the diffuse H$\alpha$ emission.  \citet{Elmegreen+2025} study the pixel--to--pixel power spectrum of the ionized gas emission, isolating as `diffuse emission' a component with a larger scale than $\sim$1~kpc; ionizing photon leakage, however, leaves an `imprint' with this characteristic scale \citep{Liu+2011, Kumari+2020}.  \citet{Belfiore+2023} defines the sizes of their HII regions based on surface brightness limits, thus drawing contours that `hug' each HII region. Conversely, we use a 60~pc radius aperture irrespective of the H$\alpha$ luminosity or surface brightness of the source, as we are limited by the angular resolution of the CO maps. This size is about twice the radius of the Str\"omgren radius of our brightest HII regions \citep[see calculation in][]{Calzetti+2025}. For apertures larger than the Str\"omgren radius, the HII region luminosity does not grow, but the contribution of the diffuse emission increases linearly with the area. To estimate the magnitude of this effect, we consider the case of HII regions with H$\alpha$ luminosity L(H$\alpha$)$_{HII, corr}\sim$10$^{37.5}$~erg~s$^{-1}$. For these regions, the diffuse H$\alpha$ luminosity is comparable to that of the HII region (Figure~\ref{fig:diffuseHa}). Of this luminosity, the fraction that can be attributed to leakage from the region itself is $\sim$1.5\%, assuming that the leaked photons have diffused homogeneously and isotropically in an area 500~pc in radius. Thus, the diffuse H$\alpha$ is likely to be contributed by ionizing regions located outside of the 60~pc aperture radius. The luminosity of HII regions is correlated with their sizes, as expected for ionization--bound regions \citep{Osterbrock+2006} and confirmed observationally \citep{Kennicutt+1988, Liu+2013}. A region with L(H$\alpha$)$_{HII, corr}\sim$10$^{37.5}$~erg~s$^{-1}$ has a radius R$\sim$11~pc. Rescaling the diffuse H$\alpha$ emission from R=60~pc to R=11~pc brings down its contribution from $\sim$100\% to $\sim$3.5\%. The selection performed by \citet{Belfiore+2023} is close in nature to a Str\"omgren radius selection, which explains the small values of the diffuse emission contamination found by these authors. Thus, the differences found by different authors on the impact of diffuse ionized gas emission on their measurements can be reconciled when the different approaches used in performing the measurements are considered. 

\section{Stellar Mass Surface Densities}\label{sec:appendixB}

The stellar mass surface density of the galaxy enters in the CO--to--H2 conversion factor, to quantify the disk pressure on the molecular clouds \citep{Bolatto+2013}. We use the JWST/NIRCam F277W/F300M bands to trace the stellar mass because these wavelengths are dominated by stellar emission \citep{Draine+2007, Dale+2023} and, at the same time, the effects of dust attenuation are minimized: A$_V$=1~mag corresponds to A$_{2.8\mu m}$=0.07~mag. NGC\,628 has been observed in both F277W and F300M, while the other two galaxies have F300M only. For NGC\,5236 coverage is available for 38 out of 56 regions, and we adopt the median stellar mass of the existing regions for the regions not observed in F300M. 

The calculation of the luminosity--to--mass ratio is performed via the models presented in \citet{Calzetti+2025}, which include constant star formation, exponentially decreasing star formation with an e--folding time of 4.3~Gyr, and exponentially increasing star formation. We use the first two models as more appropriate for the general case of disk galaxies, although results remain unchanged if the model of increasing star formation is used. We note that the model for decreasing star formation shown in Figure~\ref{fig:diffuse} are slightly different from the one used here; in Figure~\ref{fig:diffuse} we mimic the SFH of NGC\,5194 from \citet{Martinez+2018} to capture the effects of dust heating on the $\Sigma(21)_{diffuse}$ emission. This SFH consists of a slow increase until $\sim$5~Gyr ago, a broad peak between $\sim$5~Gyr and 1.5~Gyr ago, steadily decreasing since. In all cases, the stellar population models are from Starburst99 \citep{Leitherer+1999} with Padova AGB evolutionary tracks \citep{Girardi+2000}, metallicity Z=0.02 (solar) and a \citet{Kroupa+2001} IMF in the stellar mass range 0.1--120~M$_{\odot}$. Dust attenuation is added using the curve of \citet{Calzetti+2000}, and decreasing values of E(B--V) from young to old stellar populations, with the youngest ($<$5~Myr) stellar populations having E(B--V)=1~mag and the oldest populations (several Gyr) being only reddened by E(B--V)=0.1~mag. More details can be found in \citet{Calzetti+2025}.

\begin{figure}
\plottwo{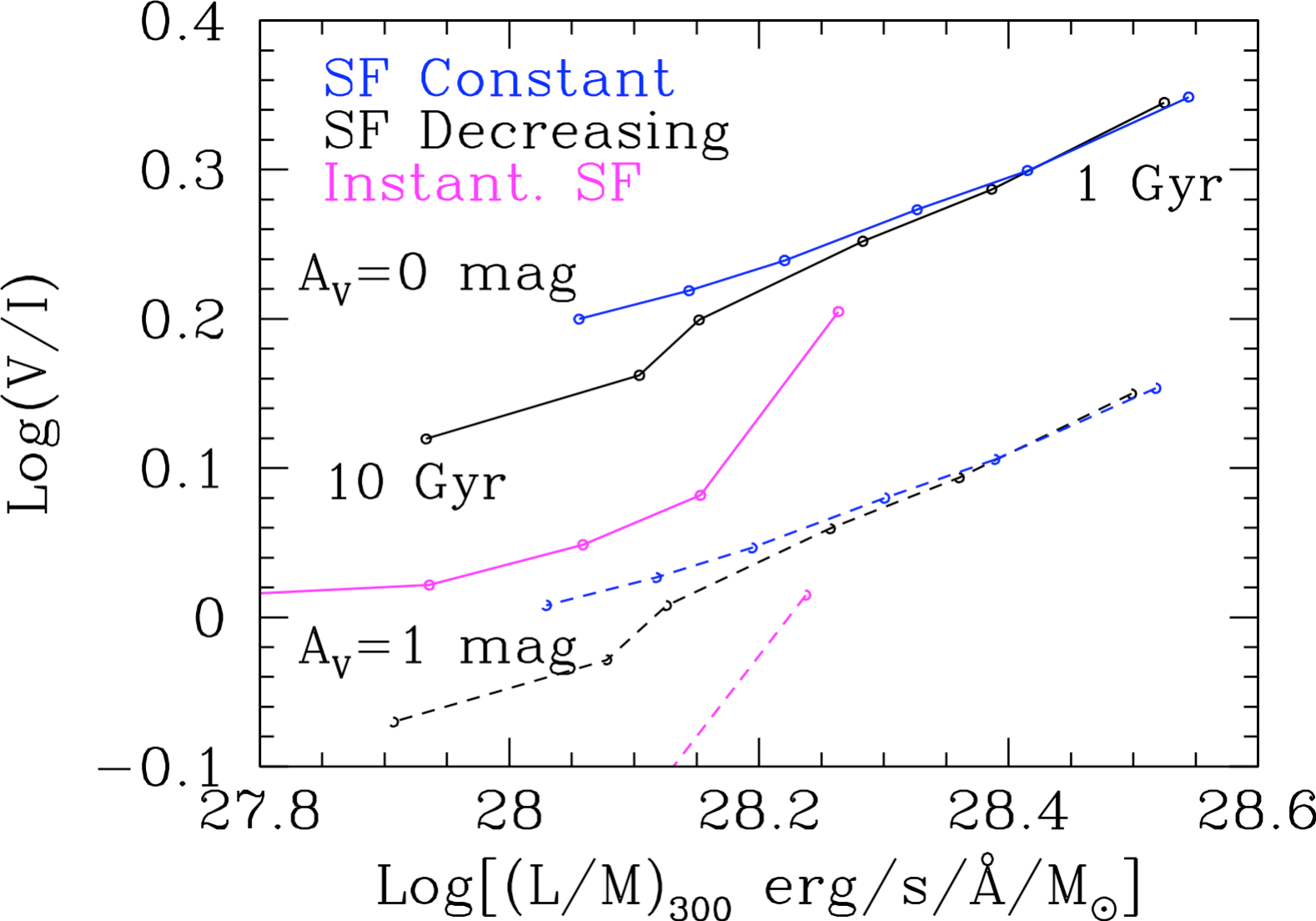}{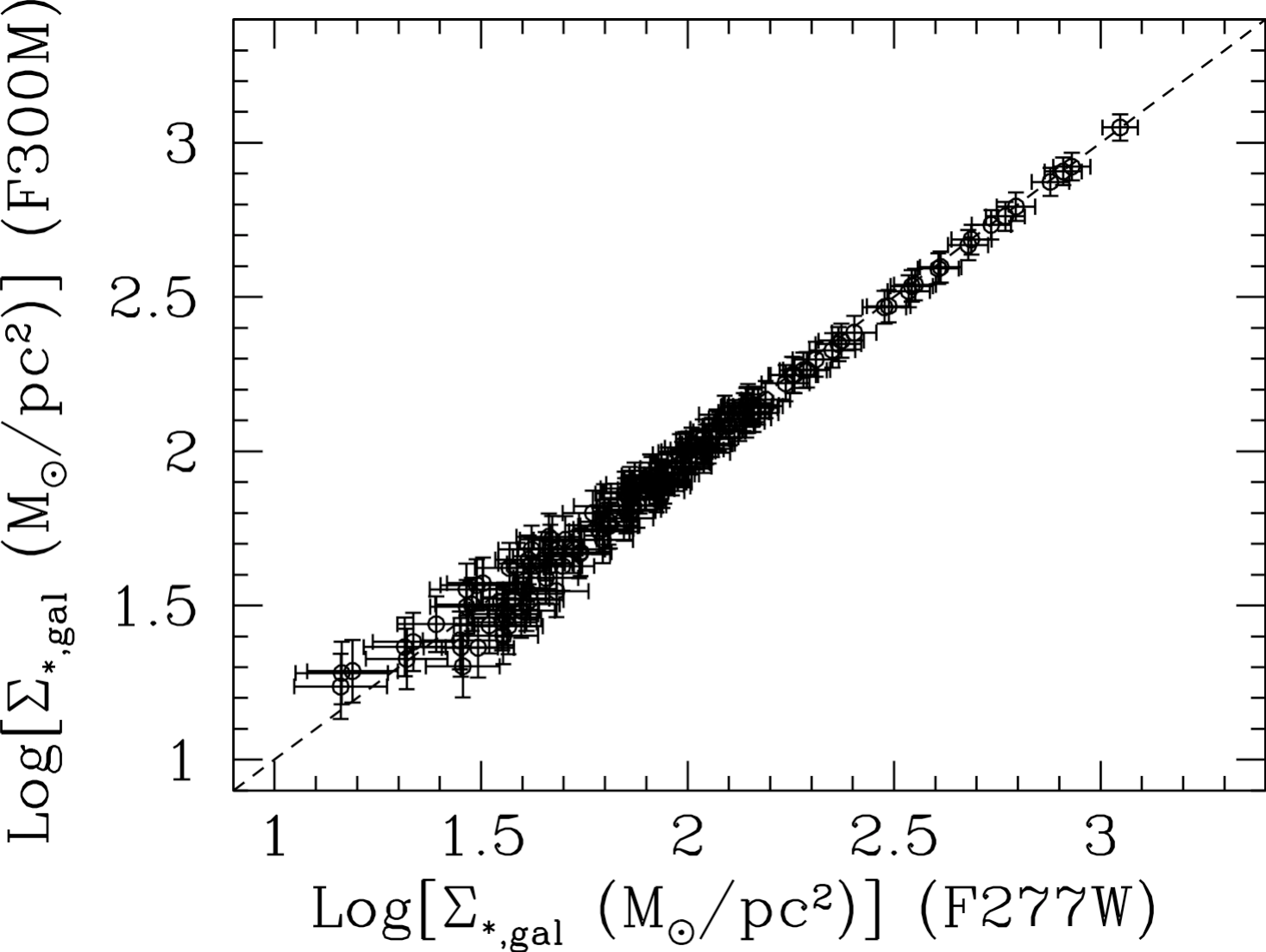}
\caption{{\bf (Left:)} V--I color versus 3~$\mu$m luminosity--to--mass ratio for population models with the following star formation histories: constant star formation (blue curves), exponentially decreasing star formation with 4.3~Gyr e--folding time (black curves) and instantaneous burst (magenta curves). The models are shown for the two cases of A$_V$=0~mag (continuous lines) and A$_V$=1~mag (dashed lines), between ages 1~Gyr and 10~Gyr. While the change in attenuation has large impact on the V--I color, it has minimal impact on the 3~$\mu$m luminosity--to--mass ratio. {\bf (Right:)} The mass surface densities of the stellar populations underlying the star--forming regions in NGC\,628, derived from measurements in F277W and F300M, are compared with each other.}
 \label{fig:massl}
\end{figure}

Figure~\ref{fig:massl}, left, shows the expectations for the luminosity--to--mass ratio in the F300M filter as a function of the V-I color for the two models of extended star formation (black and blue lines), together with the case of instantaneous star formation (magenta lines) for reference. In all cases, the duration range of the models is 1~Gyr to 10~Gyr (this is age for the instantaneous model), which we expect to be representative of the star formation histories for galaxies across most environments. The V-I color is used only to show the effect of dust attenuation. For both constant and exponentially decreasing star formation, the maximum change in the 3~$\mu$m luminosity due to changes in star formation duration is $\sim$0.6~dex, with negligible impact from dust attenuation. For the comparison case of an instantaneous burst population, the impact of age is significantly larger, but we do not use this model for our galaxies. By choosing an average value L$_{300}$/M=10$^{28.25}$~erg~s$^{-1}$~\AA$^{-1}$~M$_{\odot}^{-1}$ for all three galaxies analyzed in this work, the scatter in the luminosity--to--mass conversion is $\sim$0.3~dex. The luminosity--to--mass ratio in the F277W filter shows a similarly minimal impact from both age changes and dust attenuation, with an average value L$_{277}$/M=10$^{28.35}$~erg~s$^{-1}$~\AA$^{-1}$~M$_{\odot}^{-1}$. The comparison of luminosity--to--mass ratios derived from measurements in F300M and F277W for NGC\,628 yields virtually identical results (Figure~\ref{fig:massl}, right). 

The stellar mass surface density, $\Sigma_{*, gal}$, of the galaxy underlying each region is derived from photometry measured in the F277W/F300M filters in 200~pc areas surrounding each region, to average out small fluctuations. The flux of the star-forming region itself is subtracted out; other compact sources present in each area are also removed via iterative sigma--clipping. This procedure is similar to the one employed to determine the diffuse emission in H$\alpha$/Pa$\alpha$/21~$\mu$m (section~\ref{sec:selection}). The resulting $\Sigma_{*, gal}$ underlying the star--forming regions within each galaxy are shown in Figure~\ref{fig:diffuse}, right. 

For NGC\,628, \citet{Querejeta+2015} presented a stellar mass map derived from the Spitzer/IRAC 3.6~$\mu$m images treated with the Independent Component Analysis (ICA) method of \citet{Meidt+2014}. The scope of the ICA approach is to remove the dust emission from the 3.6~$\mu$m band and isolate the stellar--only component of each galaxy. As the Spitzer images are at a significant lower resolution than the JWST maps ($\sim$2$^{\prime\prime}$ vs. $\sim$0\farcs09), we derive $\Sigma_{*, gal}$ in several kpc--sized regions in common between the NIRCam/F300M and the IRAC/3.6~$\mu$m images. We find that for bright regions ($\Sigma_{*,gal}\gtrsim$300~M$_{\odot}$~pc$^{-2}$) the measurements in the Spitzer/IRAC image are only 10\% higher than those in the JWST/NIRCam one, while they are 1.5--1.9 times higher in fainter regions ($\Sigma_{*, gal}\lesssim$30~M$_{\odot}$~pc$^{-2}$). Uncertainties in the determination of the sky background in the Spitzer images as well as in the ICA method itself may account for the discrepancies at low luminosities. \citet{Heyer+2022} derived $\Sigma_{*, gal}$ at 600~pc resolution in NGC\,5194, using the $g-i$--vs.--H band method of \citet{Zibetti+2009}, and finding values in the range Log[$\Sigma_{*, gal}$ (M$_{\odot}$~pc$^{-2}$)]$\sim$(1.8--3.6). We find a similar range in $\Sigma_{*, gal}$ for the same galaxy ($\sim$2.0--3.4, Figure~\ref{fig:diffuse}, right), although the spatial coverage of our maps is significantly smaller than the coverage of the maps in \citet{Heyer+2022}, 3.5~kpc versus 9~kpc galactocentric radius around the nucleus. The agreements between stellar mass surface densities derived with methods completely independent of each other lend support to our approach for calculating $\Sigma_{*, gal}$.

\section{Residuals about the Best Fit of the Law of Star Formation}\label{sec:appendixC}

\begin{figure}
\plottwo{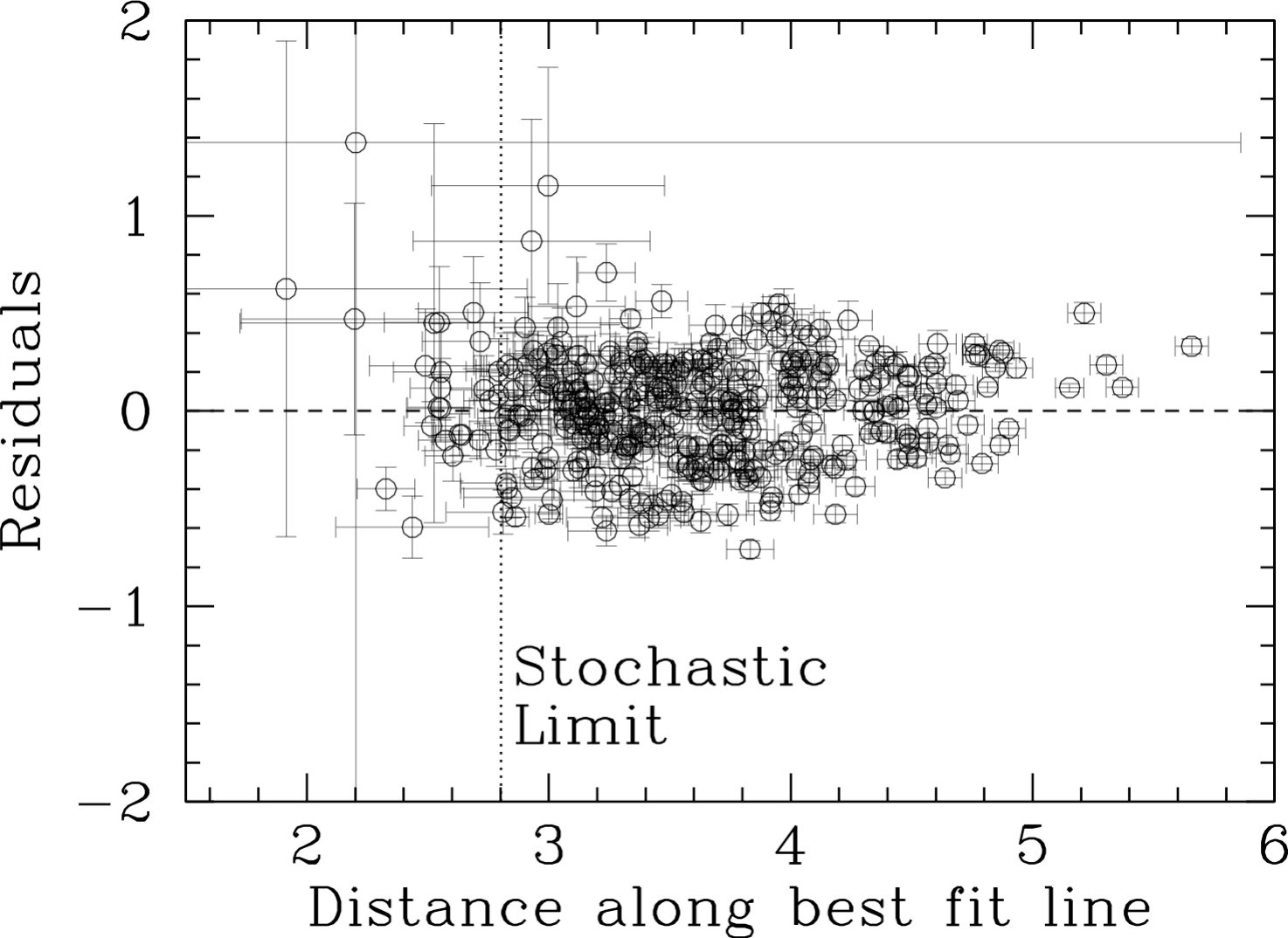}{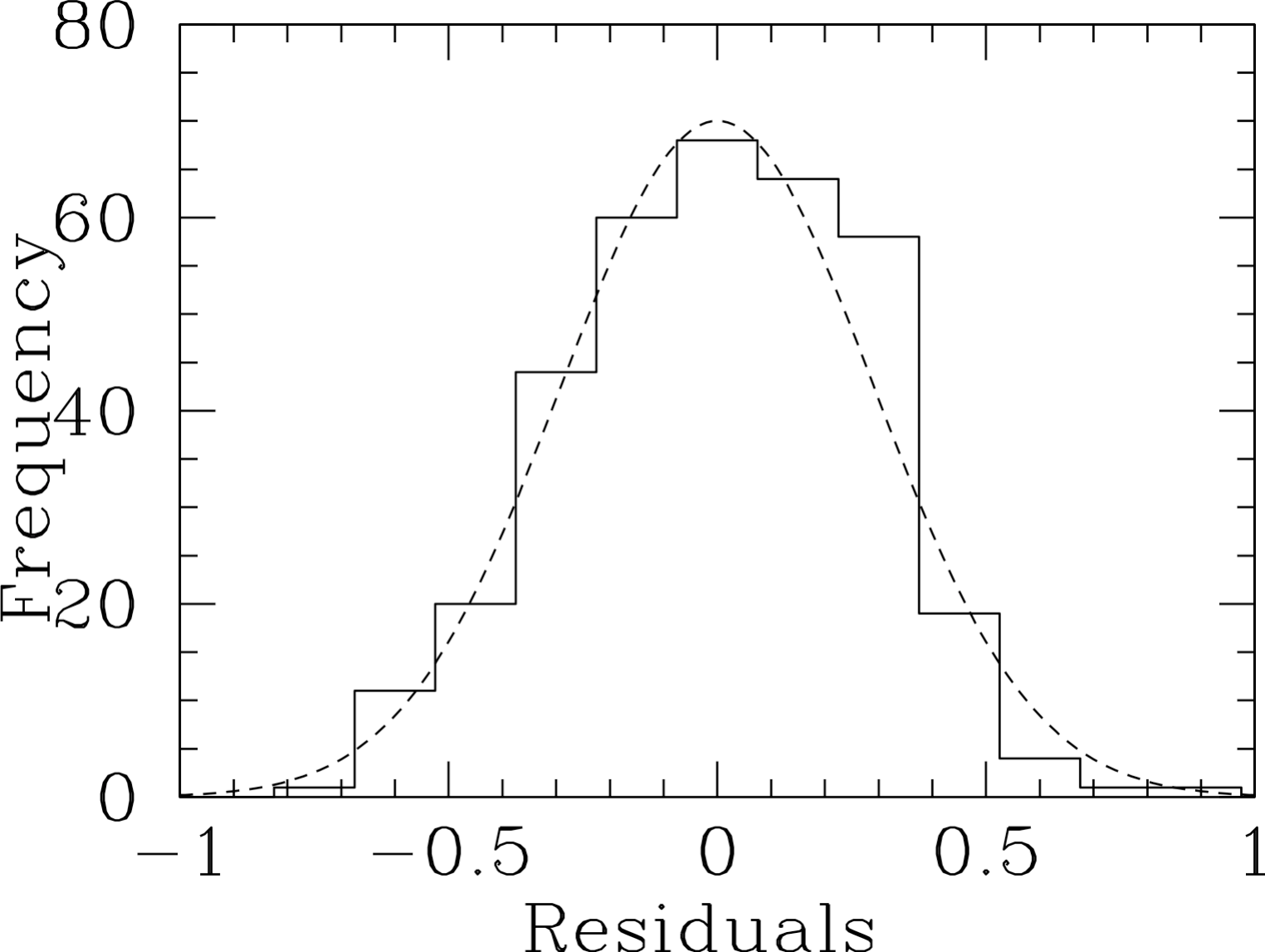}
\caption{The scatter plot {\bf (Left)} and histogram {\bf (Right)} of the residuals about the best fit line obtained with the LINMIX algorithm \citep{Kelly+2007} for the $\Sigma_{SFR}$--$\Sigma_{mol}$ data in 
Figure~\ref{fig:sklaw}, left. {\bf (Left):} The scatter plot is shown as a function of the distance along the best fit line, from an arbitrary point ($\Sigma_{mol}$=0). Positive residuals correspond to data to the top--left of the best fit line. The vertical dotted line marks the location of the stochastic sampling limit.  
The location of zero residuals is shown as a dashed horizontal line to help visualization. {\bf (Right):} A Gaussian function (dashed line) overplotted on the histogram (solid black) to guide the eye; it is not a fit to the data. The Gaussian's standard deviation is $\sigma$=0.29. 
}
\label{fig:residuals}
\end{figure}

The $\Sigma_{SFR}$--$\Sigma_{mol}$ data for the 353 HII regions in our sample scatter almost symmetrically about the LINMIX best--fit line from Table~\ref{tab:fits}, with a hint of a deviation towards high $\Sigma_{SFR}$ at a given $\Sigma_{mol}$ for bright regions (Figure~\ref{fig:residuals}, left). The excess is visible also in the histogram (Figure~\ref{fig:residuals}, right), although the overall shape is well approximated by a Gaussian function. The standard deviation of the Gaussian function (which is not a fit to the data) is $\sigma$=0.29, closer to the scatter determined from the Extended Bi--Regression and Symmetric BCES fitting algorithms than the one from the LINMIX algorithm. This larger scatter is still smaller than  the scatter measured for $\sim$2~kpc galaxy regions by \citet{Sun+2023}. 

\section{Fitting procedure for censored data}\label{sec:appendixG}

When fitting the dependence of $\epsilon_\mathrm{ff}$ on $\Sigma_\mathrm{mol}$ or other heavily censored data, we must take care to account for the effects of observational selection, particularly the lower limit on $\Sigma_\mathrm{SFR}$, which introduces a minimum measurable $\epsilon_\mathrm{ff}$ that varies as a function of the independent variable. To fit a relationship between $\epsilon_\mathrm{ff}$ and $\Sigma_\mathrm{mol}$ properly accounting for this censorship, we consider a linear model with a Gaussian scatter of the form $y = m x + b + k r_G$, where for convenience we have defined $x =\log[\Sigma_\mathrm{mol}/(\mathrm{M}_\odot\,\mathrm{pc}^{-2})]$ and $y=\log\epsilon_\mathrm{ff}$, $r_G$ is a random deviate drawn from a Gaussian distribution with zero mean and unit variance, and the slope $m$, offset $b$, and scatter $k$ are parameters to be fit from the data. In the absence of censorship we would then have the usual Gaussian likelihood function for a given $(x,y)$ measurement, given by $\mathcal{L} = (2\pi k^2)^{-1/2} \exp[-(y - mx - b)^2/2k^2]$. Censorship introduces a minimum value $y_\mathrm{min}(x)$ below which we cannot measure, indicated by the gray shaded regions in Figure~\ref{fig:eff}; the functional form describing this exclusion zone is $y_\mathrm{min} = -1.5x+0.055$. The exclusion means $y$ is limited to taking on values in the range$(y_\mathrm{min}, \infty)$ rather than $(-\infty,\infty)$, and the likelihood function at $y>y_\mathrm{min}$ must be renormalized to ensure that the integral over all $y$ for a given $x$ remains unity. It is straightforward to show that the required renormalization factor is
\begin{equation}
F = 2\left[\mathrm{erfc}\left(\frac{y_\mathrm{min}(x) - mx - b}{\sqrt{2} k}\right)\right]^{-1},
\end{equation}
and thus the likelihood function for a set of measurements $(x_i, y_i)$ with censorship is
\begin{equation}
\mathcal{L} \propto \prod_i\frac{F(x_i)}{\sqrt{2\pi k^2}}\exp\left[-\frac{\left(y_i - m x_i - b\right)^2}{2k^2}\right].
\end{equation}

We use this likelihood to fit $m$, $b$, and $k$ using the Markov Chain Monte Carlo code \textsc{emcee} (Foreman-Mackey et al.~2013); we adopt Jeffreys priors on $m$, $b$, and $k$, corresponding to $p_\mathrm{prior} \propto 1/k^2 (1+m^2)$, with the dependence on $m$ equivalent to assuming that all angles made by the line are equally likely. For each set of measurements we run the MCMC for 5000 steps using 100 walkers; we measure the autocorrelation time of the resulting chains using \texttt{emcee}'s built-in \texttt{get\_autocorr\_time} estimator, taking the longest time for any variable as our estimate, $t_\mathrm{acorr}$. We then discard the first $50t_\mathrm{acorr}$ steps for burn-in and thin the remaining steps by a factor of $t_\mathrm{acorr}/2$ to obtain our final samples. To account for the measurement uncertainties on $x$ and $y$, we generate 50 Monte Carlo realizations of the  data by drawing from the Gaussian uncertainties intervals on $x$ and $y$, and repeat the fitting procedure for each realization; we then combine the samples produced from the 50 realizations. The confidence intervals on $m$ and on the final fit line we report in section~\ref{subsec:efficiency} and Figure~\ref{fig:eff} are derived from these samples.

\section{Dependency of $\epsilon_{ff}$ on $\Sigma_{mol}$ and $\Sigma_{SFR}$}\label{sec:appendixF}

The observed dependency of $\epsilon_{ff}$ with both $\Sigma_{mol}$ and $\Sigma_{SFR}$, discussed in section~\ref{subsec:efficiency}, deviates from the expected trends that would result by simply propagating the best fit relation between $\Sigma_{SFR}$ and $\Sigma_{mol}$ (Table~\ref{tab:fits}). We find $\epsilon_{ff}\propto\Sigma_{SFR}^{0.44}$ (equation~\ref{fit_epsilon}), while the best fit in Table~\ref{tab:fits} implies $\epsilon_{ff}\propto\Sigma_{SFR}^{0.19}$, and there is no measurable correlation between $\epsilon_{ff}$ and $\Sigma_{mol}$, while the best fit implies $\epsilon_{ff}\propto\Sigma_{mol}^{0.35}$. 

In this Appendix, we investigate whether the scatter about the best fit relation (Appendix~\ref{sec:appendixC}) and the uncertainties on individual measurements play a role in the above, puzzling, result. This stems from the fact that $\epsilon_{ff}$ is a combination of  $\Sigma_{SFR}$ and $\Sigma_{mol}$, which imparts strong covariance on the data and associated uncertainties and may, therefore, affect the observed trends.

A series of 1,000 mock catalogs, each containing 400 pairs of Log($\Sigma_{mol}$) and Log($\Sigma_{SFR}$) and their uncertainties, are generated with a Monte Carlo routine, from which Log($\epsilon_{ff}$) is then calculated and the slope of each mock catalog fitted as a function of both Log($\Sigma_{mol}$) and Log($\Sigma_{SFR}$). The mock catalogs are generated with the following prescription: (a) pairs of Log($\Sigma_{SFR}$) and Log($\Sigma_{mol}$) are sampled from a distribution that follows the best fit line $y=1.85 x-4.27$ from Table~\ref{tab:fits} with {\em perpendicular (to the best fit line)} Gaussian scatter  having standard deviation $\sigma_{scatter}$=0.3 (Figure~\ref{fig:residuals}); (b) each pair is then assigned an uncertainty in both x and y, according to the distributions of measurement uncertainties in the data (including that lower luminosity data have larger uncertainties). The scatter about the best fit line and the measurement uncertainties are treated as independent contributions to the (x,y) pairs in the mock catalogs. This is because the measurement uncertainties of bright sources are several sigma smaller than the scatter of the data about the best fit line (Figure~\ref{fig:residuals}, left), indicating that the scatter does not originate entirely from the measurement uncertainties. From these pairs of Log($\Sigma_{mol}$) and Log($\Sigma_{SFR}$), Log($\epsilon_{ff}$) is then calculated via equation~\ref{eff2} and the uncertainties propagated. 

We limit the range of the fits to [$0.5; 2.6$] for Log($\Sigma_{mol}$) when determining the slopes of Log(($\epsilon_{ff}$) versus Log($\Sigma_{mol}$) and to  [$-1.8; 1$] for Log($\Sigma_{SFR}$) when the slopes are derived for Log($\epsilon_{ff}$)--vs--Log($\Sigma_{SFR}$), in agreement with the dynamical range of the data. We do not impose the combined limits to the fits, to prevent loss of dynamical range in the Log($\epsilon_{ff}$)--vs--Log($\Sigma_{mol}$) relations, which are heavily affected by the Log($\Sigma_{SFR}$)$\ge -1.8$ cutoff (Figure~\ref{fig:eff}, top--right). With these prescriptions, we can compare the slopes recovered from the mock catalogs with those derived from the data using forward modeling. The slopes recovered from the mock catalogs are shown as histograms in Figure~\ref{fig:500pc}, left.

For  both the Log($\epsilon_{ff}$)--vs--Log($\Sigma_{SFR}$) and the Log($\epsilon_{ff}$)--vs--Log($\Sigma_{mol}$) relations, the slopes recovered from the mock catalogs (labeled Exp+Scatter in Figure~\ref{fig:500pc}, left) overlap with the 1$\sigma$ range of the slopes measured from the data and are very different from the intrinsic slopes (marked as Exp and shown as downward arrows in Figure ~\ref{fig:500pc}, left).
In particular, the recovered slopes are steeper than the intrinsic slopes when Log($\Sigma_{SFR}$) is the independent variable and shallower when Log($\Sigma_{mol}$) is the independent variable. There are, therefore, major discrepancies between the intrinsic slopes and the recovered ones. 

Decreasing $\sigma_{scatter}$ from 0.3 to zero brings the recovered slopes into agreement with the intrinsic ones, while increasing  $\sigma_{scatter}$ has the opposite effect, making the slopes of Log($\epsilon_{ff}$)--vs--Log($\Sigma_{SFR}$) more positive and those of Log($\epsilon_{ff}$)--vs--Log($\Sigma_{mol}$) more negative. The uncertainties in the data, however, have a secondary impact on the recovered slopes and decreasing them to zero or increasing them by 50\% does not change the histograms of Figure~\ref{fig:500pc}, left. 

Therefore, the discrepancy between expected and recovered slopes for $\epsilon_{ff}$ is due to the scatter in the data about the best fit line (Figure~\ref{fig:residuals}), likely because the covariance between $\epsilon_{ff}$  and the independent variables alters the correlations. This can also explain why the difference between expected and recovered slopes is larger for the Log($\epsilon_{ff}$)--vs--Log($\Sigma_{mol}$) relation  than for the Log($\epsilon_{ff}$)--vs--Log($\Sigma_{SFR}$) relation:  the projection of $\sigma_{scatter}$ on the Log($\Sigma_{mol}$) axis is larger than on the Log($\Sigma_{SFR}$) one because $n > 1$. Furthermore, the dependency of $\epsilon_{ff}$ on $\Sigma_{mol}$ is stronger (slope of 1.5) than that on $\Sigma_{SFR}$ (slope of 1), exacerbating the effects of scatter/uncertainties associated with the former. 

In summary, the intrinsic relations between $\epsilon_{ff}$ and the two variables $\Sigma_{mol}$ and $\Sigma_{SFR}$ for our star--forming regions are  consistent with originating from the expectations of equation~\ref{eff2} combined with the best fits of Table~\ref{tab:fits}, i.e., to be  power laws with slightly positive exponents. 

\section{The Molecular Law of Star Formation at 500~pc Scale}\label{sec:appendixD}

The observed displacement between peaks of H$\alpha$+21~$\mu$m emission and peaks of CO emission raises the concern that ionized--gas--centered apertures of 60~pc radius, albeit larger than the typical molecular cloud's size, may miss a significant portion of the molecular gas emission. Studies of nearby galaxies by \citet{Kruijssen+2019} and \citet{Kim+2022} suggest that region sizes $\sim$0.5--1~kpc are sufficiently large to encompass both star formation and molecular gas peaks, and average out effects of separation. This is in agreement with \citet{Calzetti+2012}, who find that galaxy regions with sizes $\gtrsim$1~kpc sample the full cloud function and simply probe the clouds' filling factor.

We repeat our analysis selecting regions 500~pc in diameter as a compromise between securing a large enough area for each source and maintaining a large enough sample of sources within our HST+JWST+CO footprints to fit a SF relation. The new sources have about 70 times larger area than those in our default analysis, thus enabling us to evaluate whether region size may induce biases. The selection is performed following the same criteria used for the 60~pc radius sources. As a first step, we grow the radii at the location of the original sources, merging into a single one all sources with overlapping areas; afterwards, we add new sources as appropriate. The main difference between the 60~pc radius and the 250~pc radius sources is that while the former are often single emitting regions, the latter are most often collections of a few distinct ionized gas emitting regions. For NGC\,5194, the low S/N areas of the publicly--available CO emission map are set to zero, thus limiting the area where 250~pc radius regions can be selected to high S/N areas. We isolate a total of 128 sources within the three galaxies: 74 in NGC\,628, 41 in NGC\,5194, and 13 in NGC\,5236 (Tables~\ref{tab:NGC628large}, \ref{tab:NGC5194large} and \ref{tab:NGC5236large}). Thus, 58\% of the large area sources are from one galaxy: NGC\,628. Photometry and the derivation of physical quantities is performed as described in the main text, except that we do not apply aperture corrections to the photometric fluxes; the aperture radii selected (5\farcs55 for NGC\,628,  6\farcs80 for NGC\,5194, and 11\farcs45 for NGC\,5236) are all $>$5 times the FWHM of the CO maps, the lowest resolution maps in our sample, making aperture corrections negligible. 

\begin{figure}
\plottwo{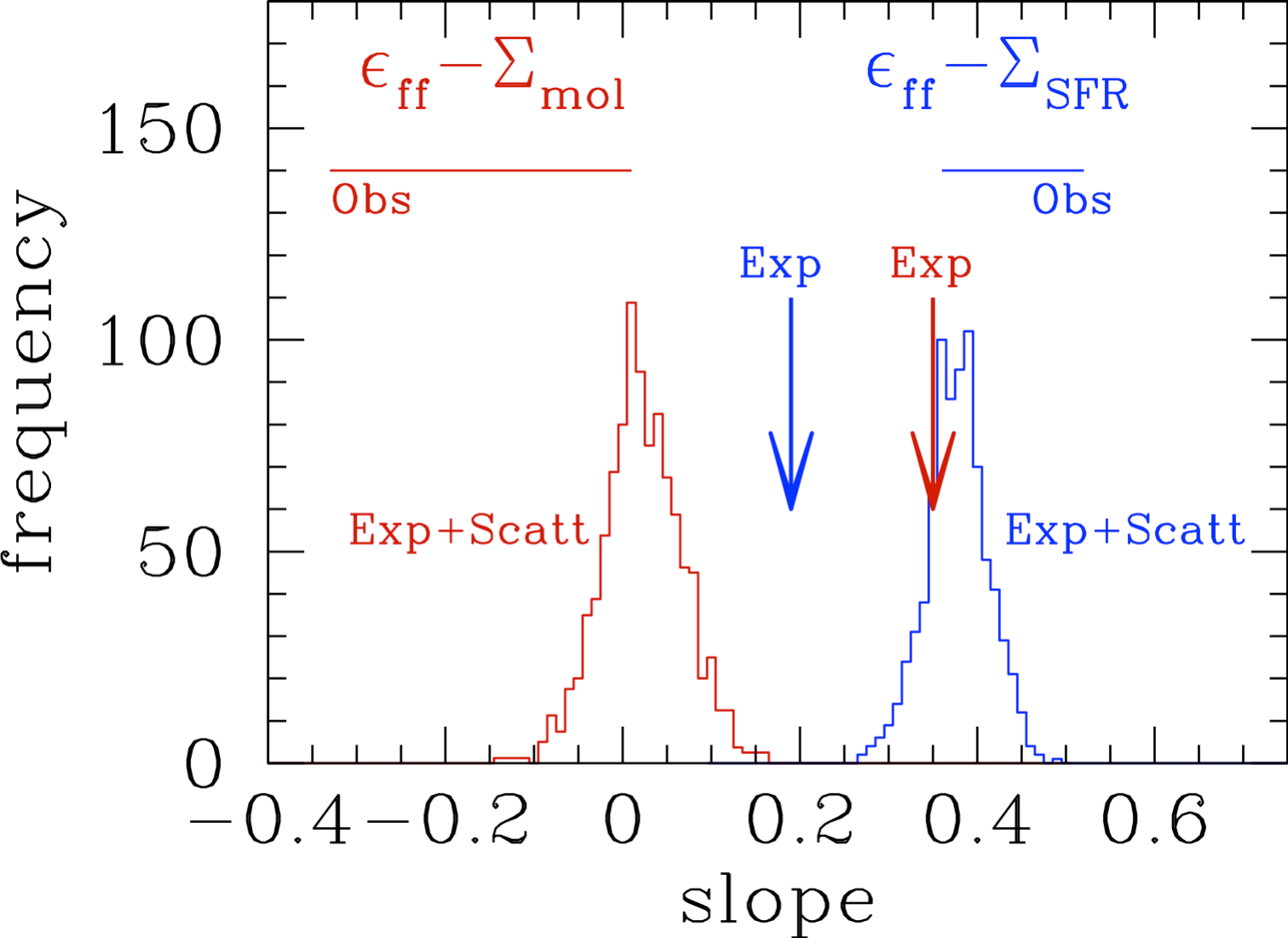}{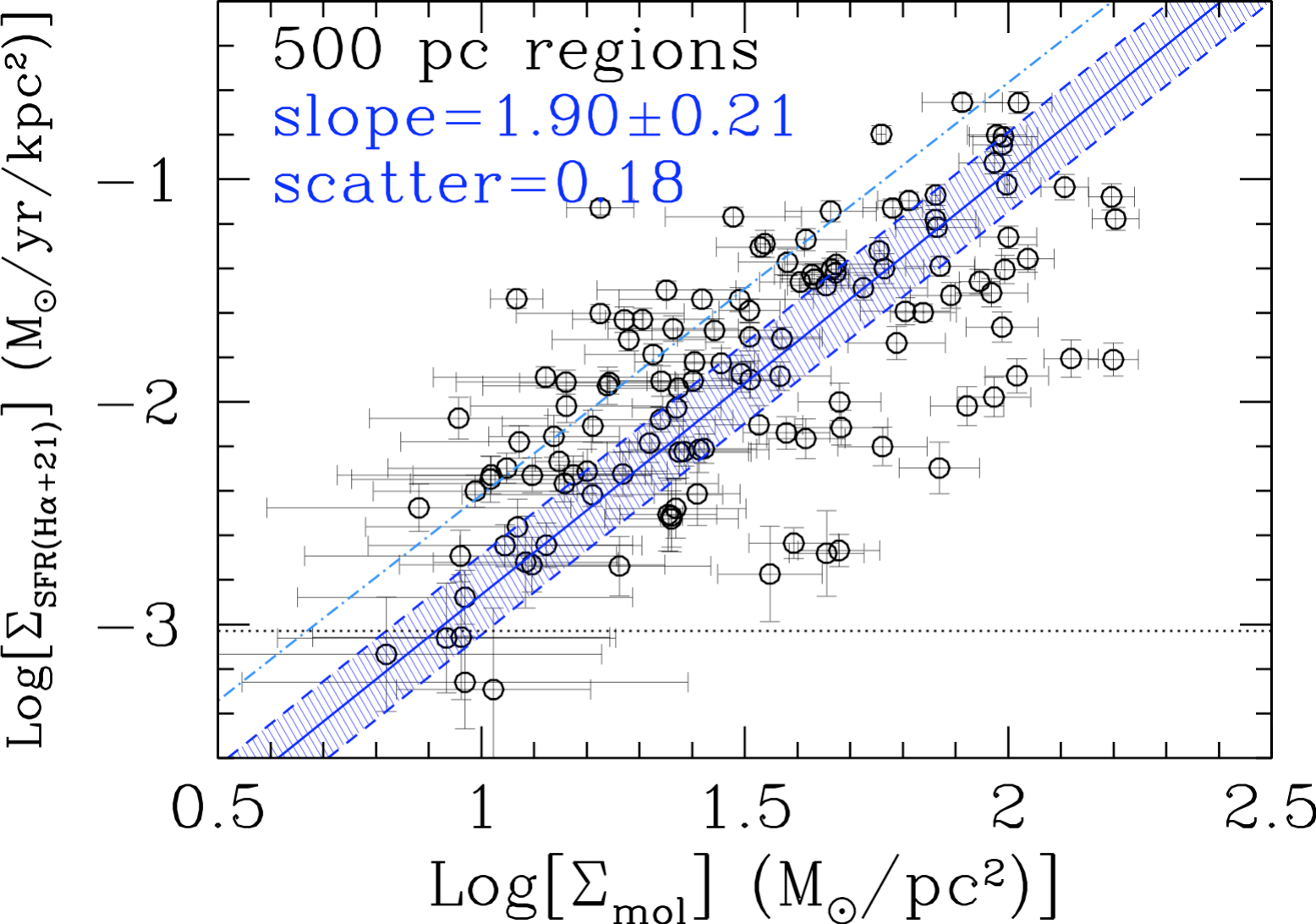}
\caption{{\bf (Left):} Histograms of the recovered slopes (Exp+Scatt) from mock catalogs of Log($\epsilon_{ff}$) as a function of Log($\Sigma_{SFR}$) (blue) and Log($\Sigma_{mol}$) (dark--red) with {\em intrinsic} slopes indicated by the downward arrows (Exp) for 1,000 realizations of $\sim$400 data each. The blue and dark--red horizontal lines (labelled Obs) mark the 1$\sigma$ range for the slopes measured from the data using the forward modeling approach (Section~\ref{sec:appendixG}). The distribution of recovered slopes from the simulations overlaps with the range of observed slopes. The deviations between the intrinsic and recovered slopes are mainly due to the scatter in the mock data, which is included to mimic the  scatter in the actual data (Figure~\ref{fig:residuals}). {\bf (Right):} The SFR surface density as a function of the molecular gas surface density for the 128 500~pc--diameter sources in our sample (black circles with 1$\sigma$ uncertainties) together with the best fit through the data (blue lines with shaded area showing the scatter). Only regions above the dotted black line, marking the location below which SFR indicators are affected by stochastic sampling of the stellar IMF, are used in the fits. The teal dashed line shows the mean value of the best fit through the 60~pc radius sources. 
}
\label{fig:500pc}
\end{figure}

The molecular gas and SF surface densities are derived applying to the 250~pc radius regions the same procedures used for the 60~pc radius regions, with the exception of the proportionality constant between H$\alpha$ and 21~$\mu$m emission. For this we use the value 0.031$\pm$0.006 instead of 0.077$\pm$0.022 as the former has been shown to be appropriate for regions of several hundred pc in size, which capture populations with SFHs of $\approx$100~Myr in duration \citep{Calzetti+2007, Calzetti+2025}. A fit of the SF relation for the 250~pc radius region yields:
\begin{equation}
Log[\Sigma_{SFR}] = (1.88\pm0.20) Log[\Sigma_{mol}] - (4.70\pm0.45),
\label{equa:sklaw500}
\end{equation}
with scatter=0.21 (Figure~\ref{fig:500pc}, right). All regions with Log[$\Sigma_{SFR}$]$> -$3.0 are used in the fit, since this is the limit above which the effects of stochastic sampling of the stellar IMF on SFR determinations is mitigated. The slope is virtually identical to that of the SF relation for the 60~pc radius region (Table~\ref{tab:fits}). The intercept is slightly lower, by a factor $\sim$2.8--2.9, than the one for the 60~pc SF relation but also consistent  within the combined 1$\sigma$ uncertainties; it is also consistent in value to the offset one expects by simply rescaling the two relations for the differences in geometry and PSF corrections between the 120~pc and 500~pc regions, as already discussed in section~\ref{subsec:nobck_removal}.

\section{The Impact of Different Choices for $\alpha_{CO}$}\label{sec:appendixE}

The choice of the expression for the CO--to--H2 conversion factor can have important effects on the resulting slope of the SF relation, which we test by adopting in this Appendix two separate assumptions for $\alpha_{CO}$: (1) the prescription by \citet{Chiang+2024}, instead of that by \citet{Bolatto+2013}, as done in \citet{Leroy+2025}; and (2) a constant $\alpha_{CO}$ value, the latter being a common assumption in the literature. 

\begin{figure}
\plottwo{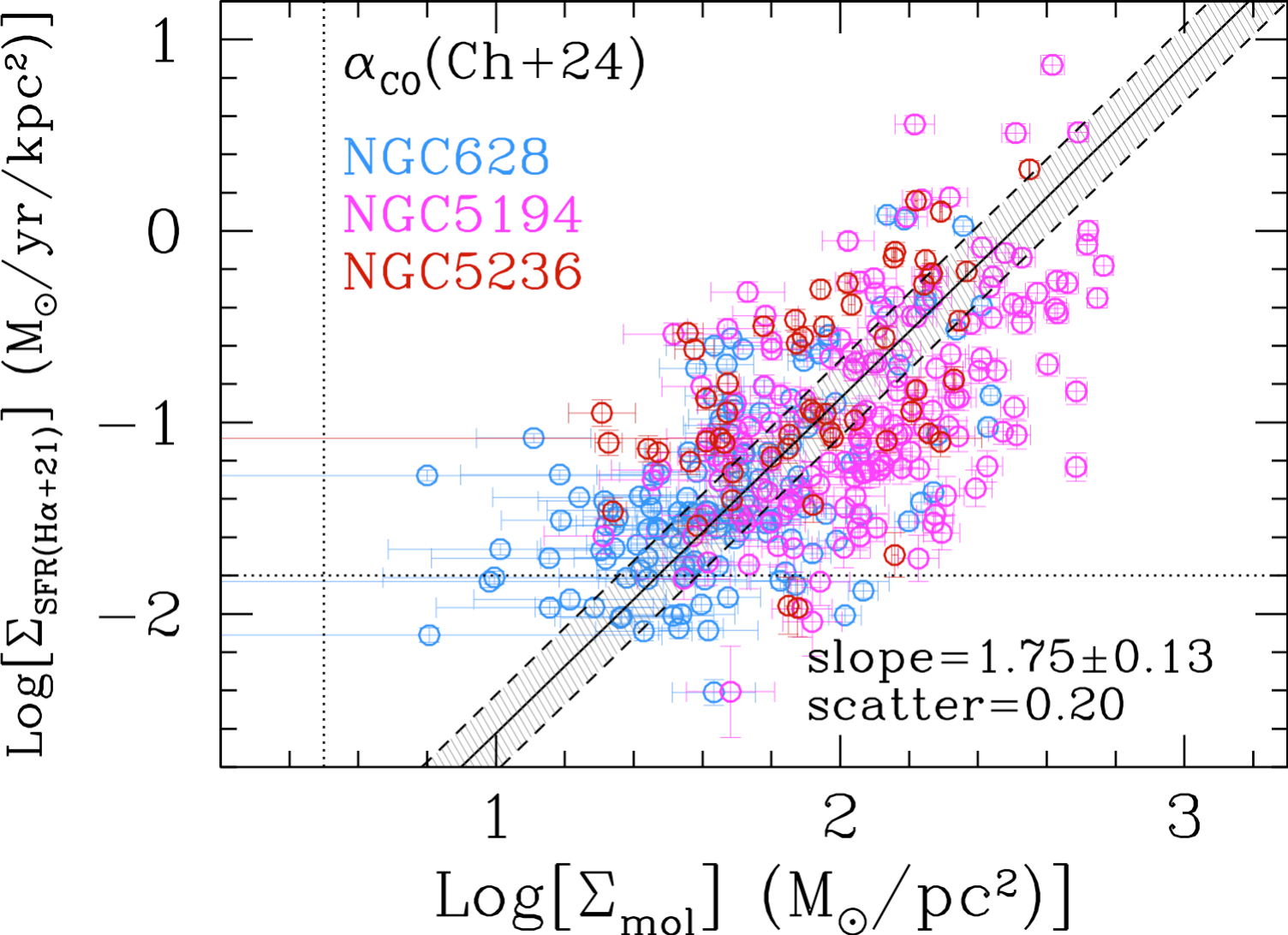}{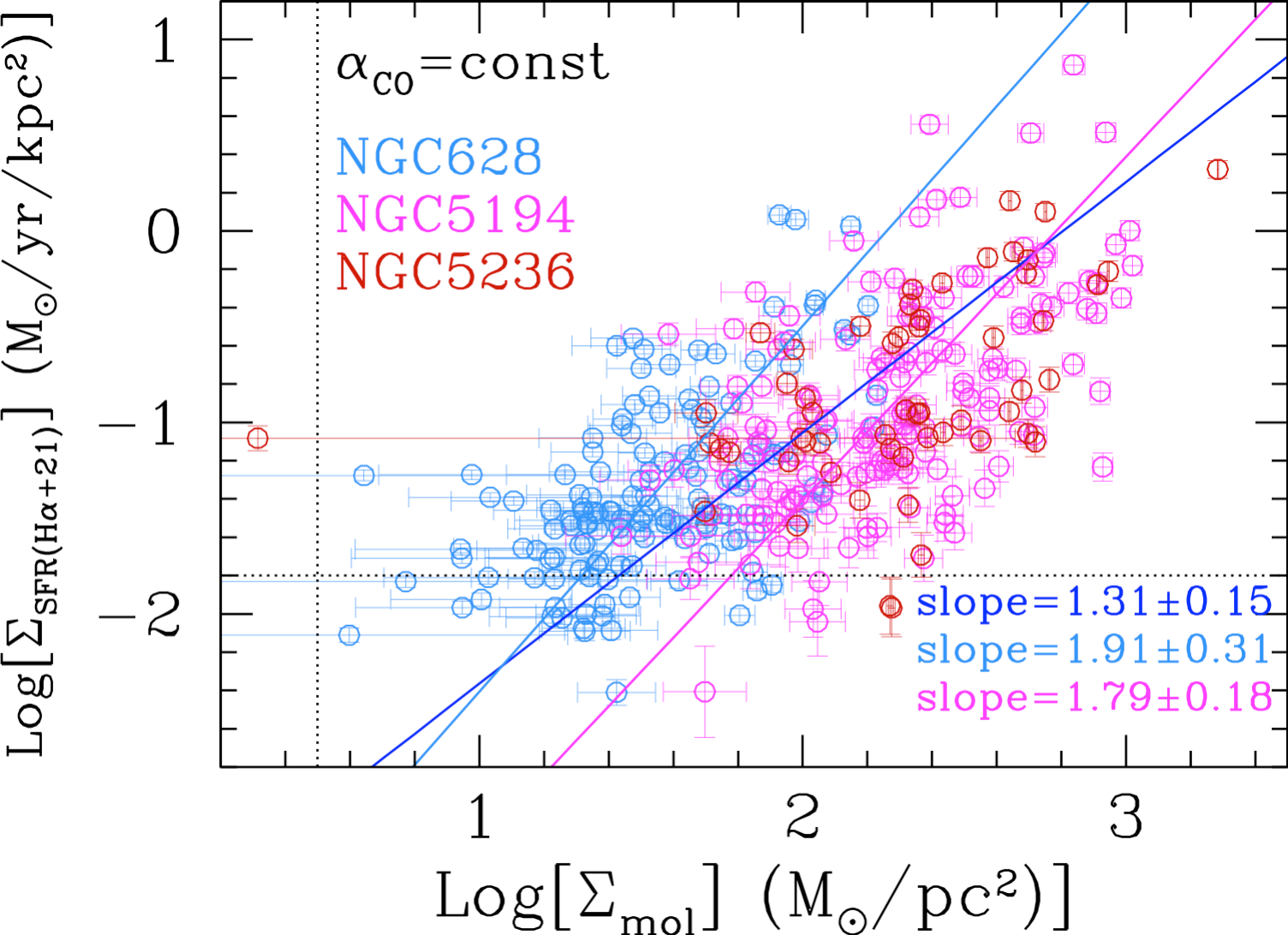}
\caption{{\bf (Left:)} The SFR surface density as a function of the molecular gas surface density for the 353 regions in our sample using the $\alpha_{CO}$ formulation of \citet{Chiang+2024}. The data for the three galaxies are shown as color circles with 1$\sigma$ uncertainties: teal for NGC\,628, magenta for NGC\,5194 and dark~red for NGC\,5236, to highlight the locus occupied by the sources in each galaxy.  The best fit through the data is shown as a black line with shaded area showing the scatter. Only regions to the top--right of the two dotted lines are used in the fits. 
{\bf (Right):} The same as the left panel, but choosing a constant, MW value for $\alpha_{CO}$ for all HII regions. The continuous lines mark: the best fits through the data for all three galaxies (blue line), the best fits through the data of NGC\,628 (teal line) and of NGC\,5194 (magenta line). The fit through the data of NGC~5236 is not reported because there are not enough sources in this galaxy for a robust fit (see text for more discussion). The fits through the sources of individual galaxies are significantly steeper than the fit through all data taken together. 
}
\label{fig:Chiang}
\end{figure}

\citet{Chiang+2024} analyze regions $\sim$2~kpc in size in a sample of nearby galaxies and derive a smaller CO(2--1)--to--CO(1--0) ratio, R$_{(21/10)}$=0.43, than the default 0.65 we adopt, and separate dependencies of $\gamma$ (see equation~\ref{alphaco}) for the two CO transitions they investigate: $\gamma$=0.48 for CO(2--1) and $\gamma$=0.22 for CO(1--0). This difference may be attributed to the large, environment--dependent variations observed for R$_{(21/10)}$ \citep{Koda+2012, Koda+2020, Koda+2025}, since each $\sim$2~kpc area averages across several environments within the galaxies. With the \citet{Chiang+2024}'s prescription for  $\gamma$ in equation~\ref{alphaco} applied to our sample of 353 HII regions, we obtain the fit:
\begin{equation}
Log[\Sigma_{SFR}] = (1.72\pm0.11) Log[\Sigma_{mol}] - (4.33\pm0.26),
\label{equa:sklawchiang24}
\end{equation}
with scatter=0.20 (Figure~\ref{fig:Chiang}, left). The slope is slightly shallower, by $\Delta n\sim$0.13 than the value n=1.85 derived with the original formulation of equation~\ref{alphaco}. This outcome is to be expected since 44\% of the HII regions in the sample come from NGC\,5194. This galaxy has been observed in CO(1--0) and the shallower $\gamma$ value for this transition in \citet{Chiang+2024} results larger $\Sigma_{mol}$ values relative to the values obtained with the formula by \citet{Bolatto+2013}. Figure~\ref{fig:Chiang}, left, identifies with colors the galaxy each HII region is drawn from, showing that the regions from NGC\,5194 tend to have slightly larger values of $\Sigma_{mol}$ than the regions from, e.g., NGC\,5236 which is a similarly active galaxy (Table~\ref{tab:properties}). Ultimately, the value of the slope we recover, $n=1.72$, is still consistent with n=1.85 within 1$\sigma$. Furthermore, we do not expect HII regions  to be the result of an average of different environments like 2~kpc galaxy regions, which imply that the value of $\gamma$ may be more uniform for different CO transitions in HII regions.

Adopting a more extreme prescription for $\alpha_{CO}$ by imposing $\alpha_{CO(1-0)}$=constant=4.35 everywhere, we get an even shallower slope, $n\sim$1.30 with a scatter=0.25, for R$_{(21/10)}$=0.65 (Figure~\ref{fig:Chiang}, right); however, the shallow slope is entirely driven by a $\sim$0.5~dex horizontal offset in $\Sigma_{mol}$ between NGC\,628 and the other two, more strongly star--forming, galaxies. The HII regions of NGC\,628 are, in fact, less bright in CO than those in both NGC\,5194 and NGC\,5236, for the same value of $\Sigma_{SFR}$. When fits are performed on the regions of individual galaxies, the resulting slopes are much steeper than the one for the cumulative fit, being $n\sim 1.9$ for NGC\,628 and $n\sim 1.8$ for NGC\,5194. The third galaxy, NGC\,5236 contains too few regions for a reliable fit, but when the sources in this galaxy are fit together with those of NGC\,5194, the slope is $n\sim$1.75. Overall, the best fit slopes for the individual galaxies are consistent with the average slope $n\sim 1.85$ for the entire sample when applying  equation~\ref{alphaco} to derive $\Sigma_{mol}$. Thus, the main effect of equation~\ref{alphaco} is to remove the bulk offset in CO brightness between different galaxies, rather than steepening the $\Sigma_{SFR}$--vs.--$\Sigma_{mol}$ relation within individual galaxies. The offset is only minimally due to the galaxy--to--galaxy metallicity differences for these targets (Table~\ref{tab:properties}), which correspond to differences in $\alpha_{CO}$, $\lesssim$0.06~dex. The majority of the offset in $\Sigma_{mol}$ is due to differences in the average weight of the galaxies' disks. This result  highlights that the CO--to--H2 conversion is highly sensitive to variations in local conditions \citep[temperature, density, disk weight, dynamical effects, etc.; e.g.,][]{Narayanan+2012, Ackermann+2012, Israel+2020} and neglecting that sensitivity flattens the $\Sigma_{SFR}$--vs.--$\Sigma_{mol}$ relation.

\newpage

\startlongtable
\begin{deluxetable}{lrrrrrrr}
\tablecolumns{8}
\tabletypesize{\small}
\tablecaption{Location and Derived Quantities for the 120~pc Regions in NGC\,628\label{tab:NGC628}}
\tablewidth{120pt}
\tablehead{
\colhead{ID} & \colhead{RA(2000),DEC(2000)}  & \colhead{Log($\Sigma_{mol}$)} & \colhead{Log($\Sigma_{SFR}$)} & \colhead{Log($\Sigma_{*,gal}$)} & \colhead{A$_V$} & \colhead{Log($\tau_{dep}$)} & \colhead{Log($\epsilon_{ff}$)} 
\\
\colhead{(1)} & \colhead{(2)} & \colhead{(3)} & \colhead{(4)} & \colhead{(5)}  & \colhead{(6)} & \colhead{(7)} & \colhead{(8)} 
\\
}
\startdata
   1 &1:36:47.1993, +15:45:50.101 &  2.0692$\pm$0.0260 & -0.3874$\pm$0.0276 &  1.6721$\pm$0.0779 &  1.467$\pm$0.035  & 8.4566$\pm$0.0379 & -1.6214$\pm$0.0421\\
   2 &1:36:47.2742, +15:45:53.541 &  2.0942$\pm$0.0260 & -1.3647$\pm$0.0306  & 1.4647$\pm$0.0889 &  1.475$\pm$0.040  & 9.4589$\pm$0.0402 & -2.6362$\pm$0.0442\\
   3 &1:36:46.9416, +15:45:47.061  & 1.3669$\pm$0.1345 & -1.5302$\pm$0.0236  & 1.4896$\pm$0.0874 &  0.545$\pm$0.034  & 8.8971$\pm$0.1365 & -1.7107$\pm$0.1664\\
\hline
\enddata
\tablenotetext{}{(1) The identification number of the source, for a total of 143 sources.}
\tablenotetext{}{(2) Right Ascension  and Declination in J2000 coordinates.}
\tablenotetext{}{(3)--(5) Logarithm (base 10) of the surface density of molecular gas (units: M$_{\odot}$~pc$^{-2}$), SFR (units: M$_{\odot}$~yr$^{-1}$~kpc$^{-2}$), and stellar mass (units: M$_{\odot}$~pc$^{-2}$) of each 120~pc--diameter region. See text for details.}
\tablenotetext{}{(6) The V--band dust attenuation, A$_V$ (units: mag), derived from the 21~$\mu$m/H$\alpha$ luminosity ratio.}
\tablenotetext{}{(7)--(8) Logarithm (base 10) of the depletion timescale, $\tau_{dep}$ (units: yr), and the efficiency per free--fall time, $\epsilon_{ff}$ (adimensional).}
\tablecomments{Values below $Log(\Sigma_{SFR})=-1.8$ and $Log(\Sigma_{mol})=0.5$ and the associated quantities should be considered not reliable. The full table is available in Machine Readable format.}
\end{deluxetable}

\startlongtable
\begin{deluxetable}{lrrrrrrr}
\tablecolumns{8}
\tabletypesize{\small}
\tablecaption{Location and Derived Quantities for the  120~pc Regions in NGC\,5194\label{tab:NGC5194}}
\tablewidth{120pt}
\tablehead{
\colhead{ID} & \colhead{RA(2000),DEC(2000)}  & \colhead{Log($\Sigma_{mol}$)} & \colhead{Log($\Sigma_{SFR}$)} & \colhead{Log($\Sigma_{*,gal}$)} & \colhead{A$_V$} & \colhead{Log($\tau_{dep}$)} & \colhead{Log($\epsilon_{ff}$)} 
\\
\colhead{(1)} & \colhead{(2)} & \colhead{(3)} & \colhead{(4)} & \colhead{(5)}  & \colhead{(6)} & \colhead{(7)} & \colhead{(8)} 
\\
}
\startdata
   1 &13:29:52.1548 +47:12:44.750   &2.0563$\pm$0.0562  & 0.1637$\pm$0.0448   &2.5306$\pm$0.0514   &2.126$\pm$0.058   &7.8926$\pm$0.0719  &-1.0672$\pm$0.0822\\
   2 &13:29:52.0449 +47:12:47.270   &1.8871$\pm$0.0755  &-0.0515$\pm$0.0437   &2.3475$\pm$0.0555   &1.728$\pm$0.058   &7.9386$\pm$0.0872  &-1.0286$\pm$0.1023\\
   3 &13:29:52.3785 +47:12:38.550   &2.2482$\pm$0.0443  &-0.2928$\pm$0.0569   &2.4121$\pm$0.0540   &3.725$\pm$0.077   &8.5410$\pm$0.0721  &-1.8115$\pm$0.0787\\
\hline
\enddata
\tablenotetext{}{(1) The identification number of the source, for a total of 154 sources.}
\tablenotetext{}{(2) Right Ascension  and Declination in J2000 coordinates.}
\tablenotetext{}{(3)--(5) Logarithm (base 10) of the surface density of molecular gas (units: M$_{\odot}$~pc$^{-2}$), SFR (units: M$_{\odot}$~yr$^{-1}$~kpc$^{-2}$), and stellar mass (units: M$_{\odot}$~pc$^{-2}$) of each 120~pc--diameter region. See text for details.}
\tablenotetext{}{(6) The V--band dust attenuation, A$_V$ (units: mag), derived from the 21~$\mu$m/H$\alpha$ luminosity ratio.}
\tablenotetext{}{(7)--(8) Logarithm (base 10) of the depletion timescale, $\tau_{dep}$ (units: yr), and the efficiency per free--fall time, $\epsilon_{ff}$ (adimensional).}
\tablecomments{Values below $Log(\Sigma_{SFR})=-1.8$ and $Log(\Sigma_{mol})=0.5$ and the associated quantities should be considered not reliable. The full table is available in Machine Readable format.}
\end{deluxetable}

\startlongtable
\begin{deluxetable}{lrrrrrrr}
\tablecolumns{8}
\tabletypesize{\small}
\tablecaption{Location and Derived Quantities for the  120~pc Regions in NGC\,5236\label{tab:NGC5236}}
\tablewidth{120pt}
\tablehead{
\colhead{ID} & \colhead{RA(2000),DEC(2000)}  & \colhead{Log($\Sigma_{mol}$)} & \colhead{Log($\Sigma_{SFR}$)} & \colhead{Log($\Sigma_{*,gal}$)} & \colhead{A$_V$} & \colhead{Log($\tau_{dep}$)} & \colhead{Log($\epsilon_{ff}$)} 
\\
\colhead{(1)} & \colhead{(2)} & \colhead{(3)} & \colhead{(4)} & \colhead{(5)}  & \colhead{(6)} & \colhead{(7)} & \colhead{(8)} 
\\
}
\startdata
   1 &13:36:59.8177 -29:52:22.717  & 2.0403$\pm$0.0036  &-0.2798$\pm$0.0451   &3.6142$\pm$0.0240   &1.888$\pm$0.055   &8.3201$\pm$0.0452  &-1.4936$\pm$0.0453\\
   2 &13:37:00.0914 -29:52:17.477  & 2.3445$\pm$0.0016  &0.3220$\pm$0.0457    &3.7394$\pm$0.0232   &3.860$\pm$0.056   &8.0225$\pm$0.0457  &-1.3481$\pm$0.0457\\
   3 &13:36:57.4739 -29:52:43.591   &1.6110$\pm$0.0110  &-1.1813$\pm$0.0663    &3.2875$\pm$0.0264   &1.462$\pm$0.082   &8.7923$\pm$0.0672  &-1.7512$\pm$0.0676\\
\hline
\enddata
\tablenotetext{}{(1) The identification number of the source, for a total of 56 sources.}
\tablenotetext{}{(2) Right Ascension  and Declination in J2000 coordinates.}
\tablenotetext{}{(3)--(5) Logarithm (base 10) of the surface density of molecular gas (units: M$_{\odot}$~pc$^{-2}$), SFR (units: M$_{\odot}$~yr$^{-1}$~kpc$^{-2}$), and stellar mass (units: M$_{\odot}$~pc$^{-2}$) of each 120~pc--diameter region. Regions with assigned $Log(\Sigma_{*,gal})$=3.05 are those outside the JWST/NIRCam/F330M footprint, and are not given an uncertainty. See text for details.}
\tablenotetext{}{(6) The V--band dust attenuation, A$_V$ (units: mag), derived from the 21~$\mu$m/H$\alpha$ luminosity ratio.}
\tablenotetext{}{(7)--(8) Logarithm (base 10) of the depletion timescale, $\tau_{dep}$ (units: yr), and the efficiency per free--fall time, $\epsilon_{ff}$ (adimensional).}
\tablecomments{Values below $Log(\Sigma_{SFR})=-1.8$ and $Log(\Sigma_{mol})=0.5$ and the associated quantities should be considered not reliable. The full table is available in Machine Readable format.}
\end{deluxetable}

\startlongtable
\begin{deluxetable}{lrrrr}
\tablecolumns{5}
\tabletypesize{\small}
\tablecaption{Location and Derived Quantities for the 500~pc Regions in NGC\,628\label{tab:NGC628large}}
\tablewidth{100pt}
\tablehead{
\colhead{ID} & \colhead{RA(2000),DEC(2000)}  & \colhead{Log($\Sigma_{mol}$)} & \colhead{Log($\Sigma_{SFR}$)}  & \colhead{Log($\Sigma_{*,gal}$)} 
\\
\colhead{(1)} & \colhead{(2)} & \colhead{(3)} & \colhead{(4)} & \colhead{(5)}  
\\
}
\startdata
   1 &1:36:37.6639  15:48:21.665    &1.6158$\pm$00.0769  &-1.2709$\pm$00.0494   &1.6092$\pm$00.0809\\
   2 &1:36:36.9323  15:48:04.704    &1.8113$\pm$00.0593  &-1.0955$\pm$00.0378   &1.6198$\pm$00.0804\\
   3 &1:36:36.0595  15:47:48.042    &1.6788$\pm$00.0776  &-2.6677$\pm$00.0724   &1.4484$\pm$00.0899\\
\hline
\enddata
\tablenotetext{}{(1) The identification number of the source, for a total of 74 sources.}
\tablenotetext{}{(2) Right Ascension  and Declination in J2000 coordinates.}
\tablenotetext{}{(3)--(5) Logarithm (base 10) of the surface density of  molecular gas (units: M$_{\odot}$~pc$^{-2}$), SFR (units: M$_{\odot}$~yr$^{-1}$~kpc$^{-2}$), and stellar mass (units: M$_{\odot}$~pc$^{-2}$)) of each 500~pc--diameter region. See text for details.}
\tablecomments{Values below $Log(\Sigma_{SFR})=-3.0$ and $Log(\Sigma_{mol})=0.5$ should be considered not reliable. The full table is available in Machine Readable format.}
\end{deluxetable}

\startlongtable
\begin{deluxetable}{lrrrr}
\tablecolumns{5}
\tabletypesize{\small}
\tablecaption{Location and Derived Quantities for the 500~pc  Regions in NGC\,5194\label{tab:NGC5194large}}
\tablewidth{100pt}
\tablehead{
\colhead{ID} & \colhead{RA(2000),DEC(2000)}  & \colhead{Log($\Sigma_{mol}$)} & \colhead{Log($\Sigma_{SFR}$)} & \colhead{Log($\Sigma_{*,gal}$)}  
\\
\colhead{(1)} & \colhead{(2)} & \colhead{(3)} & \colhead{(4)} & \colhead{(5)}  
\\
}
\startdata
   1 &13:30:01.2501 +47:12:49.757   &2.0193$\pm$0.0635  &-0.6558$\pm$0.0489   &2.3175$\pm$0.0562\\
   2 &13:30:01.4380 +47:12:37.596   &1.8048$\pm$0.0856  &-1.5930$\pm$0.0611   &2.3033$\pm$0.0565\\
   3 &13:29:53.2414 +47:12:38.574   &1.8616$\pm$0.0767  &-1.1817$\pm$0.0452   &2.3952$\pm$0.0544\\
\hline
\enddata
\tablenotetext{}{(1) The identification number of the source, for a total of 41 sources.}
\tablenotetext{}{(2) Right Ascension  and Declination in J2000 coordinates.}
\tablenotetext{}{(3)--(5) Logarithm (base 10) of the surface density of  molecular gas (units: M$_{\odot}$~pc$^{-2}$), SFR (units: M$_{\odot}$~yr$^{-1}$~kpc$^{-2}$), and stellar mass (units: M$_{\odot}$~pc$^{-2}$)) of each 500~pc--diameter region. See text for details.}
\tablecomments{Values below $Log(\Sigma_{SFR})=-3.0$ and $Log(\Sigma_{mol})=0.5$ should be considered not reliable. The full table is available in Machine Readable format.}
\end{deluxetable}

\startlongtable
\begin{deluxetable}{lrrrr}
\tablecolumns{5}
\tabletypesize{\small}
\tablecaption{Location and Derived Quantities for the 500~pc Regions in NGC\,5236\label{tab:NGC5236large}}
\tablewidth{100pt}
\tablehead{
\colhead{ID} & \colhead{RA(2000),DEC(2000)}  & \colhead{Log($\Sigma_{mol}$)} & \colhead{Log($\Sigma_{SFR}$)} & \colhead{Log($\Sigma_{*,gal}$)} 
\\
\colhead{(1)} & \colhead{(2)} & \colhead{(3)} & \colhead{(4)} & \colhead{(5)}  
\\
}
\startdata
   1 &13:36:59.4405 -29:52:22.631   &1.9975$\pm$0.0045  &-1.0237$\pm$0.0592   &3.5007$\pm$0.0248\\
   2 &13:36:58.1370 -29:52:37.102   &1.3609$\pm$0.0208  &-2.5127$\pm$0.1603   &3.3185$\pm$0.0262\\
   3 &13:36:52.8888 -29:51:50.893   &1.7591$\pm$0.0091  &-0.7976$\pm$0.0387   &3.0500$\pm$......\\
\hline
\enddata
\tablenotetext{}{(1) The identification number of the source, for a total of 13 sources.}
\tablenotetext{}{(2) Right Ascension  and Declination in J2000 coordinates.}
\tablenotetext{}{(3)--(5) Logarithm (base 10) of the surface density of  molecular gas (units: M$_{\odot}$~pc$^{-2}$), SFR (units: M$_{\odot}$~yr$^{-1}$~kpc$^{-2}$), and stellar mass (units: M$_{\odot}$~pc$^{-2}$)) of each 500~pc--diameter region. Regions with assigned $Log(\Sigma_{*,gal})$=3.05 are those outside the JWST/NIRCam/F330M footprint, and are not given an uncertainty. See text for details.}
\tablecomments{Values below $Log(\Sigma_{SFR})=-3.0$ and $Log(\Sigma_{mol})=0.5$  should be considered not reliable. The full table is available in Machine Readable format.}
\end{deluxetable}

\bibliographystyle{aasjournal}
\bibliography{bibliography_SKLaw}{}
\end{document}